\documentclass[useAMS,usenatbib,usegraphicx]{mn2e}
\usepackage{times}
\usepackage{color}
\newcommand{\co}{\mbox{$^{12}$CO}}
\newcommand{\coa}{\mbox{$^{13}$CO}}

\newcommand{\kms}{\mbox{km s$^{-1}$}}
\newcommand{\kkms}{\mbox{K km s$^{-1}$}}

\newcommand{\Msun}{\mbox{M$_\odot$}}
\newcommand{\eff}{\mbox{$\epsilon_{\rm ff}$}}

\newcommand\arcdeg{\mbox{$^\circ$}}%
\title[Outflows in The Taurus Molecular Cloud]
{Molecular Outflows Identified in the FCRAO CO Survey
  of the Taurus Molecular Cloud}
\author[Narayanan et al.]{Gopal Narayanan$^{1}$, 
Ronald Snell$^{1}$, and 
Ashley Bemis$^{1,2}$
\\ \\
$^1$ Dept. of Astronomy, Univ. of Massachusetts, Amherst MA
  01003\\
$^2$ Department of Physics and Astronomy, Bonn University,
  Wegelerstrasse 8, 53115 Bonn, Germany\\
}

\begin{document}

\date{Received 2012 June 14; in original form 2012 April 23; accepted 2012 June 22}
\maketitle
\label{firstpage}
\begin{abstract}

Jets and outflows are an integral part of the star formation
process. While there are many detailed studies of molecular outflows
towards individual star-forming sites, few studies have surveyed an
entire star-forming molecular cloud for this phenomenon. The 100
square degree FCRAO CO survey of the Taurus molecular cloud provides
an excellent opportunity to undertake an unbiased survey of a large,
nearby, molecular cloud complex for molecular outflow activity. Our
study provides information on the extent, energetics and frequency of
outflows in this region, which are then used to assess the impact of
outflows on the parent molecular cloud. The search identified 20
outflows in the Taurus region, 8 of which were previously
unknown. Both \co\ and \coa\ data cubes from the Taurus molecular map
were used, and dynamical properties of the outflows are derived. Even
for previously known outflows, our large-scale maps indicate that many
of the outflows are much larger than previously suspected, with eight
of the flows (40\%) being more than a parsec long. The mass, momentum
and kinetic energy from the 20 outflows are compared to the repository
of turbulent energy in Taurus. Comparing the energy deposition rate
from outflows to the dissipation rate of turbulence, we conclude that
outflows by themselves cannot sustain the observed turbulence seen in
the entire cloud. However, when the impact of outflows is studied in
selected regions of Taurus, it is seen that locally, outflows can
provide a significant source of turbulence and feedback. The L1551
dark cloud which is just south of the main Taurus complex was not
covered by this survey, but the outflows in L1551 have much higher
energies compared to the outflows in the main Taurus cloud. In the
L1551 cloud, outflows can not only account for the turbulent energy
present, but are probably also disrupting their parent cloud. We
conclude that for a molecular cloud like Taurus, a L1551-like episode
occuring once every $10^5$ years is sufficient to sustain the
turbulence observed.  Five of the eight newly discovered outflows have
no known associated stellar source, indicating that they may be
embedded Class 0 sources. In Taurus, 30\% of Class I sources and 12\%
of Flat spectrum sources from the Spitzer YSO catalogue have outflows,
while 75\% of known Class 0 objects have outflows. Overall, the
paucity of outflows in Taurus compared to the embedded population of
Class I and Flat Spectrum YSOs indicate that molecular outflows are a
short-lived stage marking the youngest phase of protostellar life. The
current generation of outflows in Taurus highlights an ongoing period
of active star-formation, while a large fraction of YSOs in Taurus has
evolved well past the Class I stage.

\end{abstract}

\section{Introduction}

From the emergence of a hydrostatic core, the collapse of a
proto-stellar core is accompanied by winds and mass loss
\citep{lada1985} probably driven by magnetospheric accretion
\citep[eg.][]{koenigl1991,edwards1994,hartmann1994}.  Integrated over
time, the effect of a wind from a young star is to blow away the
placental material left over from its birth and which shrouds it during
its earliest evolution. The discovery of bipolar molecular outflows
has been a key to the understanding of this process
\citep[eg.][]{snell1980}. These molecular outflows have dimensions of
up to several parsecs, masses comparable or more than than their
driving sources, and tremendous kinetic energies, typically 10$^{45}$
ergs \citep{bally1983,snell1987}. Such massive flows must represent swept-up
material as the winds, emerging from the star and/or its circumstellar
disk, interact with their ambient medium.

While jets and outflows are an integral part of the star formation
process, only a few studies out of the many detailed studies of
molecular outflows towards individual star-forming sites, have a
molecular cloud-wide view of this phenomenon. One of these is the
recent study of outflows in Perseus \citep{arce2010}. The Taurus
molecular cloud with its proximity (140 pc) and displacement from the
Galactic Plane (b$\sim -19^\circ$) affords high spatial resolution
views of an entire star forming region with little or no confusion
from background stars and gas.
%The 100 square degree FCRAO CO
%survey of the Taurus molecular cloud
%\citep{narayanan2008,goldsmith2008} 
%provides an excellent opportunity
%to undertake an unbiased survey of a large, nearby, molecular cloud
%complex for molecular outflow activity.  
The most complete inventory of the molecular gas content within the
Taurus cloud is provided by \citet{ungerechts1987}, who observed
\co\ J=1-0 emission from 750 deg$^2$ of the Taurus-Auriga-Perseus
regions. They estimate the molecular mass resident within the
Taurus-Auriga cloud to be $3.5\times 10^4$~\Msun. However, the
$30^\prime$ angular resolution of this survey precludes an examination
of the small scale structure of the cloud. The recently completed 100
square degree FCRAO CO survey with an angular resolution of
45$^{\prime\prime}$, sampled on a 20$^{\prime\prime}$ grid, and
covered in \co\ and \coa\ simultaneously, reveals a very complex,
highly structured cloud morphology with an overall mass of $2.4\times
10^4$~\Msun\ \citep{narayanan2008,goldsmith2008}. This survey provides
an excellent opportunity to perform an unbiased survey of a large,
nearby, molecular cloud complex for molecular outflow
activity. \citet{goldsmith2008} divide the Taurus cloud into eight
regions of high column density that include the L1495, B213, L1521,
Heiles Cloud 2, L1498, L1506, B18, and L1536, and tabulate the masses
and areas of these well-known regions. Of these, the L1495 and B213
clouds have been recently studied using JCMT HARP \co\ J=3-2
observations, searching for molecular outflows from young stars
\citep{davis2010}, where they have detected as many as 16 outflows.

Targeted studies with higher angular resolution of \coa\ and C$^{18}$O
emission from individual sub-clouds of Taurus reveal some of the
relationships between the molecular gas, magnetic fields, and star
formation but offer little insight to the coupling of these structures
to larger scales and features, nor do they provide an unbiased search
for molecular outflows
\citep{schloerb1984,heyer1987,mizuno1995,onishi1996}. A list of $\sim
300$ Young Stellar Objects (YSOs) derived from multi-wavelength
observations have been compiled by \citet{kenyon1995}, and this list
was compared to the column density distribution of molecular gas by
\citet{goldsmith2008}. This list of young stars has been recently
complemented by observations from the Spitzer Space Telescope
\citep{rebull2010,luhman2010}. A comprehensive review of the entire
Taurus region is provided by \citet{kenyon2008}. In this paper, we use
the up-to-date list of young stars from the Spitzer observations, and
the list from \citet{kenyon2008}. A list of Molecular Hydrogen
emission-line Objects (MHOs) that includes the Taurus region has also
been recently compiled \citep{davis2010b}, and this list is also
compared against our data in this paper.

It should be noted that the FCRAO Taurus Molecular Cloud survey has an
overall {\it rms} uncertainty (in T$_A^*$~K units) of $\sim 0.58$~K
and $\sim 0.26$~K in in \co\ and \coa\ transitions respectively
\citep{narayanan2008}. Detecting all outflow sources in Taurus was
never a goal of this survey, nevertheless, the {\it rms} sensitivity
levels reached is indeed lower than the original goals of the project,
and this allows us to use this survey to probe for outflows. In this
paper, we provide an unbiased search for outflows in the FCRAO Taurus
survey, identify their driving sources, evaluate outflow properties,
and compare these properties with those of the driving sources, and
associated molecular cloud environments. The structure of this paper
is as follows. In \S\ref{observations}, we summarise the observations
and describe our data processing and analysis with respect to the
search criteria to detect outflows, and subsequent analysis of their
properties. In \S\ref{results}, we summarise the main results, and
tabulate the properties of the detected outflows. In
\S\ref{discussion}, we compare the outflows against the energetics of
the driving sources, and the cloud environment. In
\S\ref{conclusions}, we summarise our results.

\section{Observations and Data Processing}
\subsection{Observations}
\label{observations}

Details of the observations, data collection and data processing steps
of the original Taurus Molecular Cloud survey are presented in
\citet{narayanan2008}. Here, for completeness, we describe the salient
details of the dataset. The observations were taken with the 14~meter
diameter millimetre-wavelength telescope and the 32 pixel focal plane
array SEQUOIA \citep{erickson1999} of the Five College Radio Astronomy
Observatory.  The FWHM beam sizes of the telescope at the observed
frequencies are 45\arcsec\ (115.271202 GHz) and 47\arcsec\ (110.201353
GHz). The main beam efficiencies at these frequencies are 0.45 and
0.50 respectively as determined from measurements of Jupiter. Previous
measurements of the extended error beam of the telescope and radome
structure were performed by measuring the disks of the sun and moon,
and indicate that there can be $\sim 25$\% net contribution from
extended emission outside the main beam from a region $\sim
0.5\arcdeg$ in diameter. The shape of this error beam is approximately
circular, but the amount of contribution of emission at any given
point from this error beam pattern depends on details of the
distribution of the emission from the source, however this
contribution is expected to be negligible for outflows. All data
presented here are in T$_A^*$ (K), uncorrected for telescope beam
efficiencies. The backends were comprised of a system of 64
auto-correlation spectrometers each configured with 25 MHz bandwidth
and 1024 spectral channels.  No smoothing was applied to the
auto-correlation function so the spectral resolution was 29.5 kHz per
channel corresponding to 0.076 \kms\ (\co) and 0.080 \kms (\coa) at
the observed frequencies.
%However, we apply some smoothing spectrally
%to increase our sensitivity to low-lying line wing emission in some
%objects. When such smoothing is performed, it will be referred to at
%appropriate place, when the results are presented in
%\S\ref{results}. 
The total coverage in velocity is 65 \kms\ (\co) and 68 \kms\ (\coa)
respectively. The spectrometers were centred at a v$_{\rm LSR}$ of 6
\kms.

The Taurus Molecular Cloud was observed over two observing seasons
starting in November 2003 and ending in May 2005.  The \co\ and
\coa\ lines were observed simultaneously enabling excellent positional
registration and calibration.  System temperatures ranged from 350-500
K for the \co\ line and 150-300 K for the \coa\ line.  
%The fiducial
%center position of the map was $\alpha$(2000) = $04^h32^m44^s.6$,
%$\delta$(2000) = $24\arcdeg 25\arcmin 13\arcsec.08$.

\subsection{Previously Known Outflows and YSOs in Taurus}

%% In Table~\ref{known_outflow_table}, the list of 17 previously known
%% outflow candidates in the area covered by the FCRAO Taurus survey is
%% presented. The last column of the table lists the references that
%% cover the outflow detections. Table~\ref{known_outflow_table} does not
%% include the new outflows recently reported in \citet{davis2010} from
%% JCMT HARP \co\ J=3-2 emission observations of a subset of the FCRAO Taurus
%% survey. While \citet{davis2010} confirm the detection of several
%% well-known sources, some of their other claimed outflow detections are
%% based on wing emission at only low-velocities from the ambient
%% cloud. We will discuss our results with respect to the
%% \citet{davis2010} results in \S~\ref{harp}.

The list of outflows compiled by \citet{wu2004} contains
13 outflows in the area covered by the FCRAO Taurus survey.
In addition, we found three previously identified outflows 
that were missed by \citet{wu2004}.  Two of these outflows 
(IRAS 04169+2702 and IRAS 04302+2247) were from the survey 
of \citet{bontemps1996}.  The other outflow (IRAS 04240+2559) 
was discovered by \citet{mitchell1994} and is associated with
DG Tau.  Thus the number of previously known outflows candidates is
16, and these are given in Table~\ref{known_outflow_table}. In the 
last column of this table we describe the nature of the outflow,
whether bipolar or with only red or blue outflow, and give 
appropriate references.  

Recently, \citet{davis2010} surveyed an approximately one square
degree region of the L1495 region of Taurus in the CO J=3-2 line using
HARP on the JCMT.  They identify as many as 16 molecular
outflow candidates in this region.  They confirm the detection of
several well-known outflows and several of their outflows we believe
are part of much larger outflows that will be discussed later.
We note that many of their outflow candidates are identified based on wing
emission at relatively low-velocities (2.5 km s$^{-1}$) from the
ambient cloud line centre velocity.  The velocity field in Taurus is
very complex, which can be seen in the channel maps presented in
\citet{narayanan2008}, and often there are secondary velocity
components that might mimic outflows.  Thus the identification of
outflows based on a simple formula without examining the large-scale
environment can be fraught with difficulties.  We have not included the
results of \citet{davis2010} in Table~\ref{known_outflow_table}, but
instead, we have integrated their results into the discussion in our
Results section.

Based on observations of mid and far-infrared bands with the Spitzer
Space Telescope, \citet{rebull2010} tabulate the properties of
pre-main-sequence objects in Taurus. Their survey covers $\sim 44$
deg$^2$, and encompasses most of the highest column density regions of
the \citet{narayanan2008} study. This Spitzer Survey lists 215
previously known members in Taurus, and they identify and tabulate
properties of 148 new candidate members. All 363 of these YSOs are
over-plotted in applicable areas in the plots presented below, with
the previously known candidates plotted as red stars, and new
candidate members as red triangles.

%% High spatial resolution observations of many molecular cloud cores
%% have been obtained by the Nobeyama 45-m telescope in H$^{13}$CO$^+$ at
%% an angular resolution of 20 arcsec \citep{onishi2002}. These
%% observations, because they are optically thin, trace the quiescent,
%% high-density gas in regions of Taurus. We will use this
%% \citet{onishi2002} data throughout the paper.

We also use the data from the General Catalogue of Herbig-Haro (HH)
objects, 2nd edition compiled by Bo
Reipurth\footnote{http://casa.colorado.edu/hhcat/}.  In addition, we
plot the molecular hydrogen emission-line objects (the so-called
MHOs), which are the shock-excited IR counterparts to the HH objects
from the list compiled by \citet{davis2010b}.

\begin{table*}
%\tablewidth{0pt} 
\centering
\begin{minipage}{120mm}
\caption{\label{known_outflow_table}List of Known Outflows in Taurus} 
\begin{tabular}{lllll}
\hline
Names&RA (J2000)& Dec (J2000)& Comments\\
\hline
IRAS04166+2706 & 04:19:42.6 & 27:13:38 & Bipolar$^{a,b}$\\
IRAS04169+2702 & 04:19:58.4 & 27:09:57 & Bipolar$^{a}$\\
IRAS04181+2655 (CFHT-19) &  04:21:10.53 & 27:02:06.0 & Bipolar$^{a}$\\
IRAS04239+2436	& 04:26:56.9 &	24:43:36 & Red Lobe, HH300 flow$^{c}$\\
IRAS04240+2559 (DG Tau) & 04:27:04.7 & 26:06:17 & Red lobe, optical jet$^{d}$\\ 
IRAS04263+2426 (Haro 6-10) & 04:29:23.7 & 24:32:58 & Bipolar flow$^{e}$\\ 
IRAS04278+2435 (ZZ Tau) & 04:30:53.0 & 24:41:40 & Red Lobe$^{f}$\\
TMC 2A & 04:31:59.9 & 24:30:49 & bipolar $^{g}$\\
L1529 & 04:32:44.7 & 24:23:13 & not confirmed$^{h,i}$\\
IRAS04302+2247 & 04:33:16.8 & 22:53:20 & bipolar$^{a}$\\
IRAS04325+2402 (L1535)& 04:35:33.5 & 24:08:15 & Red$^{f}$\\
IRAS04361+2547 & 04:39:13.9 & 25:53:21 & Bipolar$^{a,j}$\\
IRAS04365+2535 (TMC1A) & 04:39:34.8 & 25:41:46 & Bipolar$^{a,k,l}$\\
IRAS04368+2557 (L1527) & 04:39:53.3 & 26:03:06 & Bipolar$^{a,l,m,n}$\\
IRAS04369+2539 (IC 2087) & 04:39:58.9 & 25:45:06 & Red$^{f}$\\
IRAS04381+2540 (TMC1) & 04:41:13.0 & 25:46:37 & Bipolar$^{a,k}$\\
\hline
\end{tabular}

{\it a}) \citet{bontemps1996}; ({\it b}) \citet{tafalla2004}; ({\it
  c}) \citet{arce2001}; ({\it d}) \citet{mitchell1994}; ({\it e})
\citet{stojimirovic2007}; ({\it f}) \citet{heyer1987}; ({\it g}) A
complicated region mapped by \citet{Jiang2002} with a likely bipolar
outflow associated with 04292+2422 (Haro 6-13) and possible a second
outflow associated with 04288+2417 (HK Tau); ({\it h})
\citet{lichten1982}; ({\it i}) \citet{goldsmith1984}; 04292+2422 Haro
6-13; ({\it j}) \citet{terebey1990}; ({\it k}) \citet{chandler1996};
({\it l}) \citet{tamura1996}; ({\it m}) \citet{hogerheijde1998}; ({\it
  n}) \citet{zhou1996}.
\end{minipage}
\end{table*}

\subsection{Data Processing}

For all the details of the data processing of the original Taurus
survey, refer to \citet{narayanan2008}. In this paper, we start with
the 88 'hard-edge' cubes covering the full $100$ square degrees. The
88 hard-edge cubes form a grid of $11\times 8$ data cubes. They are
dubbed hard--edge cubes, as they do not have overlapping regions
between two contiguous cubes. Each 'hard-edge' cube is assembled from
a set of input regridded $30\arcmin\times 30\arcmin$ cubes, after
removing the spectral and spatially derived baselines described above
from each cube, and subsequently averaging the data together,
weighting them by $\sigma^{-2}$, where $\sigma$ is the rms in the
derived baseline. In angular offsets, the full extent of the combined
hard--edge cubes are (5.75\arcdeg, $-5.75$\arcdeg) in RA offsets and
($-2.75$\arcdeg, 5.75\arcdeg) in Dec offsets from the fiducial centre
of the map ($\alpha$(2000) = $04^h32^m44.6^s$, $\delta$(2000) =
$24\arcdeg 25\arcmin 13.08\arcsec$). Thus for the full
$11.5\arcdeg\times 8.5\arcdeg$ region spaced at 20\arcsec, there are
3,167,100 spectra in each isotopologue in the combined set of
hard--edge cubes. Most of the hard--edge cubes have a spatial size of
1 square degree, except for the cubes that lie on the four edges of
the region covered, which measure 1.25 square degrees. The hard--edge
cubes at the four corners of the Taurus map have a size of 1.5625
square degrees ($1.25\arcdeg\times 1.25\arcdeg$).

\subsubsection{Unbiased Search for Outflows}
\label{unbiased_search}

Our unbiased search for outflows from the Taurus Molecular Cloud
Survey was carried out as follows. The systemic velocity of most of
the Taurus emission (from the \coa\ data) varies from $\sim$ 6 to
7~\kms. So we initially define outflow wings to be at least
3~\kms\ offset from the systemic velocity, and define a blueshifted
outflow velocity range of $-1$ to $3$~\kms, and a redshifted outflow
velocity range of $9$ to $13$~\kms.  We make integrated intensity
images of blueshifted and redshifted emission in \co\ in the
aforementioned velocities in each of the 88 hard-edge cubes, and
overlay them both against an integrated intensity image made between
the velocities of 4 to 8~\kms\ in \coa. Due to the lower optical depth
of the isotopic transition, the integrated intensity in \coa\ at low
velocities represents the emission from the ambient cloud medium. By
overlaying the higher velocity linewing emission from \co, these plots
are meant to reveal the distribution of ambient gas, and highlight the
distribution of any high-velocity redshifted and/or blueshifted gas
that may be present. Many such examples of these kind of plots will be
presented below. When we see any extended blueshifted or redshifted
linewing emission, for any hard edge cube, we follow it up with other
tests to ascertain the presence of outflows.

In addition to the 88 hard-edged cubes, we have 88 {\it rms} images
each in \co\ and \coa, where these images are derived from evaluating
the rms noise level in spectral windows where no line emission is
present. Propagating errors, it is thus possible to obtain integrated
intensity images along with the corresponding uncertainties at every
pixel location in a map. When extended high-velocity features are
seen, we compute average spectra over these regions. To isolate
emission from the presumed outflow lobes, polygonal areas delineating
the presumed outflow lobes are interactively drawn, and our procedure
then obtains average \co\ and \coa\ spectra within these polygonal
regions. However, the presence of significant emission in the
previously defined blueshifted and redshifted velocity intervals is
not a sufficient requirement to identify outflows. The velocity
structure of the Taurus Molecular Cloud is complex, and in many
regions there are additional cloud components that appear in these
velocity intervals which are used to identify outflows. To distinguish
between outflows and secondary velocity components, we made use of our
large maps, and the emission in both isotopic species. Extended
narrow-line features seen in both \co\ and \coa\, although in the
velocity interval defined as outflow emission, are almost certainly
secondary cloud components and not outflows. We have applied a
relatively strict set of criteria in identifying outflow, thus we
believe that the outflows in our final list are in fact true outflows.

%% In addition to
%% simple averages, we also perform weighted averages in high-velocity
%% blue or redshifted emission. For example, the weighted average
%% spectrum in blue-shifted line wing is derived in a particular region,
%% by only averaging spectra in pixel locations where the integrated
%% intensity in \co\ is at least 3 times its corresponding uncertainty at
%% that location. At these same locations, the \coa\ emission is also
%% averaged to produce a single spectrum of \co\ and \coa\ which have
%% been selected specifically to have strong high velocity blue-shifted
%% \co\ gas. 
%% When such a final \co\ spectrum has extended blue linewing
%% but not much of detectable \coa\ emission in regions with detected blueshifted
%% integrated intensity emission, it can be construed as evidence for outflow. On the other
%% hand, sometimes, we detect a separate blue-shifted velocity component
%% that has a peak between $-1$ to $3$ \kms, and in some of these cases,
%% the \coa\ spectrum also shows a corresponding peak at the same
%% velocity. In this latter instance, we interpret the blue-shifted
%% feature seen in the integrated intensity map as due to an unassociated
%% foreground cloud at a different velocity, and not a molecular outflow.

When extended blueshifted or redshifted regions of emission are seen,
we follow that by constructing position-velocity (PV) diagrams that
cut through these regions. The PV cuts are chosen to cut through the
bulk of the redshifted and/or blueshifted emission, and where
available, passing through the YSO that could be the potential driving
source. In these diagrams, outflows usually show up as horizontal
or oblique linear features. Examples of these PV diagrams will be
shown below. Unassociated foreground or background turbulent cloud
components usually show up as vertical features in PV diagram often
offset from the low-velocity ambient cloud emission of the bulk
material of the Taurus Molecular Cloud.

Overlaid on these integrated intensity images, we can plot the Spitzer
sources \citep{rebull2010}, HH objects, MHO objects
\citep{davis2010b}, previously identified outflow sources in Taurus
(Table~\ref{known_outflow_table}), and the newly identified outflow
locations from the JCMT HARP survey \citep{davis2010}. In the presence
of outflows, the locations of the YSOs and other signposts of
star-formation help identify the driving sources for the outflows.

%% In some instances, we also make first moments (centroid velocity maps)
%% of \co\ emission at ambient cloud velocities in a given region that%% shows the morphological features of an outflow. In such instances,
%% centroid velocity maps can show how much the higher velocity molecular
%% ouftlows are impacting even the low-velocity components of the ambient cloud.

The analyses described in this paper were all performed entirely in
python. The hard-edged cubes were manipulated using
PyFITS\footnote{PyFITS is a product of the Space Telescope Science
  Institute, which is operated by AURA for NASA}. Plots were made by
heavily extending the FITSFigure class provided by
APLpy\footnote{APLpy (Astronomical Plotting Library in Python,
  http://aplpy.sourceforge.net)}. Astronomical utilities such as
coordinate transformations, integrated intensity calculations,
position-velocity cuts are all developed in a newly available
open-source set of python utilities called
idealpy\footnote{http://idealpy.sourceforge.net/}.

\section{Results}
\label{results}
\subsection{List of Detected Outflows}

\begin{figure*}%[hbp]
\begin{center}
\includegraphics[width=0.7\hsize]{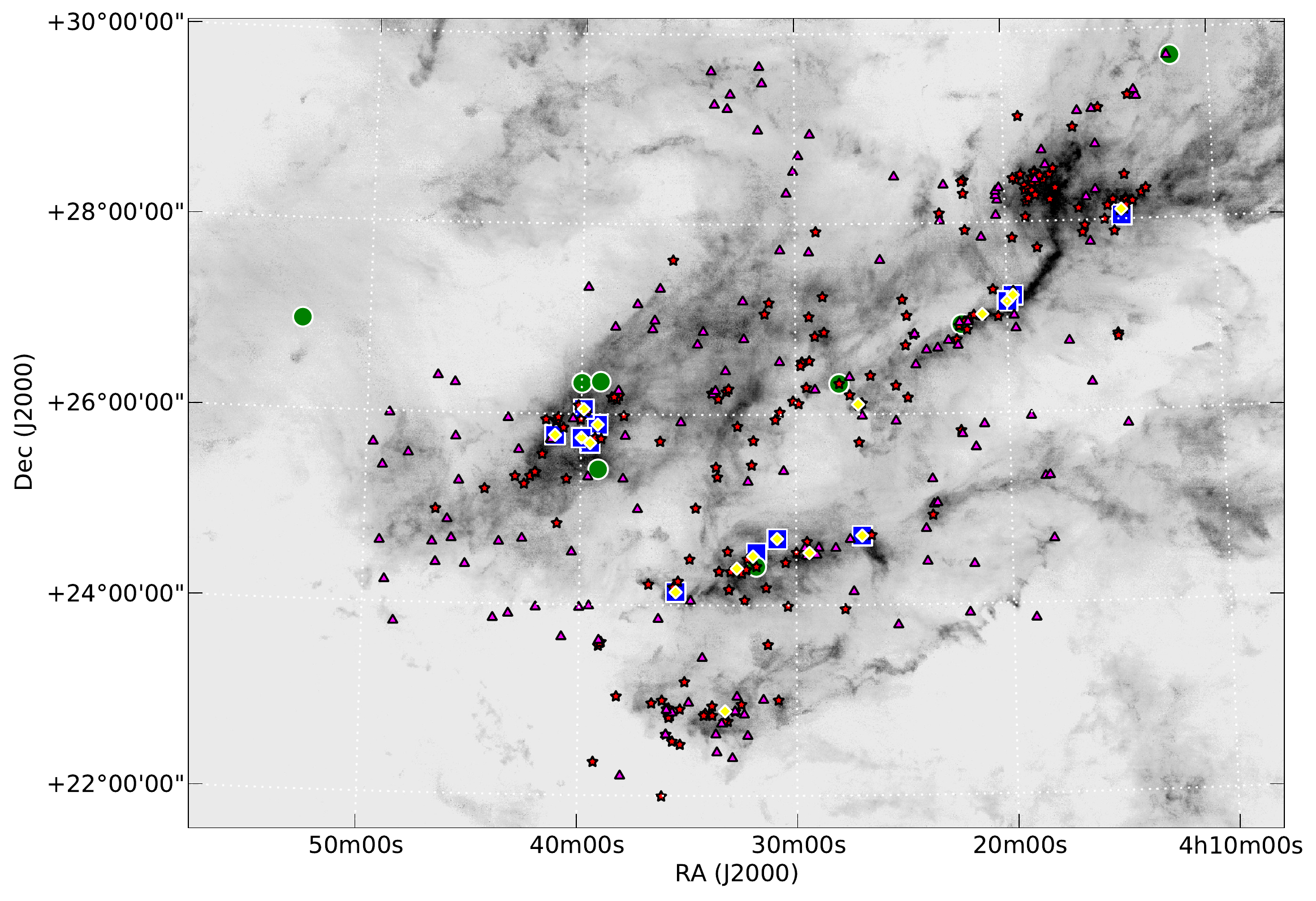}
\caption{Locations of YSOs and outflow sources in the Taurus Molecular
  Cloud. The grey scale shows the molecular hydrogen column density
  map presented in \citet{goldsmith2008}. Red stars show the location
  of previously known YSOs in Taurus (from Table 4 of
  \citealt{rebull2010}), while red triangles show the location of newly
  identified YSOs in Taurus (from Spitzer observations and Table 5 of
  \citealt{rebull2010}). Yellow diamonds represent the location of
  known outflows catalogued in Table~\ref{known_outflow_table}. Blue squares
  show the location of previously known outflows identified in this
  survey, while green circles denote new outflows that have been
  detected in this current survey. \label{overview_figure}}
\end{center}
\end{figure*}

In Figure~\ref{overview_figure}, we present the location of detected
outflows overlaid on the molecular hydrogen column density map from
\citet{goldsmith2008}. All Taurus YSOs from Spitzer observations are
marked in the plot. In addition, the list of hitherto known outflows 
is denoted in the figure, along with new outflow detections stemming
from this study. In Table~\ref{outflowlist}, we present all 20
detected outflows in this study with brief comments on each.

\begin{figure*}%[hbp]
\begin{center}
\includegraphics[width=0.6\hsize]{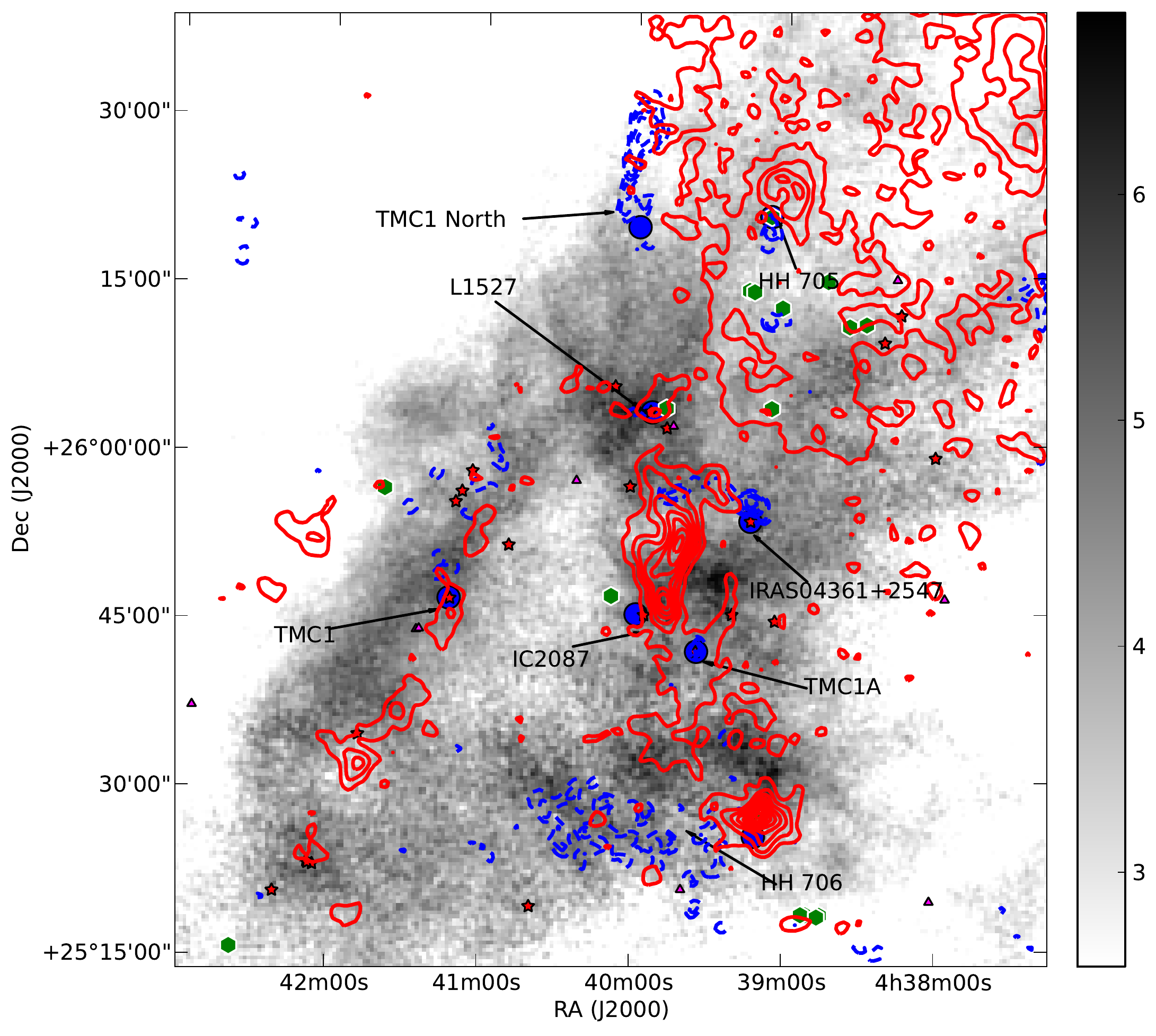}
\caption{Overview contour map of blueshifted and redshifted gas about
  a $78^\prime\times 85^\prime$ region approximately centred on
  IC2087, where eight distinct outflows can be seen (individual
  outflows are called out with arrows). The grey scale shows the
  \coa\ integrated intensity map. See Figure~\ref{041159_outflow} for
  details on symbols and markers.  Overlaid on the gray-scale
  \coa\ image, \co\ blueshifted integrated intensity (within velocity
  interval -1 to 3.9 \kms) is shown in blue dashed contours, and
  redshifted integrated intensity (8.2 to 13 \kms) is shown in red
  solid contours. Blueshifted contours range from 0.64 to 2.7 in steps
  of 0.075 \kkms, and redshifted contours range from 0.64 to 8.4 in
  steps of 0.075 \kkms.
\label{ic2087_overview}}
\end{center}
\end{figure*}

\begin{table*}

\centering
\begin{minipage}{150mm}
\caption{\label{outflowlist}List of detected Outflows}
\begin{tabular}{rlllcll}
\hline
{S.No} & {Name}&{RA} & {Dec}
& {New} &{YSO} &{Comments}\\
&&{(J2000)}& {(J2000)}& {Detection} & {Class$^{a}$}\\
\hline
1. & 041149+294226 & 04:11:49.0 & 29:42:30 & Y & -- & Bipolar, closely associated\\
&&&&&&with new Spitzer source 041159+294236\\
2. & IRAS04113+2758 & 04:14:25.2 &	28:02:21 & N & I & Red only\\
3. & IRAS04166+2706	& 04:19:42.6 & 	27:13:38 & N & I & Bipolar, much longer
than \\
&&&&&&previously known\\
4. & IRAS04169+2702	& 04:19:58.4 &	27:09:57 & N & I & Bipolar\\
5. & FS TAU B & 04:22:00.7 & 26:57:32 & Y & I & Red Lobe only and
maybe\\
&&&&&&seen in \citet{davis2010}\\
6. & IRAS04239+2436	& 04:26:56.9 &	24:43:36 & N & I & Red\\
7. & IRAS04248+2612	& 04:27:57.7 &	26:19:19 & Y & F & Red Lobe, Optical Flow
seen in\\
&&&&&&\citet{gomez1997}\\
8. & Haro 6-10 & 04:29:23.7 & 24:32:58 & N & I & Bipolar\\
%& (Haro 6-10)&&&&\\
9. & ZZ Tau IRS & 04:30:53.0 & 24:41:40 & N & F & Red Lobe, first
detected by\\
&&&&&&\citet{heyer1987}\\
10. & Haro 6-13	& 04:31:50.9 & 24:24:18 & Y & F & Red\\
%%11. & TMC2A & 04:31:59.8 & 24:30:48 & Y & Red\\
11. & IRAS04325+2402 & 04:35:33.5 &	24:08:15 & N & I & Red\\
12. & HH 706 flow & 04:39:11.4 & 25:25:16 &	Y & -- & Bipolar\\
13. & HH 705 flow & 04:39:06.7 & 26:20:29 & Y & -- & Bipolar\\
14. & IRAS04361+2547 & 04:39:13.9 & 25:53:21 & N & I & Bipolar\\
15. & TMC1A	& 04:39:34.8 & 25:41:46 & N & I & Blue\\
16. & L1527	& 04:39:53.3 &	26:03:06 & N & I & Bipolar\\
17. & IC2087 IR	& 04:39:58.9 &	25:45:06 & N & F & Red\\
18. & TMC-1 North & 04:39:59.1 & 26:19:36 &	Y & -- & Blue\\
19. & TMC1	& 04:41:13.0 & 25:46:37 & N & I & Bipolar\\
20. & 045312+265655 & 04:53:12.1 & 26:56:55 & Y & -- & Bipolar\\
%\tableline\\
%NEW2 & 04:22:10.3 &	26:55:55 &	Red\\
\hline
\end{tabular}

{\it a}) From classification of \citet{rebull2010} when driving source
is identified. F indicates Flat spectrum source, which is intermediate
between Class I and Class II.
\end{minipage}
\end{table*}

%\end{deluxetable*}

The detected outflows are mostly concentrated in regions of high
column density. Individual regions where multiple outflows can be
found can be identified in Figure~\ref{overview_figure}. However, the
Heiles Cloud 2 region in Taurus is especially notable in that the
molecular ring \citep{schloerb1984} is the seat of many YSOs. In this
region, we have detected eight outflows (three of which are new
detections; see Figure~\ref{ic2087_overview} for an overview figure of
this region). More details of the flows in the Heiles Cloud 2 region
are presented later in this section.

Below we summarise each outflow in more detail.

\subsubsection{041149+294226}

\begin{figure}%[hbp]
\begin{center}
\includegraphics[width=0.9\hsize]{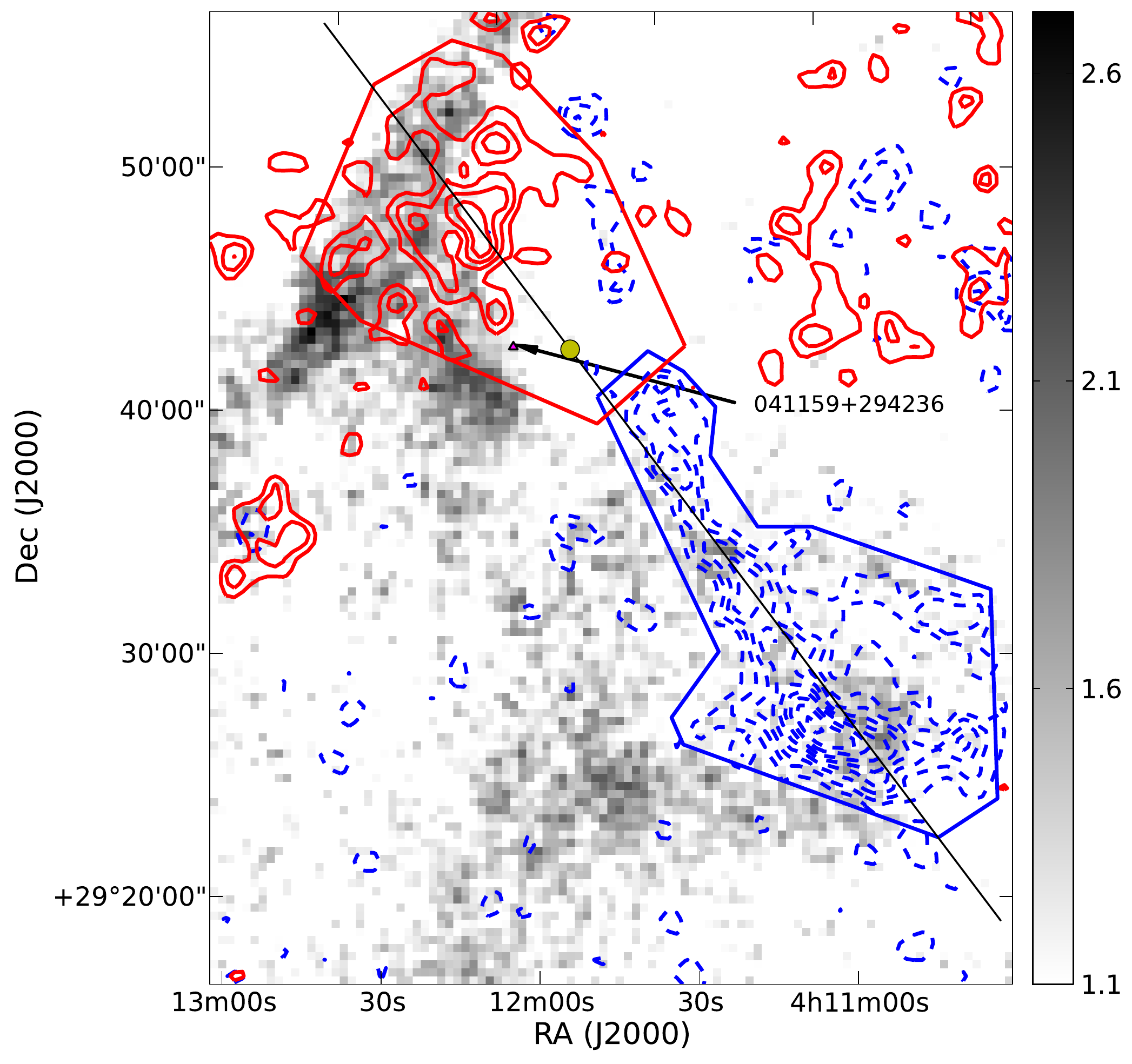}
\caption{Contour map of blueshifted and redshifted gas around the
  Spitzer source 041159+294236. The grey scale shows the
  \coa\ integrated intensity map. In this and all subsequent outflow
  maps, the \coa\ integrated intensity map is always made for a
  velocity range of 4 to 8 \kms (representing the low velocities
  associated with the ambient cloud emission). In this and all
  subsequent outflow maps, if present, red stars show the location of
  previously known YSOs in Taurus (from Table 4 of
  \citealt{rebull2010}), while red triangles show the location of
  newly identified YSOs in Taurus (from Spitzer observations and Table
  5 of \citealt{rebull2010}). When available, green hexagons represent
  locations of known HH objects, and cyan circles represent the
  locations of claimed outflows from \citet{davis2010}. Overlaid on
  the gray-scale \coa\ image, \co\ blueshifted integrated intensity
  (within velocity interval {\bf -1 to 4.2 \kms}) is shown in blue dashed
  contours, and redshifted integrated intensity ({\bf 8.75 to 13 \kms}) is
  shown in red solid contours. Blueshifted contours range from 0.45 to
  2.2 in steps of 0.075 \kkms, and redshifted contours range from 0.5
  to 2.3 in steps of 0.075 \kkms. In this and other subsequent contour
  maps, a straight line representing a cut for the position-velocity diagram
  will be shown followed by a position velocity diagram (for this
  figure, it is Figure~\ref{041159_posvel}), and the yellow circle
  represents the point representing the origin in the Y-axis of
  Fig~\ref{041159_posvel}. Again, in this figure and all subsequent
  figures, the blue and red polygons show the regions
  defined as the blueshifted and redshifted lobes of the outflow,
  whose average spectra are shown in a figure showing spectra that
  will follow (for this source, Figure~\ref{041159_spectra}).
\label{041159_outflow}}
\end{center}
\end{figure}

In the north-west corner of the region covered in our Taurus survey
(see Figure~\ref{overview_figure}), we discover a new bipolar outflow
whose driving source is unknown. We call this outflow with the
coordinates of the position of a likely origin as 041149+294226. There
is a new Spitzer source, 041159+294236 (from Table 4 of
\citealt{rebull2010}) in the vicinity of this outflow. In
Figure~\ref{041159_outflow}, which shows a region $\sim 33^\prime
\times 40^\prime$ in size, the blueshifted and redshifted integrated
intensity contours of \co\ are overlaid on the low-velocity (4 to 8
\kms) integrated intensity emission from \coa\ (shown in
greyscale). In this figure, and in all subsequent outflow plots, a
variety of symbols representing Spitzer YSOs, HH objects, and outflow
sources known in the past are marked when available (see the figure
caption of Fig~\ref{041159_outflow} for more details). The outflow has
a position angle of $\sim 127^\circ$, and appears quite collimated in
blueshifted gas, while the redshifted emission is more diffuse. From
the tip of the blueshifted lobe to the tip of the redshifted lobe, the
full extent of the outflow is 1.6 pc!

\begin{figure}%[hbp]
\begin{center}
\includegraphics[width=0.9\hsize]{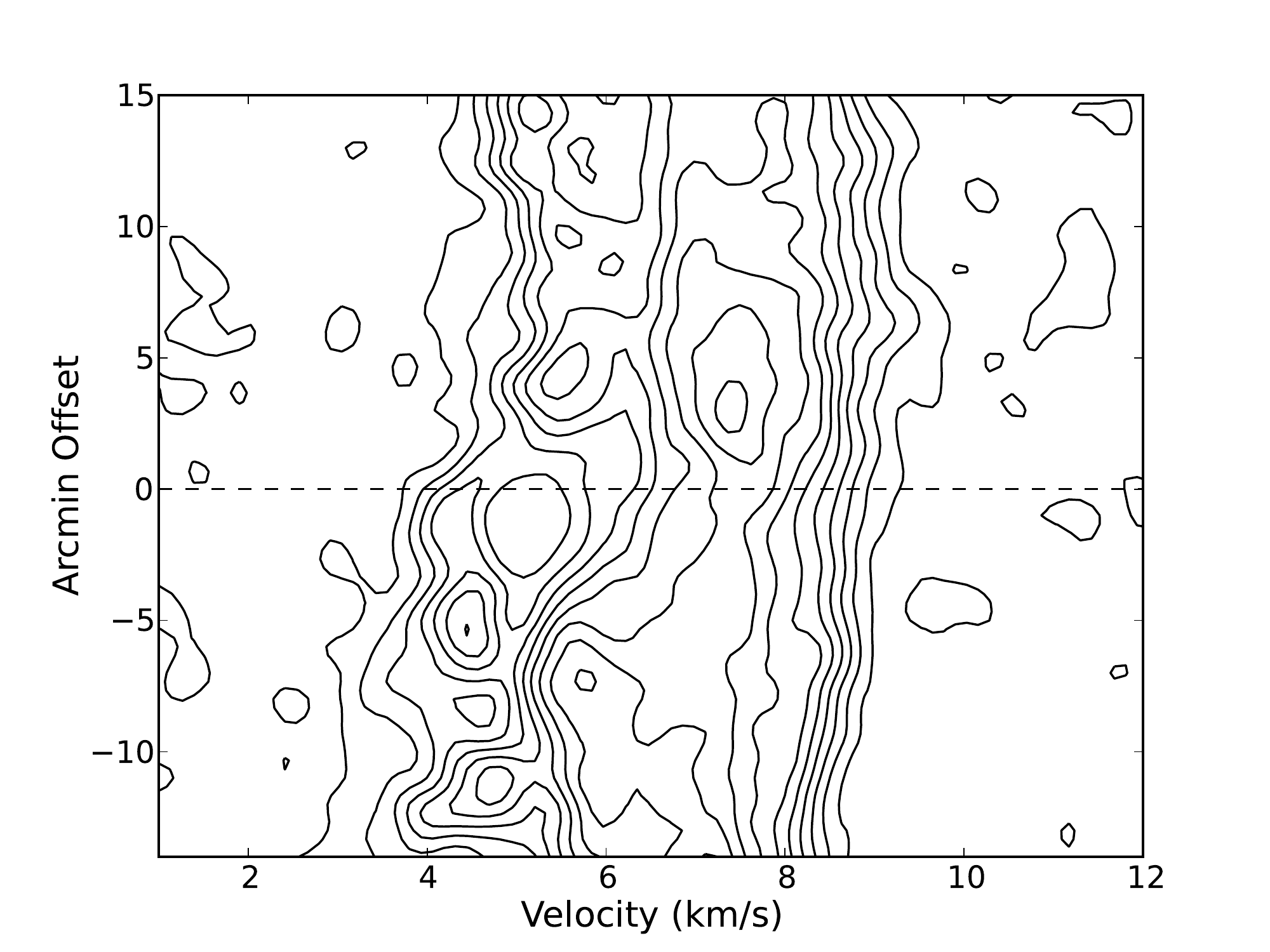}
\caption{Position velocity of \co\ emission towards 041149+294226
  region, through the slice shown in Figure~\ref{041159_outflow} at a
  position angle of $127^\circ$. The contour range is 0.125 to 2.8 K
  in steps of 0.3 K. Shown in dashed line is the position of the
  yellow circle shown in Fig~\ref{041159_outflow}.
\label{041159_posvel}}
\end{center}
\end{figure}

\begin{figure}%[hbp]
\begin{center}
\includegraphics[width=0.9\hsize]{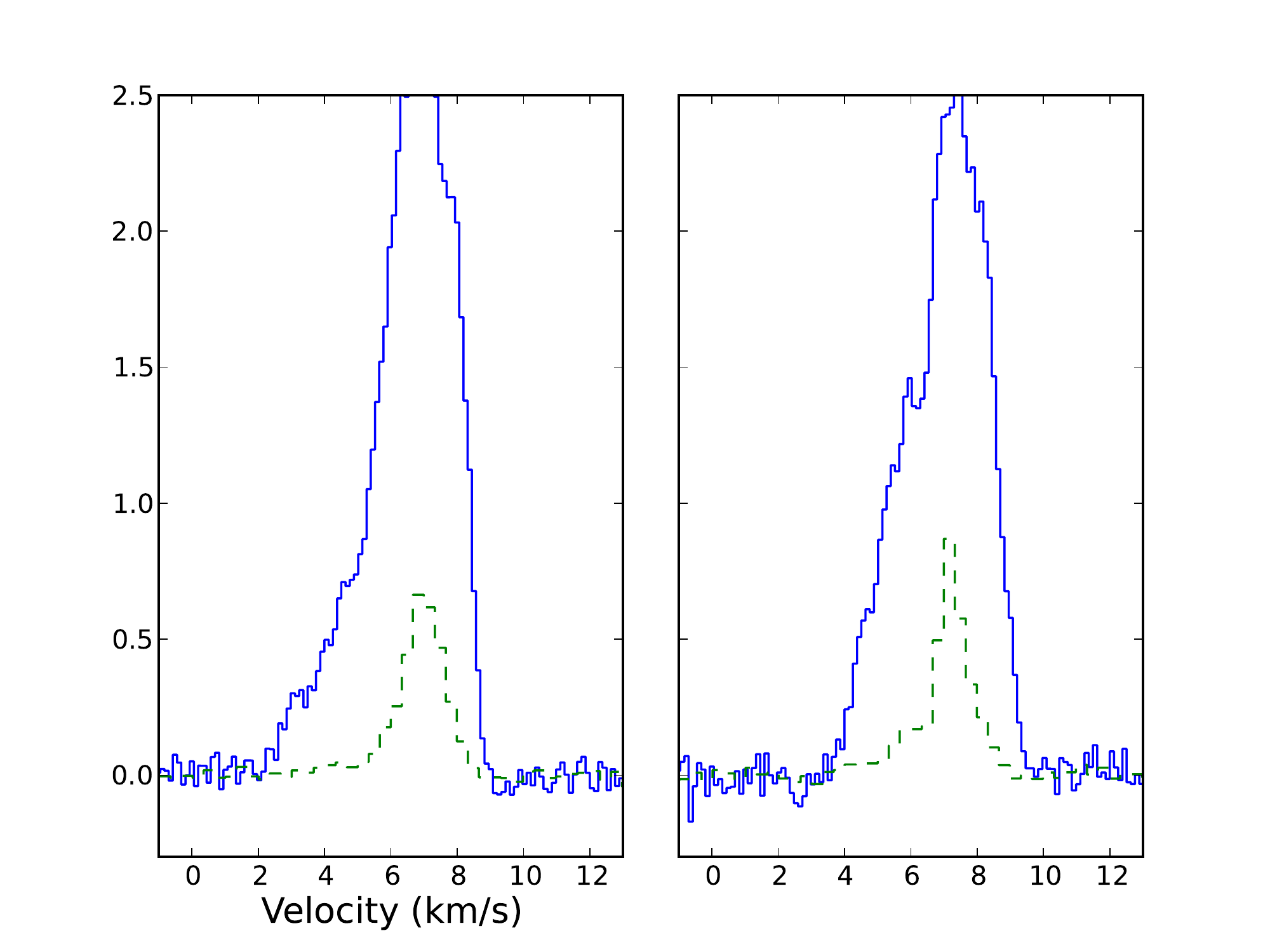}
\caption{Average Spectra of \co\ emission (blue solid lines) and
  \coa\ (green dashed lines) towards the blueshifted(left) and
  redshifted (right) lobes of the outflow in the 041149+294226 region
  shown in Fig~\ref{041159_outflow}. The temperature scale in this and
  other spectra shown in this paper is in
  T$_{A}^{*}$. \label{041159_spectra}}
\end{center}
\end{figure}

The \coa\ emission shown in Figure~\ref{041159_outflow} is likely
tracing the column density of ambient gas representing the cloud
core. If so, it appears that the blueshifted gas has broken out of the
cloud core, while the redshifted gas from the outflow may be directed
into the bulk of the ambient cloud core. The Spitzer source
041159+294236 lies close to the axis of the bipolar outflow and is the
only known YSO in this field. However, 041159+294236 is a Class III
object, thus as we discuss later, unlikely to be the driving source of
this outflow. It is more likely that an hitherto undetected embedded
object is the driving source.  A position-velocity (p-v) diagram
derived along this solid line marked in Figure~\ref{041159_outflow} is
shown in Figure~\ref{041159_posvel}, where the zero offset position is
marked by a yellow dot as the possible location of the driving source
in Figure~\ref{041159_outflow}. The p-v diagram shows the extended
blueshifted line wings due to the outflow but only weak redshifted
emission. In Figure~\ref{041159_spectra}, we show the average spectra
(in both \co\ and \coa) obtained towards the regions defined by the
red and blue polygons in Figure~\ref{041159_outflow}. Broad emission
marking the outflow can be seen in the averaged spectra, however the
high velocity blueshifted emission is much more prominent than the
high velocity redshifted emission. Figures \ref{041159_posvel} and
\ref{041159_spectra} indicate that the redshifted lobe is not as
prominent as the blueshifted one in this source. But, averaged spectra
in smaller localized regions of the redshifted lobe do show evidence
of redshifted line wings. In general, instead of using just one line
of evidence, we use the {\it combination} of integrated intensity,
position-velocity plots and averaged spectra to identify outflow
lobes.

\subsubsection{IRAS 04113+2758}

The first evidence of high velocity gas in this region was provided by
\citet{moriarty-schieven1992} who surveyed the J=3-2 CO line toward a
number of YSOs.  Toward IRAS 04113+2758 they detected a CO line with a
total velocity extent of 40.3 \kms, the largest of any
source in their survey.  Recently, \citet{davis2010}
mapped the region with the JCMT as part of the Legacy Survey of the
Gould Belt with HARP.  Their maps revealed several possible outflows
in this regions, including a bipolar outflow with an
ill-defined morphology associated with what they call
YSO 1/2.  We believe YSO 1/2 is coincident with IRAS 04113+2758
\citep{kenyon2008}. 

\begin{figure}%[hbp]
\begin{center}
\includegraphics[width=0.9\hsize]{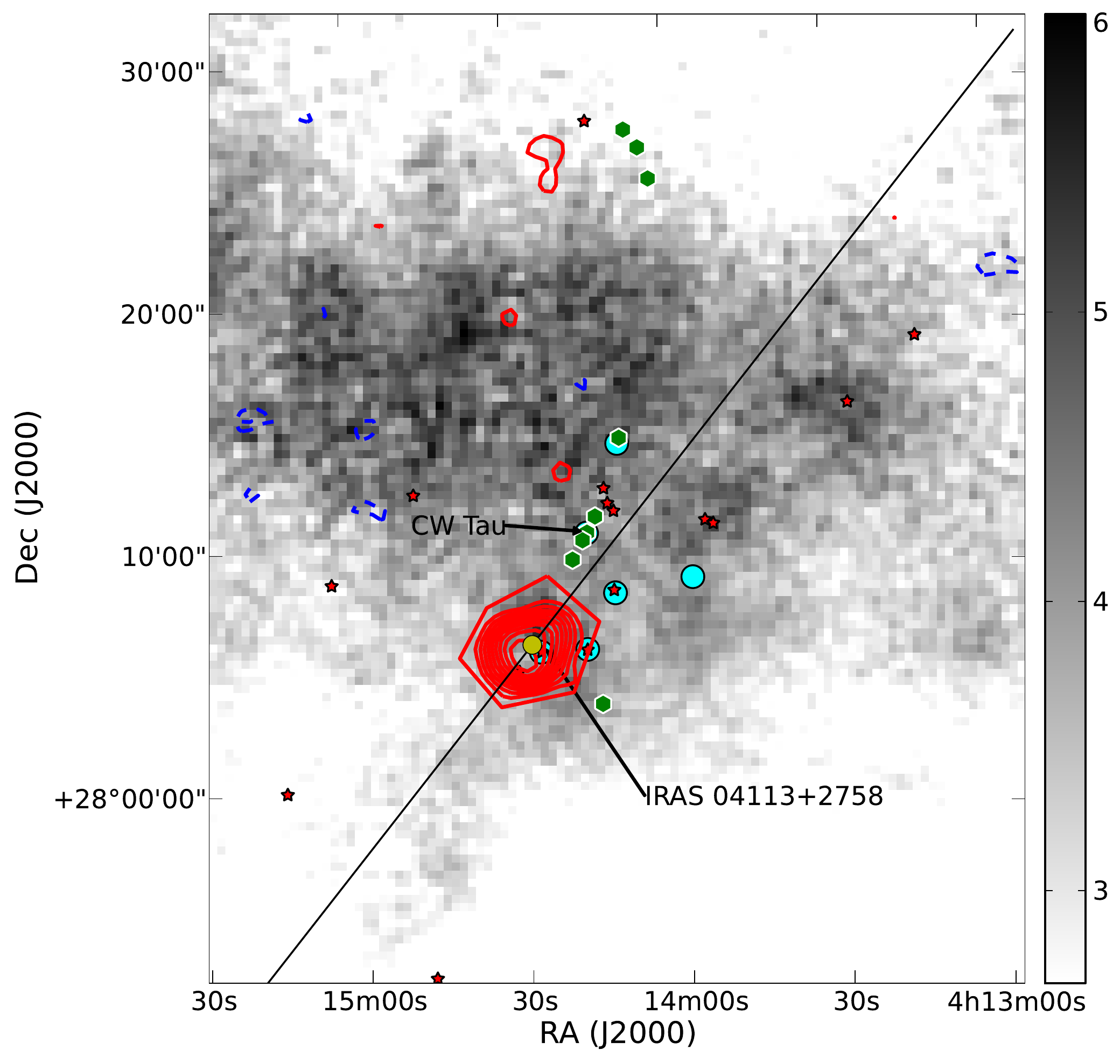}
\caption{Contour map of blueshifted and redshifted gas around IRAS
  04113+2758. See Figure~\ref{041159_outflow} for details on symbols
  and markers.  \co\ blueshifted integrated intensity is within
  {\bf -1 to 2 \kms} and redshifted integrated intensity is within
  {\bf 9 to 13 \kms}. Blueshifted contours range from 0.45 to
  1.9 in steps of 0.075 \kkms, and redshifted contours range from 0.45
  to 3.3 in steps of 0.075 \kkms. 
\label{04113_outflow}}
\end{center}
\end{figure}

\begin{figure}%[hbp]
\begin{center}
\includegraphics[width=0.9\hsize]{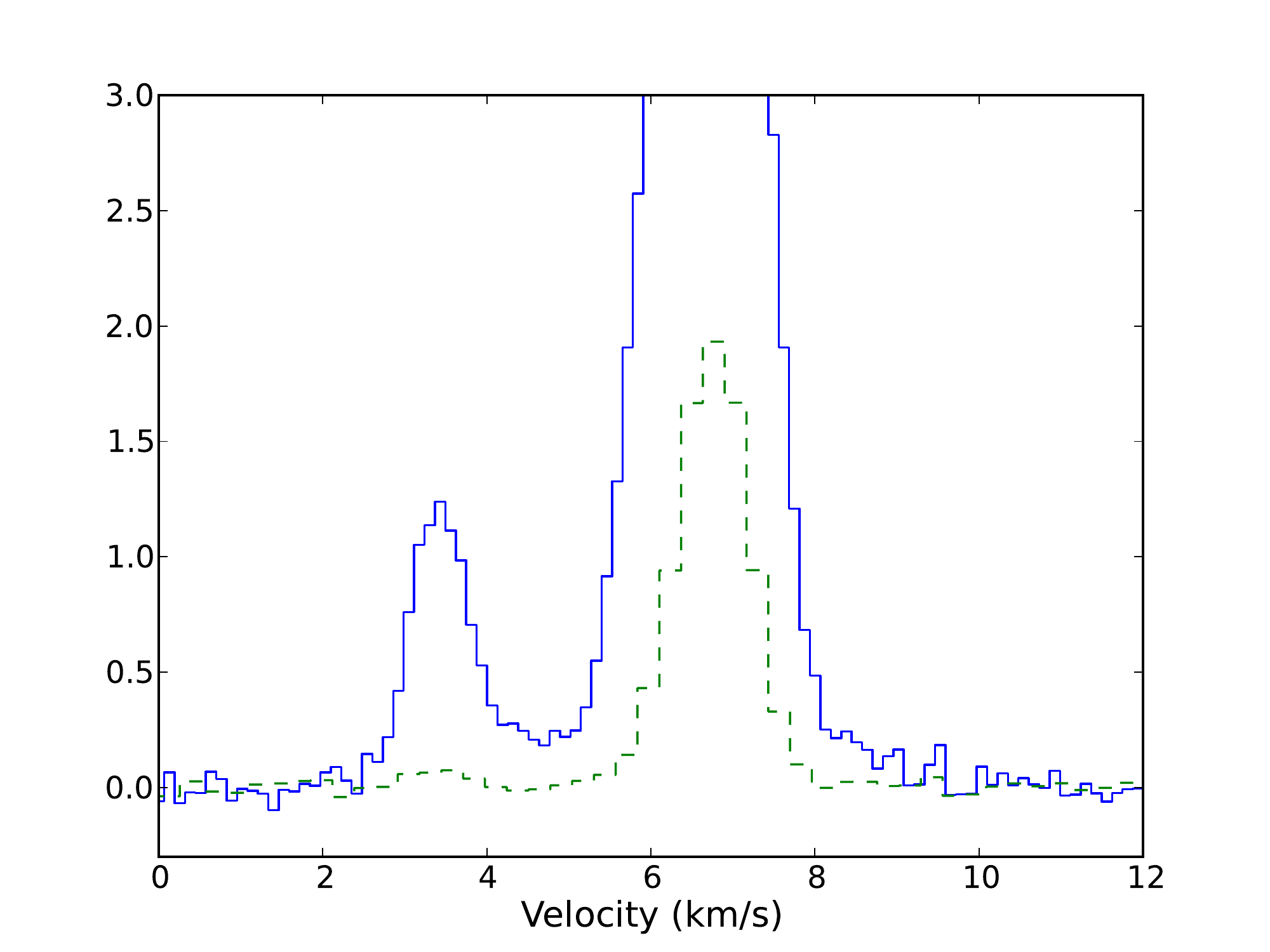}
\caption{Average Spectra of \co\ emission (blue solid lines) and
  \coa\ (green dashed lines) towards the blue-shifted component at RA of
  04:13:40 and Dec of 28:00:00 in the IRAS 04113+2758 region shown in
  Fig~\ref{04113_outflow}. The temperature scale is in
  T$_{A}^{*}$. 
\label{04113_extra_component_spectra}}
\end{center}
\end{figure}

In Figure~\ref{04113_outflow}, we show our map of the redshifted
emission in a $\sim 25^\prime \times 30^\prime$ region about IRAS
04113+2758. The locations of three other possible outflows identified
by \citet{davis2010} in this region, W-CO-B1, W-CO-flow1 and IRAS
04108+2803A, are denoted by cyan-coloured circles in this figure.  The
redshifted high velocity emission that we detect is found toward IRAS
04113+2758.  The detection of blueshifted emission is complicated by
the presence of a second velocity component at V$_{\rm LSR} \sim$ 4
\kms\ that is found extending from the south-west corner of
Figure~\ref{04113_outflow} to the centre of this Figure.  The spectra
shown in Figure~\ref{04113_extra_component_spectra} obtained to the
south-east of IRAS 04113+2758 shows this second velocity component in
both \co\ and \coa\ emission.  Little blueshifted emission is detected
blue-ward of the feature as can be seen in
Figure~\ref{04113_outflow}. We believe that the two blue only outflows
(W-CO-B1 and IRAS 04108+2803A) found by \citet{davis2010} in this
region may be result of the confusion with this widespread secondary
velocity component.  We note that IRAS 04108+2803 was part of the
\citet{moriarty-schieven1992} survey, and they detected a relatively
narrow line suggesting this is unlikely an outflow source.

The source CW Tau is located north-west of IRAS 04113+2758.
This source has a small optical emission-line jet \citep{Gomezdecastro1993,
Dougados2000} oriented north-west to south-east.  A long slit spectrum 
\citep{hirth1994} found that the south-east jet is blueshifted while 
the north-west jet is redshifted.  Thus, it is very unlikely that CW Tau
is the source of the redshifted molecular emission seen in this region.

\begin{figure}%[hbp]
\begin{center}
\includegraphics[width=0.9\hsize]{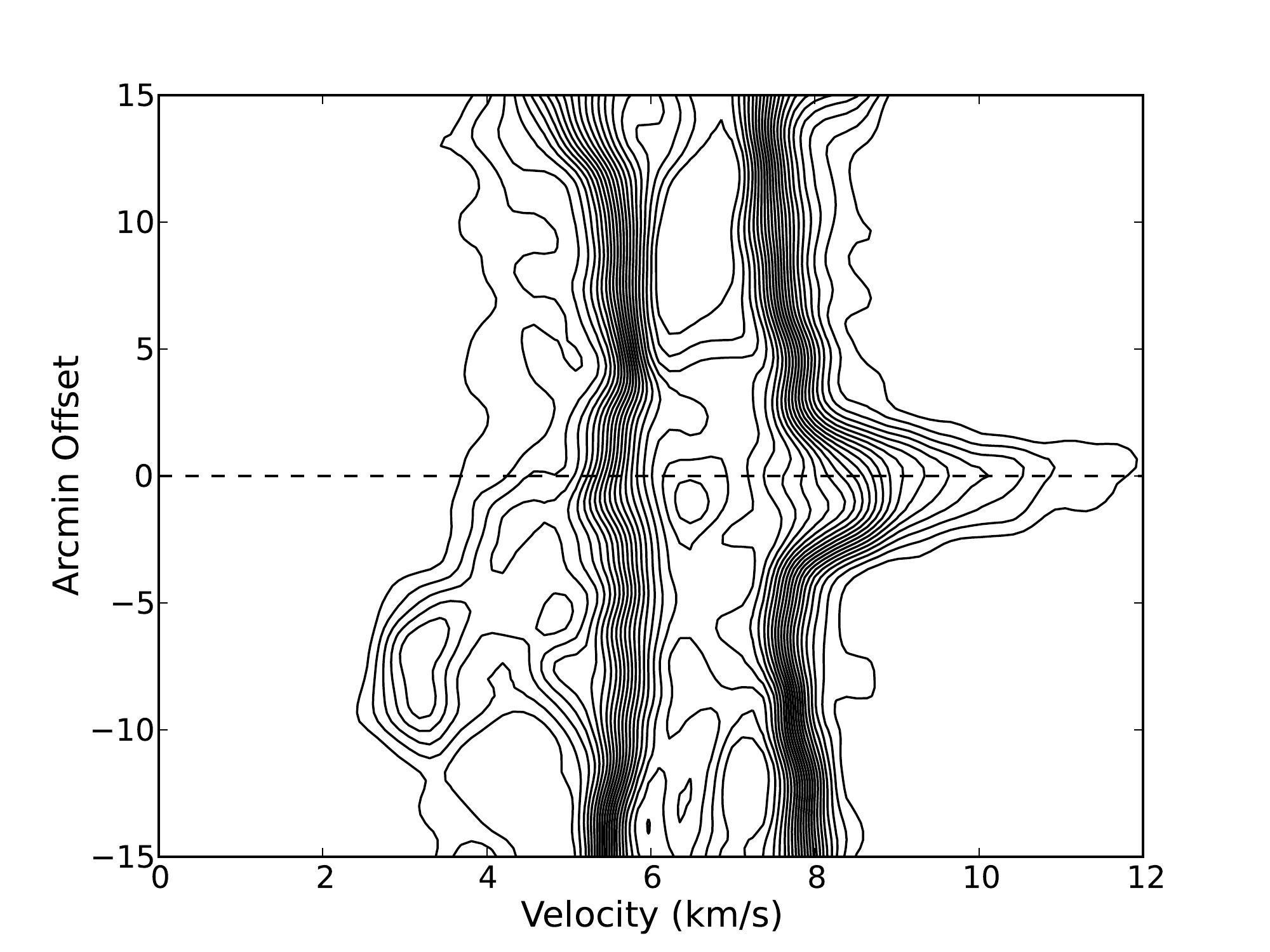}
\caption{Position velocity of \co\ emission towards the IRAS
  04113+2758 region, through the slice at position angle of $52^\circ$
  shown in Figure~\ref{04113_outflow}. The contour range is 0.3 to 4.5
  K in steps of 0.3 K. Shown in dashed line is the position of the
  yellow circle shown in Fig~\ref{04113_outflow}.
\label{04113_posvel}}
\end{center}
\end{figure}

\begin{figure}%[hbp]
\begin{center}
\includegraphics[width=0.9\hsize]{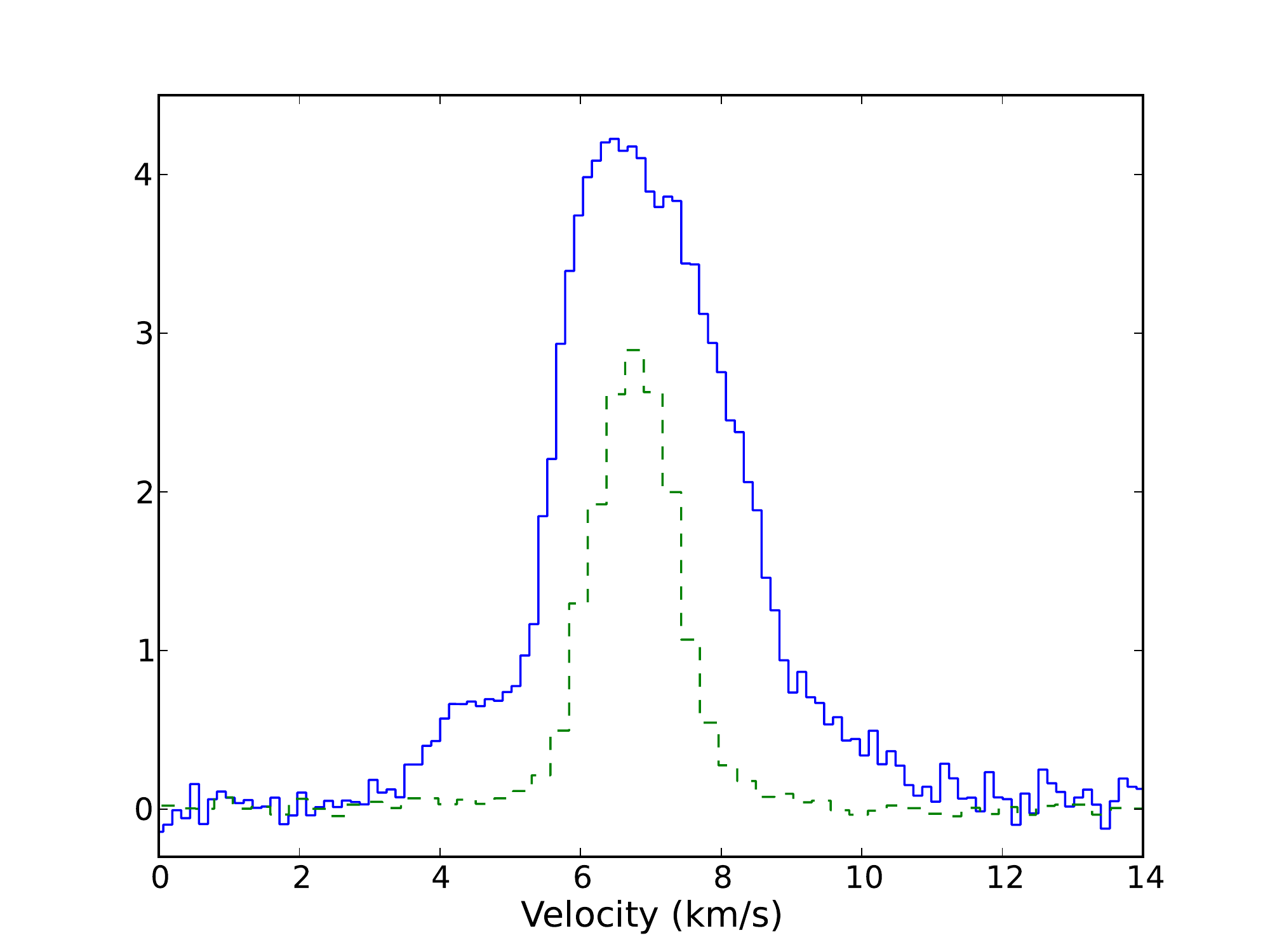}
\caption{Average Spectra of \co\ emission (blue solid lines) and
  \coa\ (green dashed lines) towards the blueshifted (left) and
  redshifted (right) lobes in the IRAS 04113+2758 region shown in
  Fig~\ref{04113_outflow}. The temperature scale is in
  T$_{A}^{*}$. \label{04113_spectra}}
\end{center}
\end{figure}

The p-v diagram shown in Figure~\ref{04113_posvel}
shows the high velocity redshifted emission as
well as secondary blue velocity component.  The averaged
spectra within the polygon marked in Figure~\ref{04113_spectra} shows
these same features.  We note that the spectrum obtained by 
\citet{moriarty-schieven1992} shows very high velocity blue shifted 
emission and the data presented in \citet{davis2010} also indicates
a small region of blueshifted emission near IRAS 04113+2758.  We 
conclude that the outflow associated with IRAS 04113+2758 is likely
bipolar, however in our data the blueshifted outflow is too confused by 
the secondary blue velocity component of the cloud to confirm.

\subsubsection{IRAS 04166+2706}

An outflow was first detected in the IRAS 04166+2706 region by
\citet{bontemps1996}. This outflow was clearly bipolar, however the
small map made by these authors, less than $100\times 100$ arcseconds,
did not show the full extent of the outflow.  A larger map was made by
\citet{tafalla2004} in the CO J=2-1 line and it showed the outflow to be
highly collimated with extremely high velocities.  The full angular
extent of the outflow as mapped by \citet{tafalla2004} was about 6
arcminutes, and emission was detected over the velocity range of
V$_{LSR}$ = -50 to 50 km s$^{-1}$.  More recently
\citet{santiagogarcia2009} mapped the outflow with the IRAM
interferometer in the CO J=2-1 and SiO J=2-1 lines.  Their images
revealed two components to the outflow, a low velocity conical shell
that surrounds a high velocity jet.  This bipolar
outflow was also detected in the survey by \citet{davis2010}.

\begin{figure}%[hbp]
\begin{center}
\includegraphics[width=0.9\hsize]{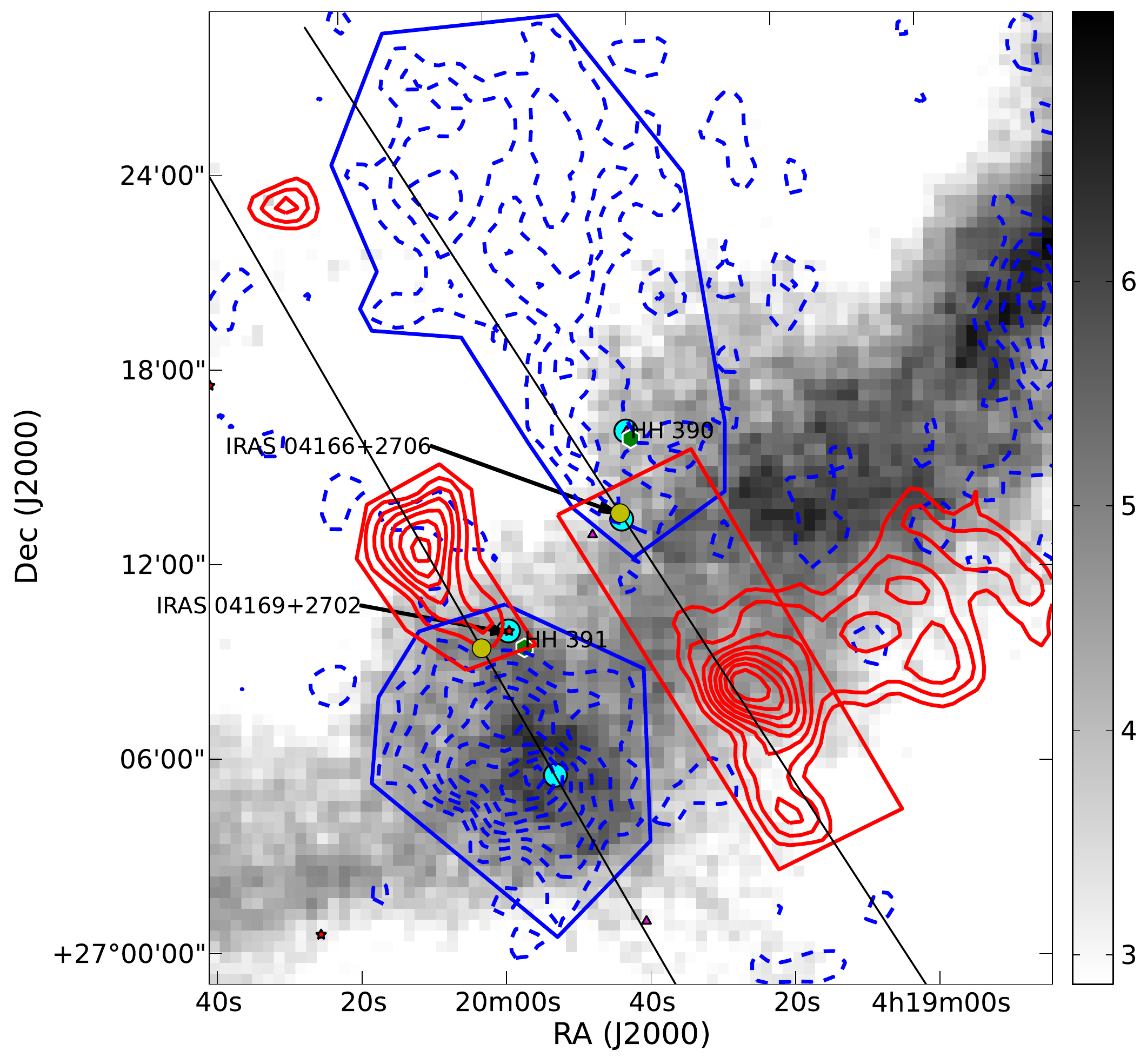}
\caption{Contour map of blueshifted and redshifted gas around IRAS
  04166+2706 and IRAS 04169+2702. See Figure~\ref{041159_outflow} for
  details on symbols and markers.  \co\ blueshifted and redshifted
  integrated intensity are for velocities of {\bf -1 to 4.0 \kms} and
  {\bf 9 to 13 \kms} respectively. Blueshifted contours range from 0.35
  to 4.3 in steps of 0.075 \kkms, and redshifted contours range from
  0.5 to 4.1 in steps of 0.075 \kkms. 
\label{04166_outflow}}
\end{center}
\end{figure}

\begin{figure}%[hbp]
\begin{center}
\includegraphics[width=0.9\hsize]{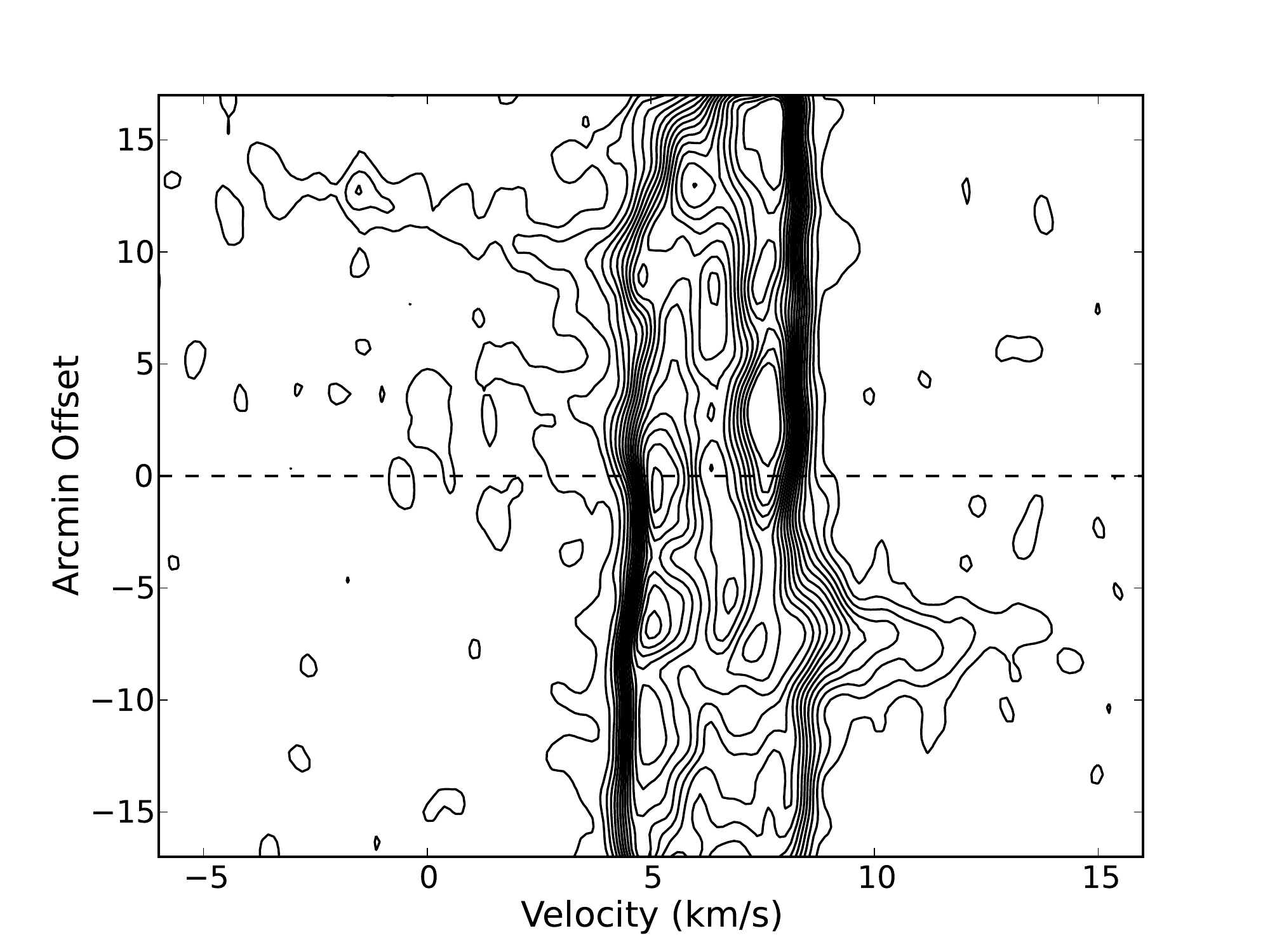}
\caption{Position velocity of \co\ emission towards the IRAS
  04166+2706 region, through the slice at position angle of
  $123^\circ$ through the location of 04166+2706 shown in
  Figure~\ref{04166_outflow}. The contour range is 0.13 to 3.5 K in
  steps of 0.3 K. Shown in dashed line is the position of the yellow
  circle shown in Fig~\ref{04166_outflow}.
\label{04166_posvel}}
\end{center}
\end{figure}

In Figure~\ref{04166_outflow}, we show blueshifted and redshifted
emission in a $26^\prime\times 30^\prime$ region around IRAS
04166+2706 region, superimposed on the low-velocity \coa\ emission in
gray-scale. The adjacent outflow associated with IRAS 04169+2702 (see
\S\ref{04169}) is also seen quite clearly. The outflow towards IRAS
04166+2706 in our data is much larger than previously detected. The
overall length of the outflow is $28^\prime$ ($\sim 1.2$ pc)! The p-v
diagram produced along the line marked in Fig~\ref{04166_outflow} is
shown in Figure~\ref{04166_posvel}.  Our observations are not
sufficiently sensitive to detect the extremely high velocity or the
intermediate high velocity emission that was detected by
\citet{tafalla2004}. We only detect the emission they labeled as
standard high velocity emission.  The p-v diagram also shows that the
highest velocity red and blueshifted emission is found to be
substantially displaced along the outflow axis from IRAS 04166+2706.
In fact, the highest velocity emission detected in our map is well
beyond the maps presented in either \citet{tafalla2004} or
\citet{santiagogarcia2009}.

\begin{figure}%[hbp]
\begin{center}
\includegraphics[width=0.9\hsize]{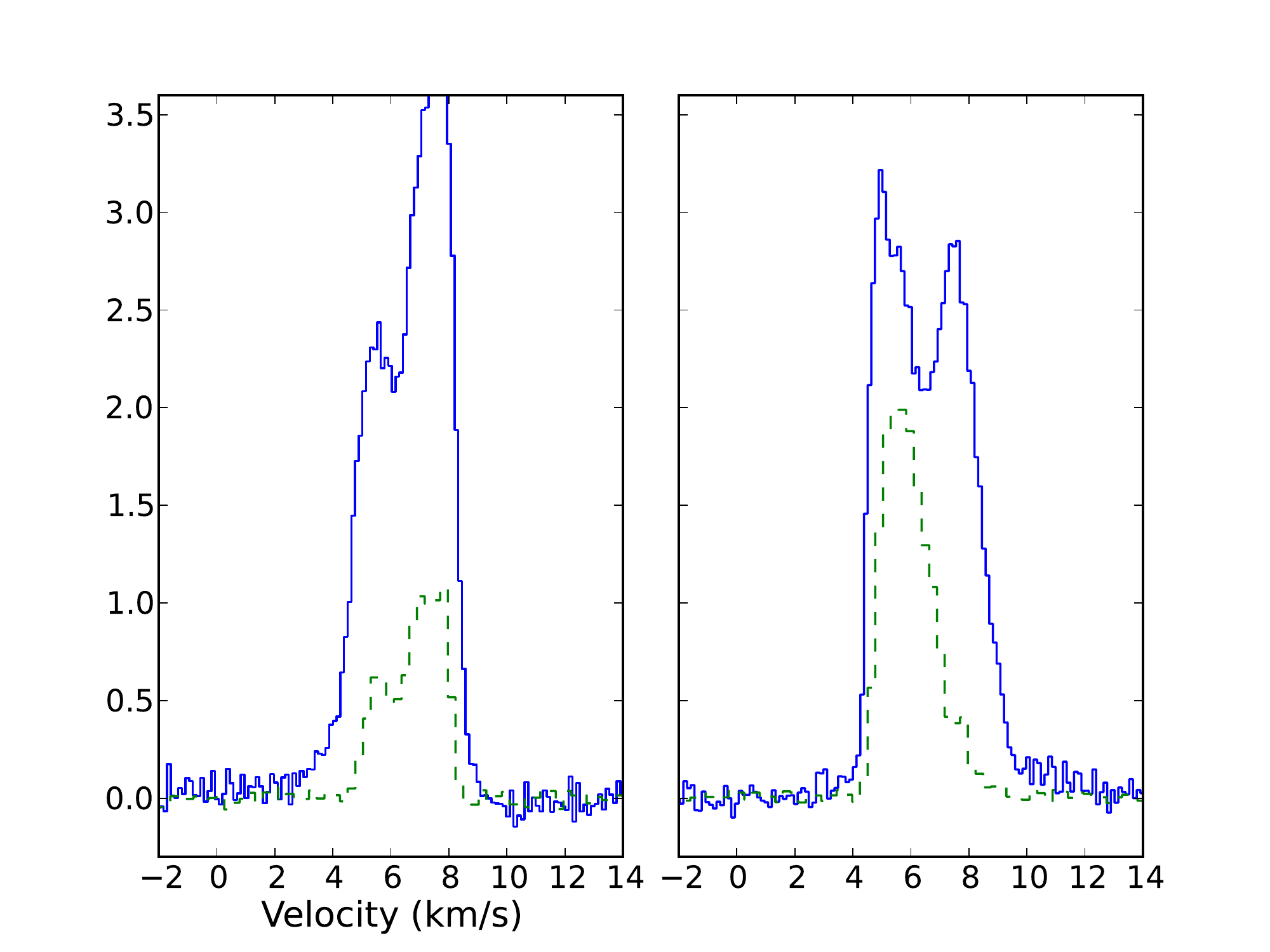}
\caption{Average Spectra of \co\ emission (blue solid lines) and
  \coa\ (green dashed lines) towards the blueshifted (left) and
  redshifted (right) lobes in the IRAS 04166+2706 region shown in
  Fig~\ref{04166_outflow}. The temperature scale is in
  T$_{A}^{*}$. \label{04166_spectra}}
\end{center}
\end{figure}

Averaged spectra within the polygons defining the redshifted and
blueshifted outflow emission associated with the IRAS 04166+2706
outflow (see Figure~\ref{04166_outflow}) are shown in
Figure~\ref{04166_spectra}.  Both redshifted and blueshifted emission
are readily seen in these averaged spectra, however even in the
averaged spectra, the velocity extent corresponds to what
\citet{tafalla2004} labeled as standard high velocity emission.

%% A outflow was first detected in the IRAS 04166+2706 region by
%% \citet{bontemps1996}. This outflow is clearly bipolar, however the
%% small map made by these authors, less than $100\times 100$ arcseconds,
%% did not show the full extent of the outflow.  A larger map was made by
%% \citet{tafalla2004} who showed the outflow to be highly collimated
%% with extremely high velocities.  The full angular extent of the
%% outflow as mapped out by \citet{tafalla2004} is about 6 arcminutes,
%% and emission was detected over the velocity range of V$_{LSR}$ = -50
%% to 50 km s$^{-1}$.  More recently \citet{santiagogarcia2009} mapped
%% the outflow with the IRAM interferometer in the CO J=-2-1 and SiO
%% J=2-1 lines.  Their images revealed two components to the outflow, a
%% low velocity conical shell component that surrounds a high velocity
%% jet component. This bipolar outflow was also detected in the survey
%% by \citet{davis2010}.

%% In Figure~\ref{04166_outflow}, we show blueshifted and redshifted
%% emission in a $26^\prime\times 30^\prime$
%% region in the IRAS 04166+2706 region, superimposed on the low-velocity
%% \coa\ emission in gray-scale. The adjacent outflow in IRAS 04169+2702
%% (see \S\ref{04169}) is also seen quite clearly. The outflow towards
%% IRAS 04166+2706 in our data is much larger than previously
%% detected. The overall length of the outflow is $28^\prime$ ($\sim 1.2$
%% pc)! The p-v diagram measured along the line shown in
%% Fig~\ref{04166_outflow} is shown in Figure~\ref{04166_posvel}.

\subsubsection{IRAS 04169+2702}
\label{04169}

The CO J=3-2 survey of \citet{moriarty-schieven1992} provided the
first evidence of high velocity gas towards IRAS 04169+2702.  They
measured a total velocity extent of 25.4 \kms.  \citet{bontemps1996}
showed this to be a bipolar outflow in the small map they made in the
CO J=2-1 line.  The small region they surveyed only partially mapped
the extent of this outflow.  The survey of \citet{davis2010} covered
this region, but they only detected the redshifted emission. However
the angular extent of their redshifted emission was much larger than
that shown in \citet{bontemps1996}.  \citet{davis2010} suggested a
possible second red only outflow about 4 arcminutes south of IRAS
04169+2702 which they labeled SE-CO-R1.

Figure~\ref{04166_outflow} shows the full extent of this outflow
detected in our survey. This bipolar outflow is approximately 12
arcminutes in extent, much larger than previously measured.
\citet{gomez1997} identified three HH knots (HH 391A, B, and C) that
extend approximately 4.5 arcminutes south of IRAS 04169+2702, these
lie well beyond the extend of the outflow shown in
\citet{bontemps1996}, but lie within the blueshifted molecular outflow
emission defined by our data.  We see no evidence for the redshifted
outflow SE-CO-R1 found by \citet{davis2010}.

\begin{figure}%[hbp]
\begin{center}
\includegraphics[width=0.9\hsize]{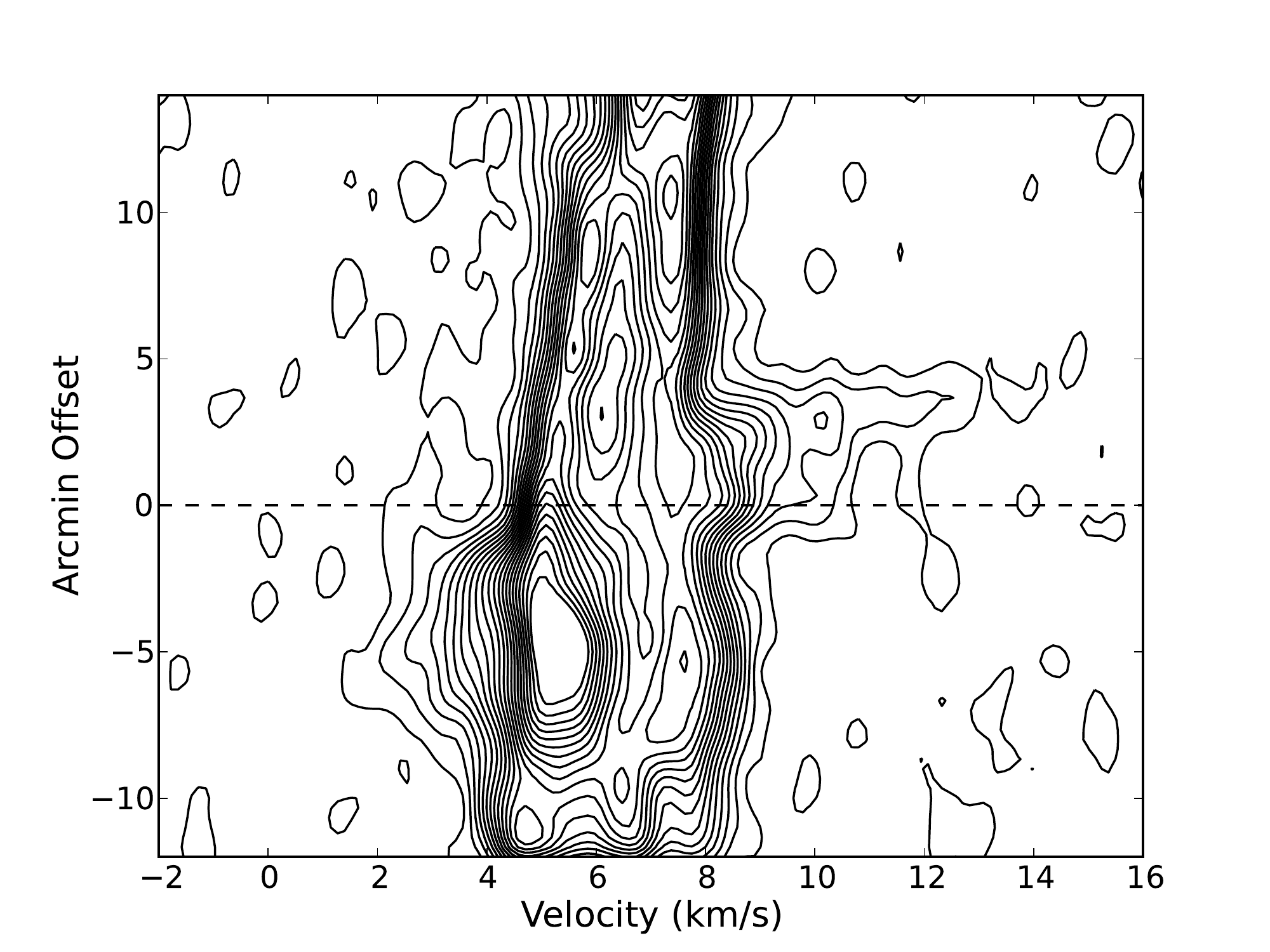}
\caption{Position velocity of \co\ emission towards the IRAS
  04169+2702 region, through the slice at p.a. of $120^\circ$ through
  the location of 04169+2702 shown in Figure~\ref{04166_outflow}. The
  contour range is 0.1 to 4.3 K in steps of 0.3 K. Shown in dashed
  line is the position of the yellow circle shown in
  Fig~\ref{04166_outflow}.
\label{04169_posvel}}
\end{center}
\end{figure}

The p-v diagram along a line passing through IRAS 04169+2702 is shown
in Figure~\ref{04169_posvel} and shows clearly the bipolar nature of
this outflow.  The averaged spectra obtained within the polygons
associated with this outflow also show clearly the high velocity
redshifted and blueshifted emission.

%% The CO J=3-2 survey of \citet{moriarty-schieven1992} provided the
%% first evidence of high velocity gas towards IRAS 04169+2702.  They
%% measured a total velocity extent of 25.4 \kms.  \citet{bontemps1996}
%% showed this to be a bipolar outflow in the small map they made in the
%% CO J=2-1 line.  The small regions surveyed only partially mapped the
%% extent of this outflow.  The survey of \citet{davis2010} covered this
%% region, however they only detected the redshifted emission, although
%% the angular extent of their redshifted emission was much larger than
%% that shown in \citet{bontemps1996}.  \citet{davis2010} detected a
%% second outflow about 4 arcminutes sourth of IRAS 04169+2702 which they
%% labeled SE-CO-R1.

%% Figure~\ref{04166_outflow} shows the full extent of the outflow
%% detected in our survey. The p-v diagram of the outflow taken about the
%% line passing through IRAS 04169+2702 is shown in Figure~\ref{04169_posvel}.

\begin{figure}%[hbp]
\begin{center}
\includegraphics[width=0.9\hsize]{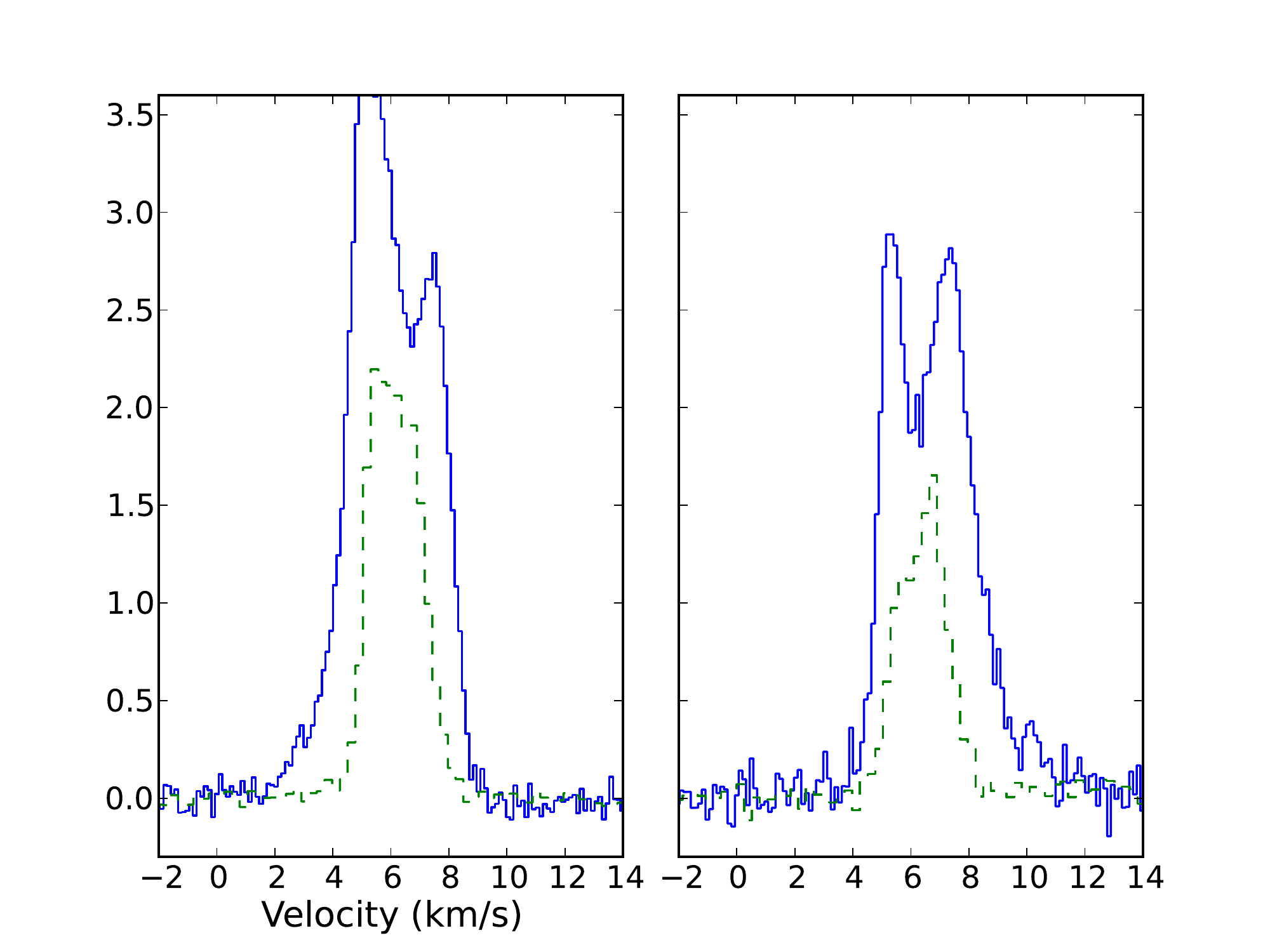}
\caption{Average Spectra of \co\ emission (blue solid lines) and
  \coa\ (green dashed lines) towards the blueshifted (left) and
  redshifted (right) lobes in the IRAS 04169+2702 region shown in
  Fig~\ref{04166_outflow}. The temperature scale is in
  T$_{A}^{*}$. \label{04169_spectra}}
\end{center}
\end{figure}

\subsubsection{FS Tau B}

This molecular outflow was first identified by \citet{davis2010}.
They detected redshifted only emission near FS Tau A/B.  Previously,
images of H$\alpha$ and [S II] emission obtained by
\citet{eisloffel1998} revealed a striking bipolar outflow associated
with FS Tau B. Besides the presence of two conical reflection nebulae,
a jet and counter jet were detected originating from FS Tau B.  The
blueshifted jet was directed toward the northeast, while the
redshifted counter jet was directed toward the southwest.  In
addition, a bow-shock feature was found associated with the blueshifted
jet and a string of blueshifted HH-objects that extend approximately 6
arcminutes to the north-east of FS Tau B.  \citet{davis2010} assumed
FS Tau B as the source of the molecular outflow they detected.

\begin{figure}%[hbp]
\begin{center}
\includegraphics[width=0.9\hsize]{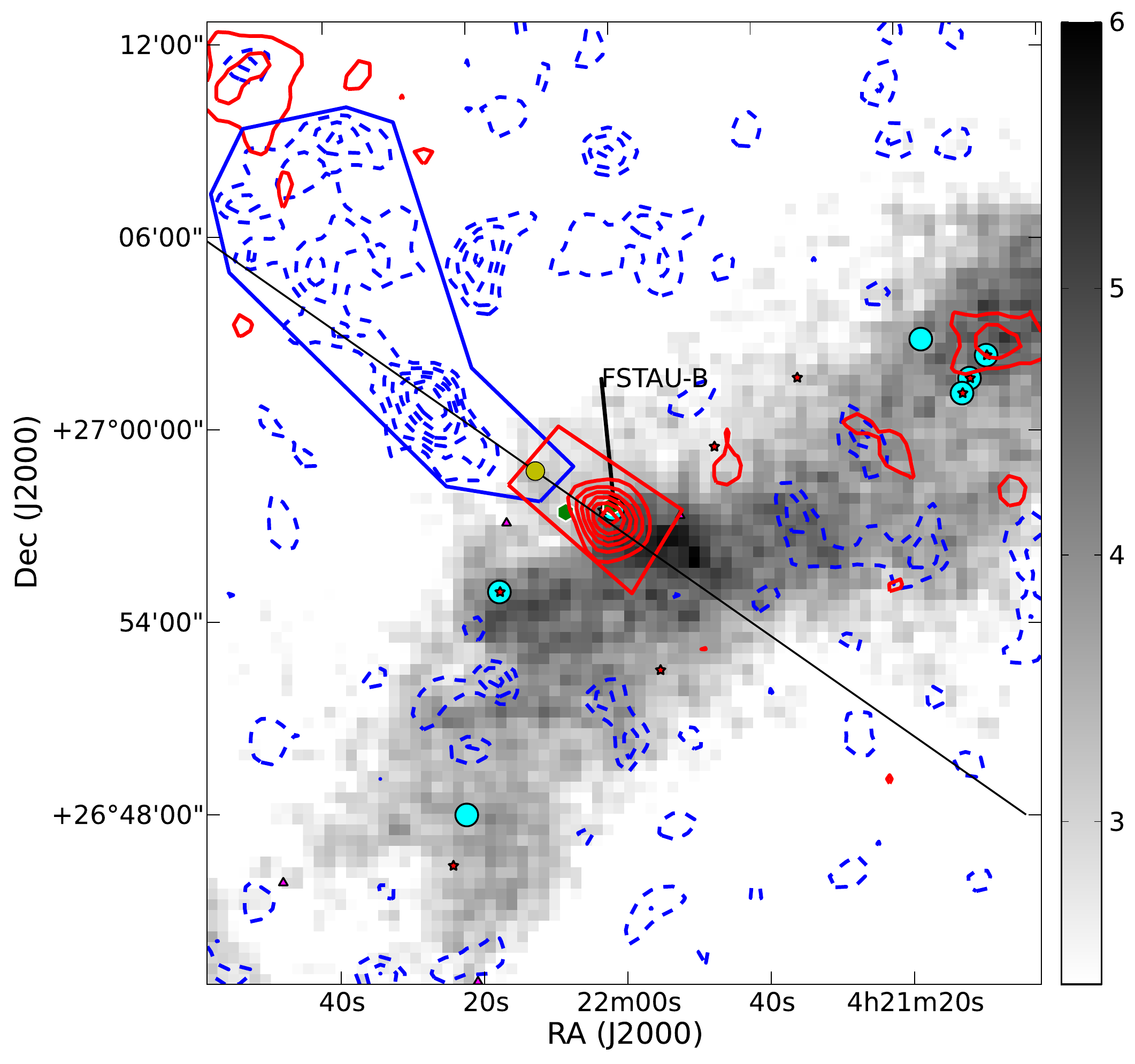}
\caption{Contour map of blueshifted and redshifted gas about a
  $30^\prime\times 30^\prime$ region near FS-Tau~B. See
  Figure~\ref{041159_outflow} for details on symbols and markers.
  \co blueshifted and redshifted integrated intensity is for
  velocities of {\bf -1 to 4.1 \kms} and {\bf 7.8 to 12 \kms} respectively.
  Blueshifted contours range from 0.45 to 3.1
  in steps of 0.075 \kkms, and redshifted contours range from 0.45 to
  3.1 in steps of 0.075 \kkms.
\label{nbs2_outflow}}
\end{center}
\end{figure}

Our map of the redshifted and blueshifted emission in this region is
presented in Figure~\ref{nbs2_outflow}. In addition to the redshifted
emission centred on FS Tau B, seen by \citet{davis2010}, we find
blueshifted emission to the north-east.  The blueshifted emission is
located along the string of HH-objects found by
\citet{eisloffel1998}. We believe the red and blueshifted emission to
be all part of a large bipolar molecular outflow which shares the same
position angle and velocity bipolarity as the optical jets from FS Tau
B.

\begin{figure}%[hbp]
\begin{center}
\includegraphics[width=0.9\hsize]{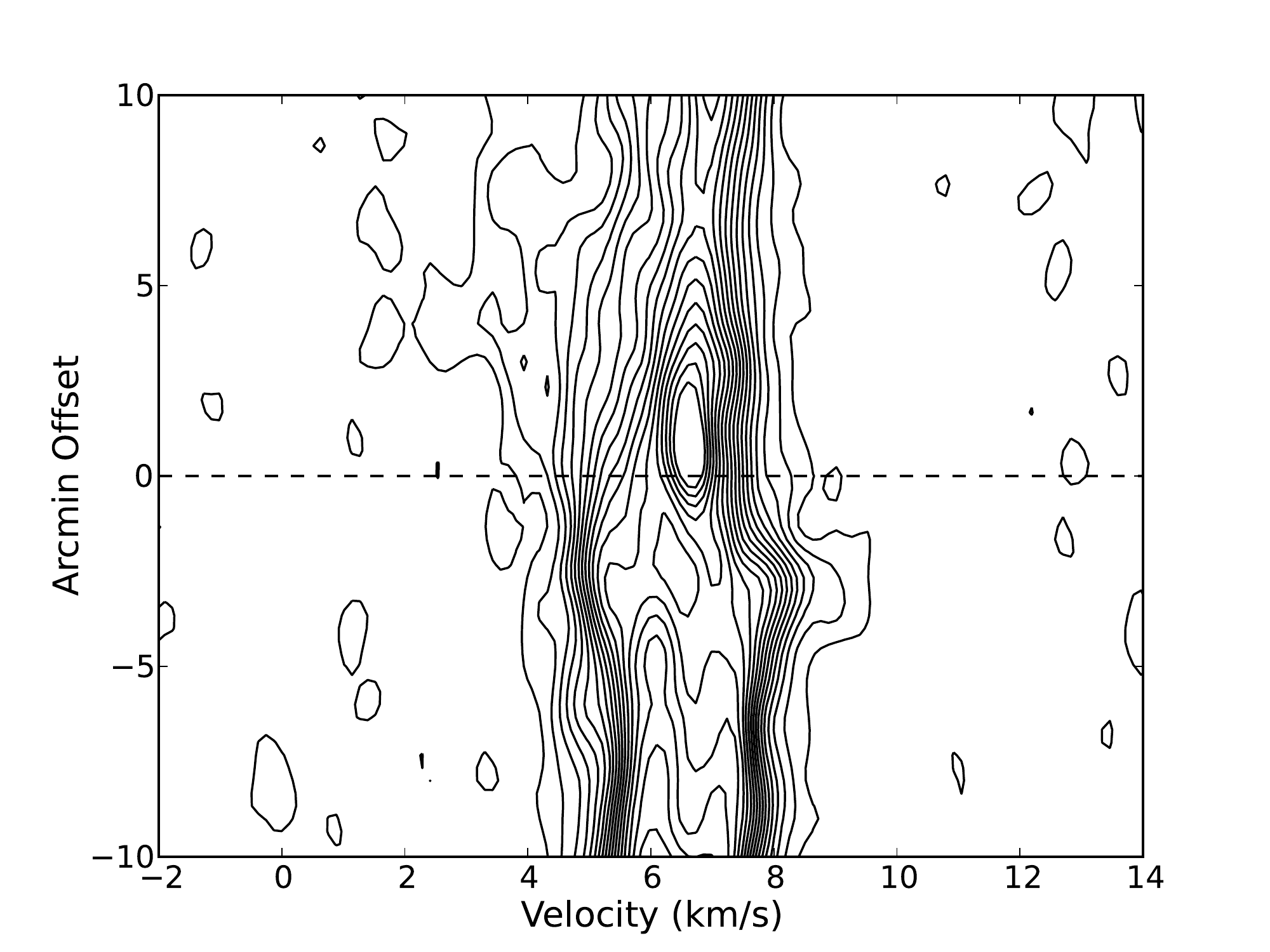}
\caption{Position velocity of \co\ emission towards the FS~Tau~B
  region, through the slice at p.a. of $146^\circ$ shown in
  Figure~\ref{nbs2_outflow}. The contour range is 0.12 to 3.5 K in
  steps of 0.3 K. Shown in dashed line is the position of the yellow
  circle shown in Fig~\ref{nbs2_outflow}.
\label{nbs2_posvel}}
\end{center}
\end{figure}

\begin{figure}%[hbp]
\begin{center}
\includegraphics[width=0.9\hsize]{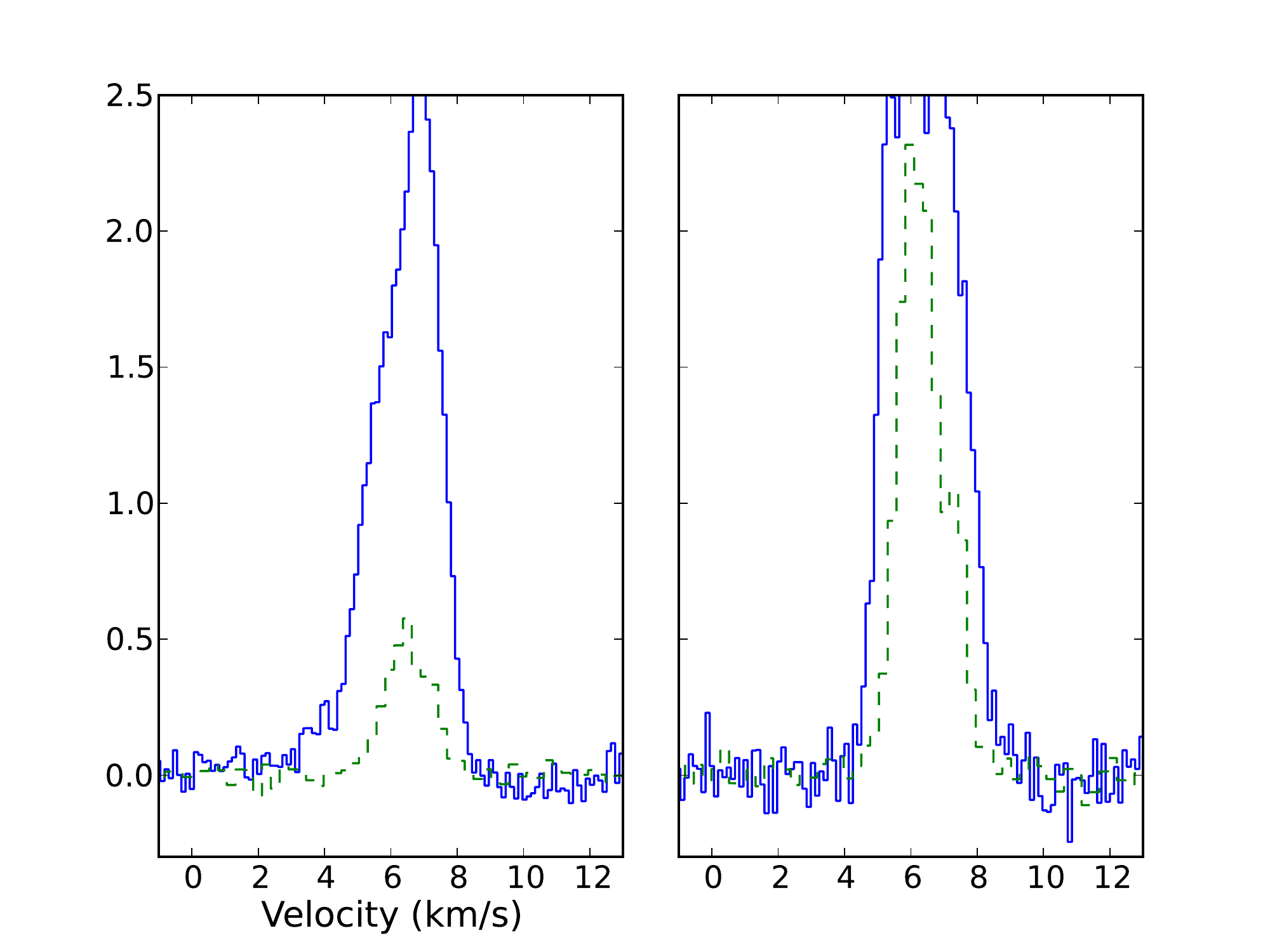}
\caption{Average Spectra of \co\ emission (blue solid lines) and
  \coa\ (green dashed lines) towards the blueshifted (left) and
  redshifted lobes (right) in the FS~Tau~B region shown in
  Fig~\ref{nbs2_outflow}. The temperature scale is in T$_{A}^{*}$.
\label{nbs2_spectra}}
\end{center}
\end{figure}

A p-v diagram of the outflow along a line through FS Tau B is shown in
Figure~\ref{nbs2_posvel}.  This p-v diagram shows much more prominent
blueshifted emission than redshifted emission.  In
Figure~\ref{nbs2_spectra} averaged spectra within the polygons marked
in Figure~\ref{nbs2_outflow} are shown and again in these
averaged spectra the blueshifted emission is much more obvious than
the redshifted emission.

\subsubsection{IRAS04239+2436}

\begin{figure}%[hbp]
\begin{center}
\includegraphics[width=0.9\hsize]{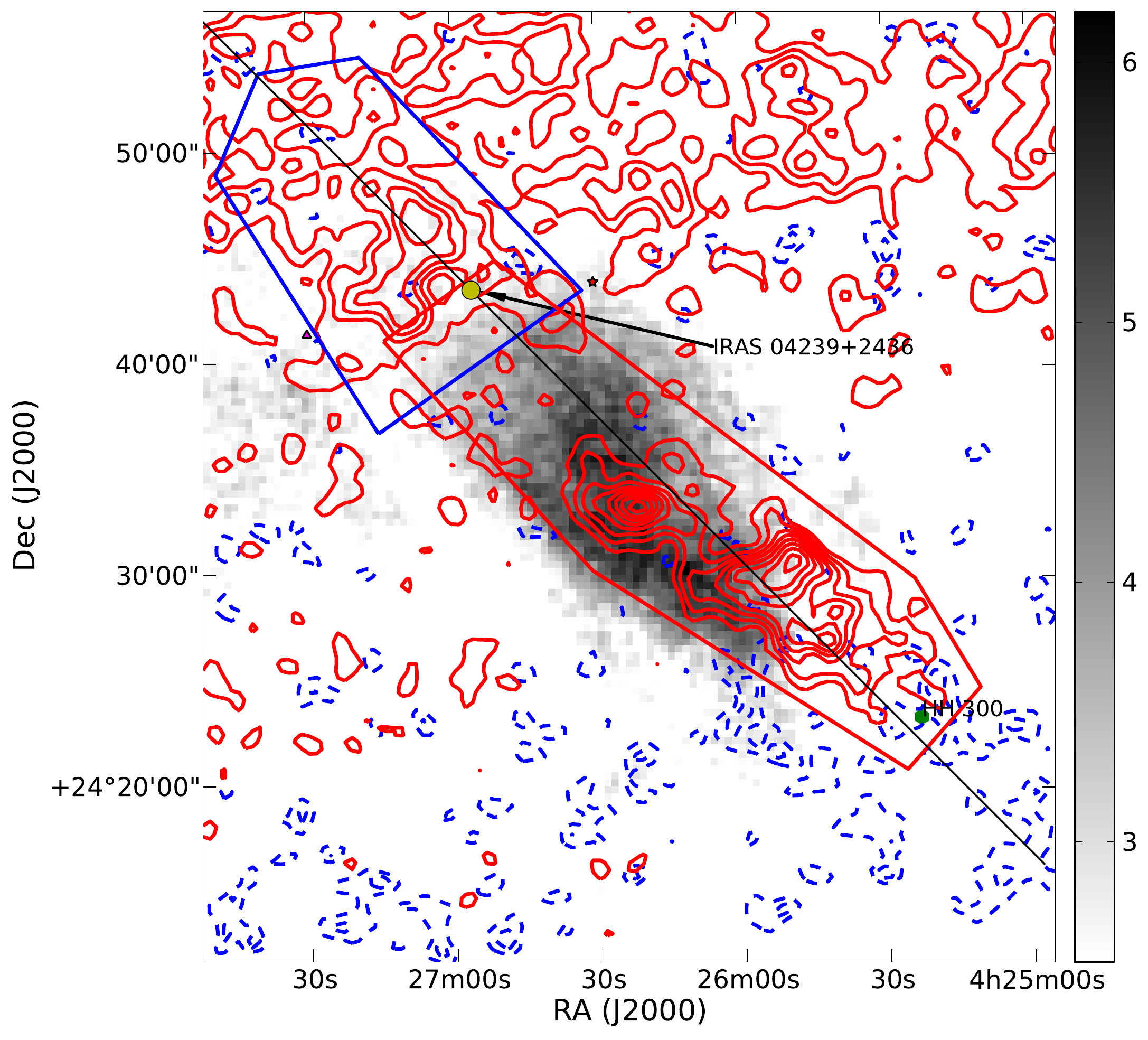}
\caption{Contour map of blueshifted and redshifted gas about a
  $40^\prime\times 45^\prime$ region near IRAS04239+2436. See
  Figure~\ref{041159_outflow} for details on symbols and markers.
  \co\ blueshifted and redshifted integrated intensity is for
  velocities of {\bf 0 to 3.4 \kms} and {\bf 7.9 to 12 \kms}
  respectively. Blueshifted contours range from 0.36 to 1.8 in steps
  of 0.075 \kkms, and redshifted contours range from 0.36 to 3.72 in
  steps of 0.075 \kkms.
\label{04239_outflow}}
\end{center}
\end{figure}

\begin{figure}%[hbp]
\begin{center}
\includegraphics[width=0.9\hsize]{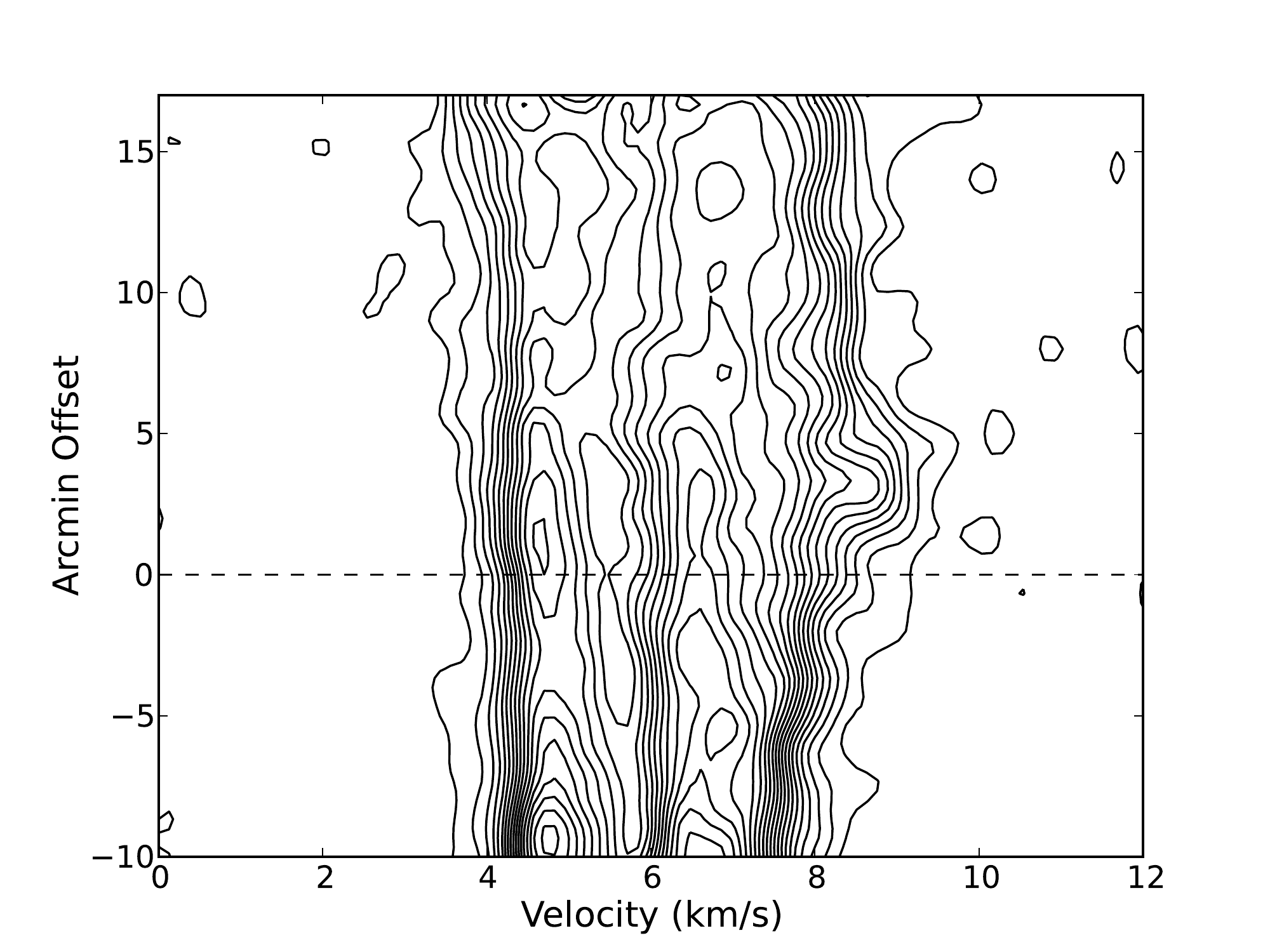}
\caption{Position velocity of \co\ emission towards the IRAS-4239+2436
  region, through the slice at p.a. of $135^\circ$ shown in
  Figure~\ref{04239_outflow}. The contour range is 0.15 to 3.95 K in
  steps of 0.3 K. Shown in dashed line is the position of the yellow
  circle shown in Fig~\ref{04239_outflow}.
\label{04239_posvel}}
\end{center}
\end{figure}

\begin{figure}%[hbp]
\begin{center}
\includegraphics[width=0.9\hsize]{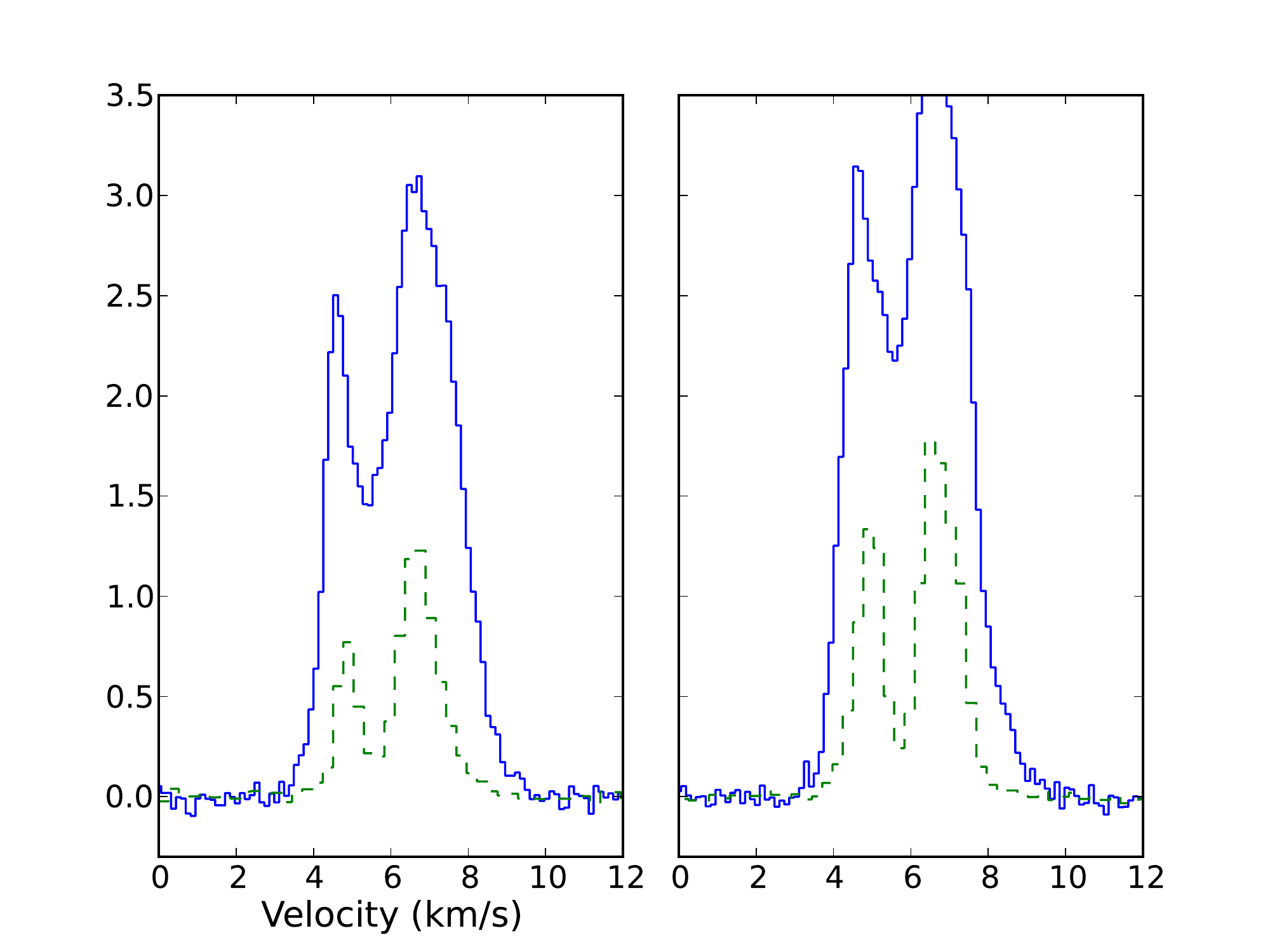}
\caption{Average Spectra of \co\ emission (blue solid lines) and
  \coa\ (green dashed lines) towards the blueshifted lobe (left) and
  redshifted lobe (right) towards the outflow in the IRAS04239+2436
  region shown in Fig~\ref{04239_outflow}. The temperature scale is in
  T$_{A}^{*}$. \label{04239_spectra}}
\end{center}
\end{figure}

The CO J=3-2 spectrum obtained by \citet{moriarty-schieven1992} toward
IRAS 04239+2436 had a full velocity width of 24.7 \kms, suggesting the
presence of a molecular outflow.  This region was mapped by
\citet{arce2001} and they detected a redshifted only molecular outflow
that is associated with the large HH 300 optical outflow
\citep{reipurth1997}.  Our map of the outflow emission shown in
Figure~\ref{04239_outflow} shows a structure very similar to that
found by \citet{arce2001}.  The p-v diagram in Fig~\ref{04239_posvel}
shows a broadening due to the outflow. The averaged spectra in this
region presented in Figure~\ref{04239_spectra} shows high velocity
redshifted emission and little evidence for any high velocity
blueshifted emission.

\citet{arce2001} suggest that the blue-shifted complement of this
outflow is obscured due to contamination from emission from another
molecular cloud along the line of sight. We can confirm the presence
of a strong \co\ and \coa\ peak at $\sim 5$ \kms\ (see
Fig~\ref{04239_spectra}), which might obscure the detection of a
blue-shifted wing towards this source.

\subsubsection{IRAS04248+2612}

\begin{figure}%[hbp]
\begin{center}
\includegraphics[width=0.9\hsize]{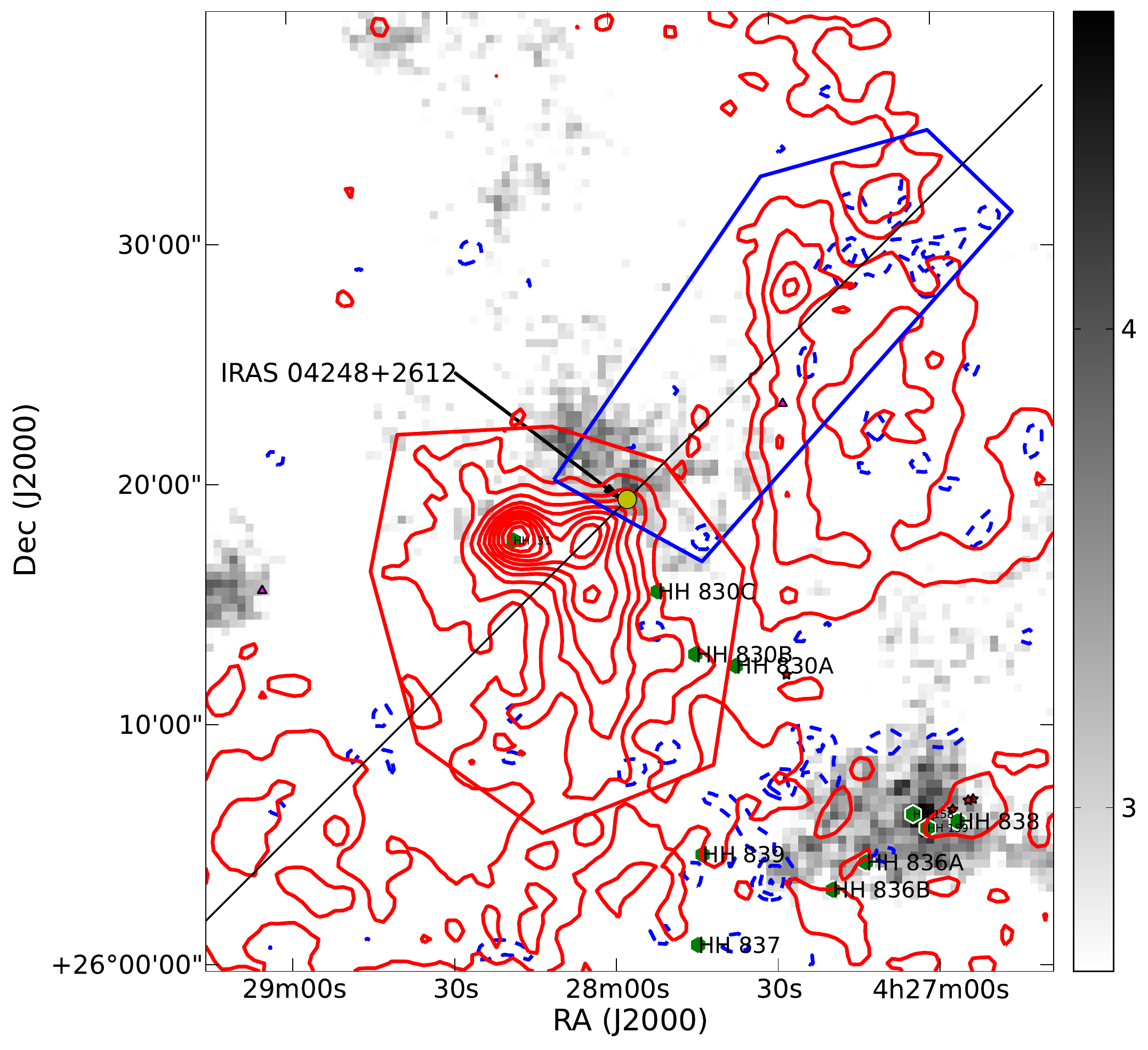}
\caption{Contour map of blueshifted and redshifted gas about a
  $40^\prime\times 45^\prime$ region near IRAS04248+2612. See
  Figure~\ref{041159_outflow} for details on symbols and markers.
  \co\ blueshifted and redshifted integrated intensity is for
  velocities of {\bf 0 to 3.6 \kms} and {\bf 8.1 to 13 \kms}
  respectively. Blueshifted contours range from 0.43 to 2.0 in steps
  of 0.075 \kkms, and redshifted contours range from 0.43 to 6.6 in
  steps of 0.075 \kkms.
\label{04248_outflow}}
\end{center}
\end{figure}

\begin{figure}%[hbp]
\begin{center}
\includegraphics[width=0.9\hsize]{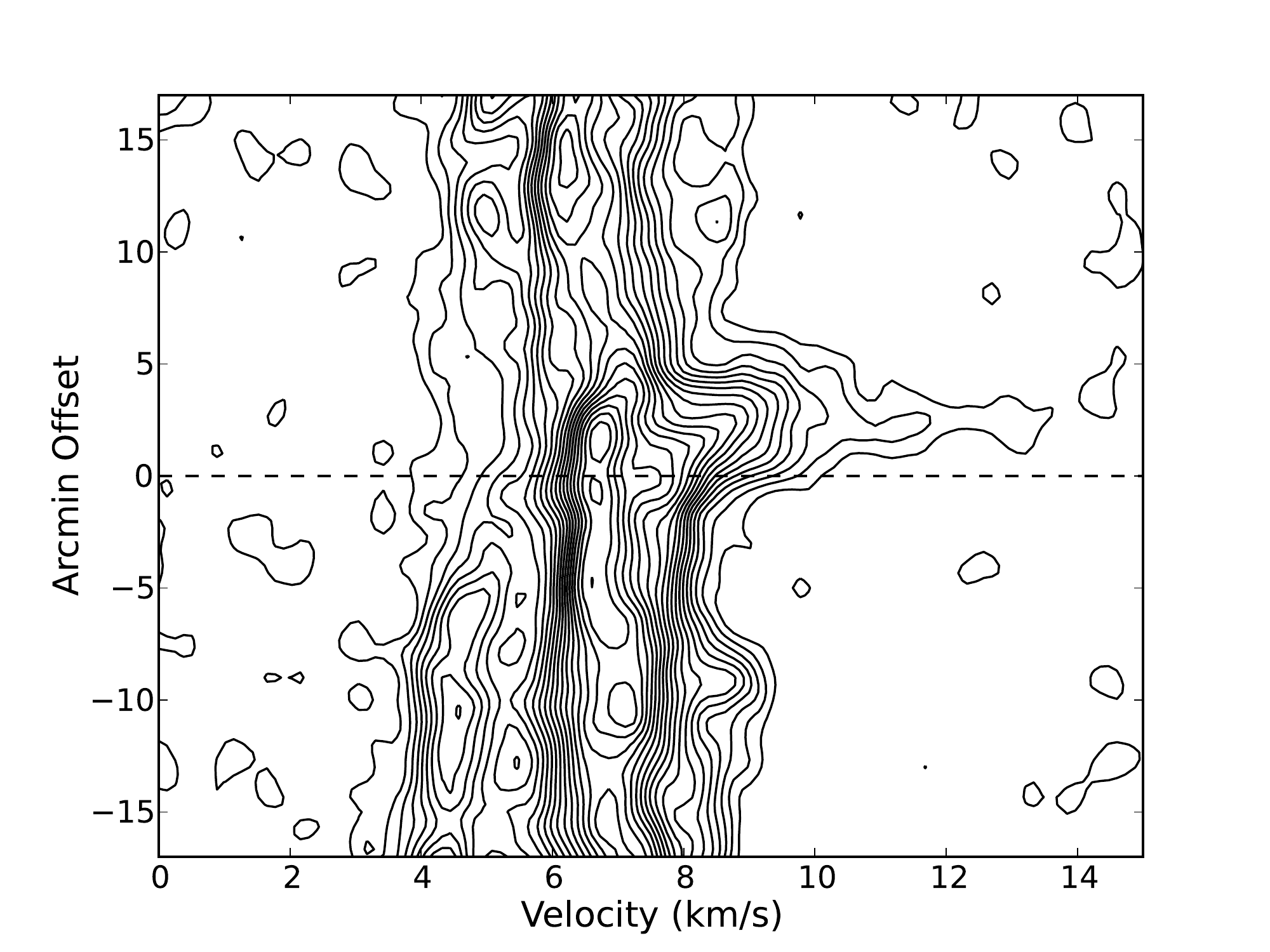}
\caption{Position velocity of \co\ emission towards the IRAS-4239+2612
  region, through the slice at p.a. of $45^\circ$ shown in
  Figure~\ref{04248_outflow}. The contour range is 0.15 to 3.95 K in
  steps of 0.3 K. Shown in dashed line is the position of the yellow
  circle shown in Fig~\ref{04248_outflow}.
\label{04248_posvel}}
\end{center}
\end{figure}

\begin{figure}%[hbp]
\begin{center}
\includegraphics[width=0.9\hsize]{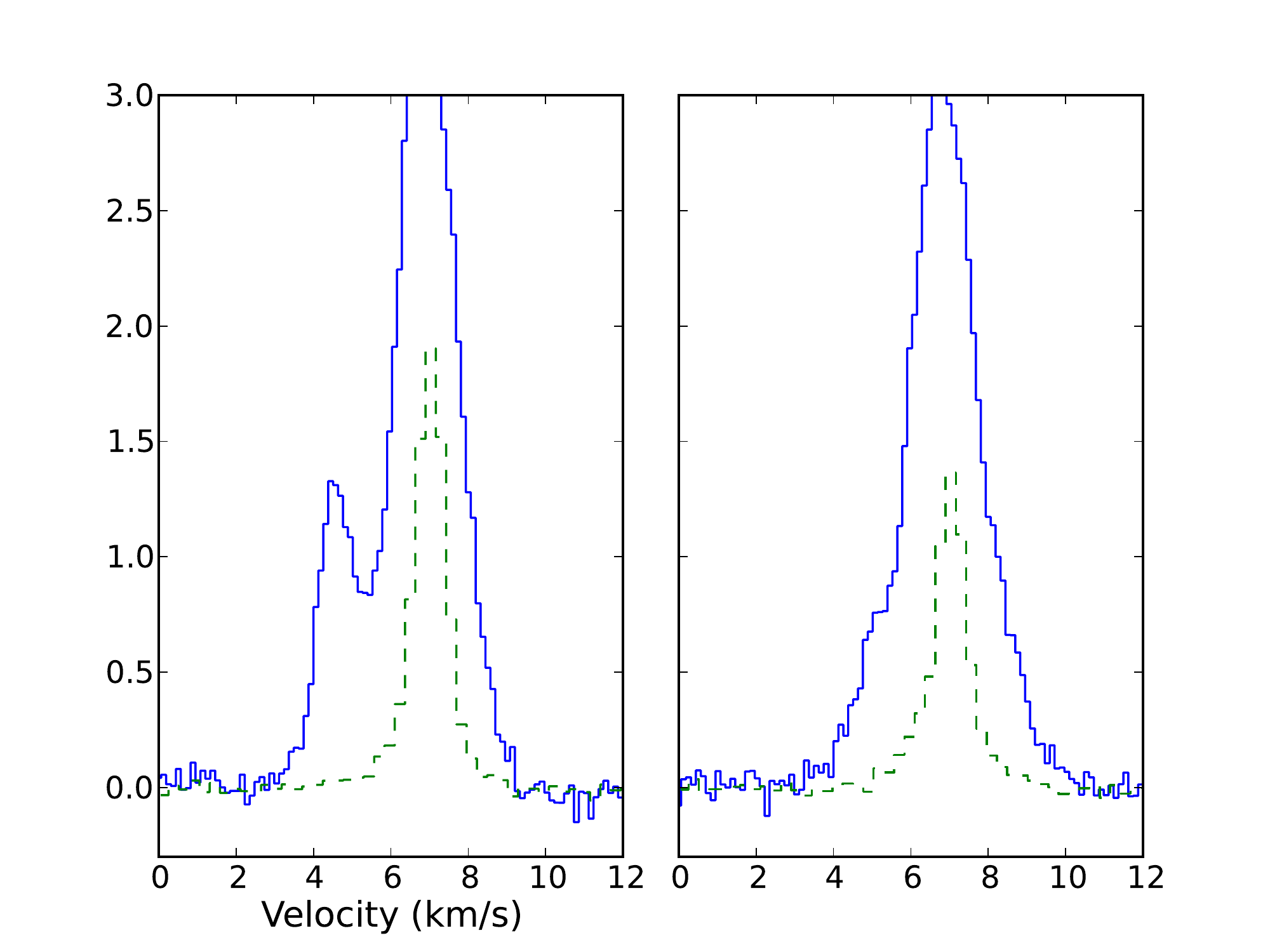}
\caption{Average Spectra of \co\ emission (blue solid lines) and
  \coa\ (green dashed lines) towards the blueshifted (left) and
  redshifted (right) lobes of the outflow in the IRAS04248+2612 region
  shown in Fig~\ref{04248_outflow}. The temperature scale is in
  T$_{A}^{*}$. \label{04248_spectra}}
\end{center}
\end{figure}

The outflow associated with IRAS 04248+2612 (also called HH31 IRS2) is
newly detected.  The distribution of high velocity CO emission is
shown in Figure~\ref{04248_outflow}.  The only prominent outflow
emission is redshifted, and this emission extends to the southeast of
IRAS 04248+2612.  Associated with the redshifted emission is a string
of HH objects (HH 31A-I) that extend to about 5 arcminutes to the
south-east of the IRAS source \citet{gomez1997}.  A position-velocity
diagram obtained along the line marked in Figure~\ref{04248_outflow}
is presented in Figure~\ref{04248_posvel} and shows the prominent
redshifted emission slightly offset from IRAS 04248+2612.  The
averaged spectra in the polygons marked in Figure~\ref{04248_outflow}
are shown in Figure~\ref{04248_spectra}.  Although the spectrum
\citet{moriarty-schieven1992} obtained toward the IRAS source does not
show very broad emission, we see clearly redshifted emission in the
averaged spectrum.  We also selected a region to the north-west of
IRAS 04248+2612 to obtain an averaged blue spectrum, and this spectrum
shows a distinct feature at approximately 4.5 \kms\ also seen in the
p-v plot.  Whether this is an unrelated velocity feature or connected
to the outflow is impossible to tell, however this feature is not
detected in \coa.

\subsubsection{Haro~6-10}

The first evidence for high velocity gas in this region was provided
by a snap-shot interferometer survey by \citet{terebey1989}.  A small
map of this region in the J=3-2 line of CO was made by
\citet{hogerheijde1998} and revealed a bipolar outflow.  This region
was subsequently mapped more extensively by \citet{stojimirovic2007}
in the CO J=1-0 line and a large bipolar outflow was detected.  This
outflow is associated with a giant Herbig-Haro flow centred on Haro
6-10 \citep{devine1999}.  Haro 6-10 is also called GV Tau.

\begin{figure}%[hbp]
\begin{center}
\includegraphics[width=0.9\hsize]{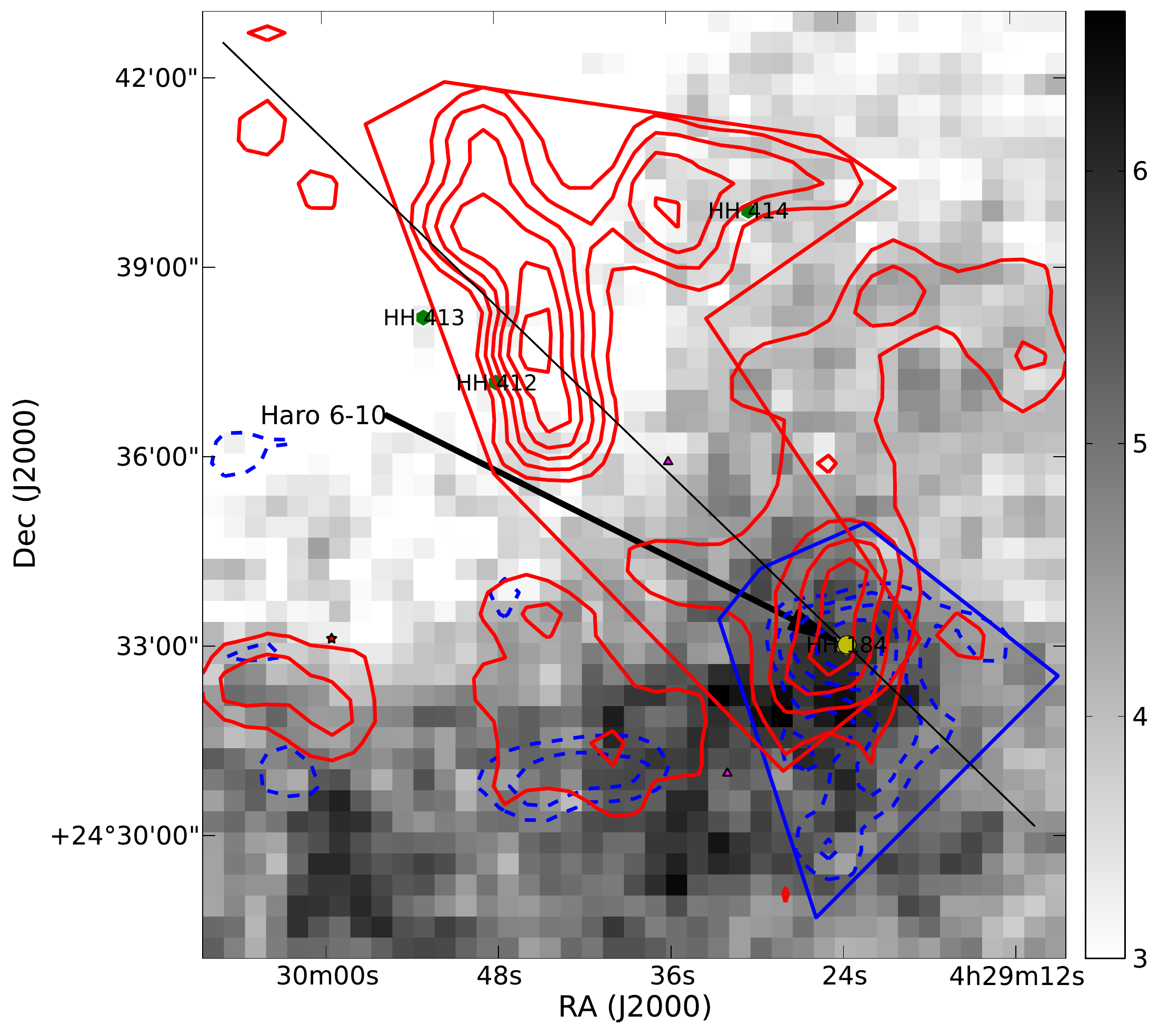}
\caption{Contour map of blueshifted and redshifted gas about a
  $15^\prime\times 15^\prime$ region near Haro~6-10. See
  Figure~\ref{041159_outflow} for details on symbols and markers.
  \co\ blueshifted and integrated intensity are for velocities of {\bf
    0 to 3.5 \kms} and {\bf 8.5 to 12 \kms} respectively. Blueshifted
  contours range from 0.39 to 1.7 in steps of 0.075 \kkms, and
  redshifted contours range from 0.39 to 2.3 in steps of 0.075
  \kkms.
  \label{haro610_outflow}}
\end{center}
\end{figure}

\begin{figure}%[hbp]
\begin{center}
\includegraphics[width=0.9\hsize]{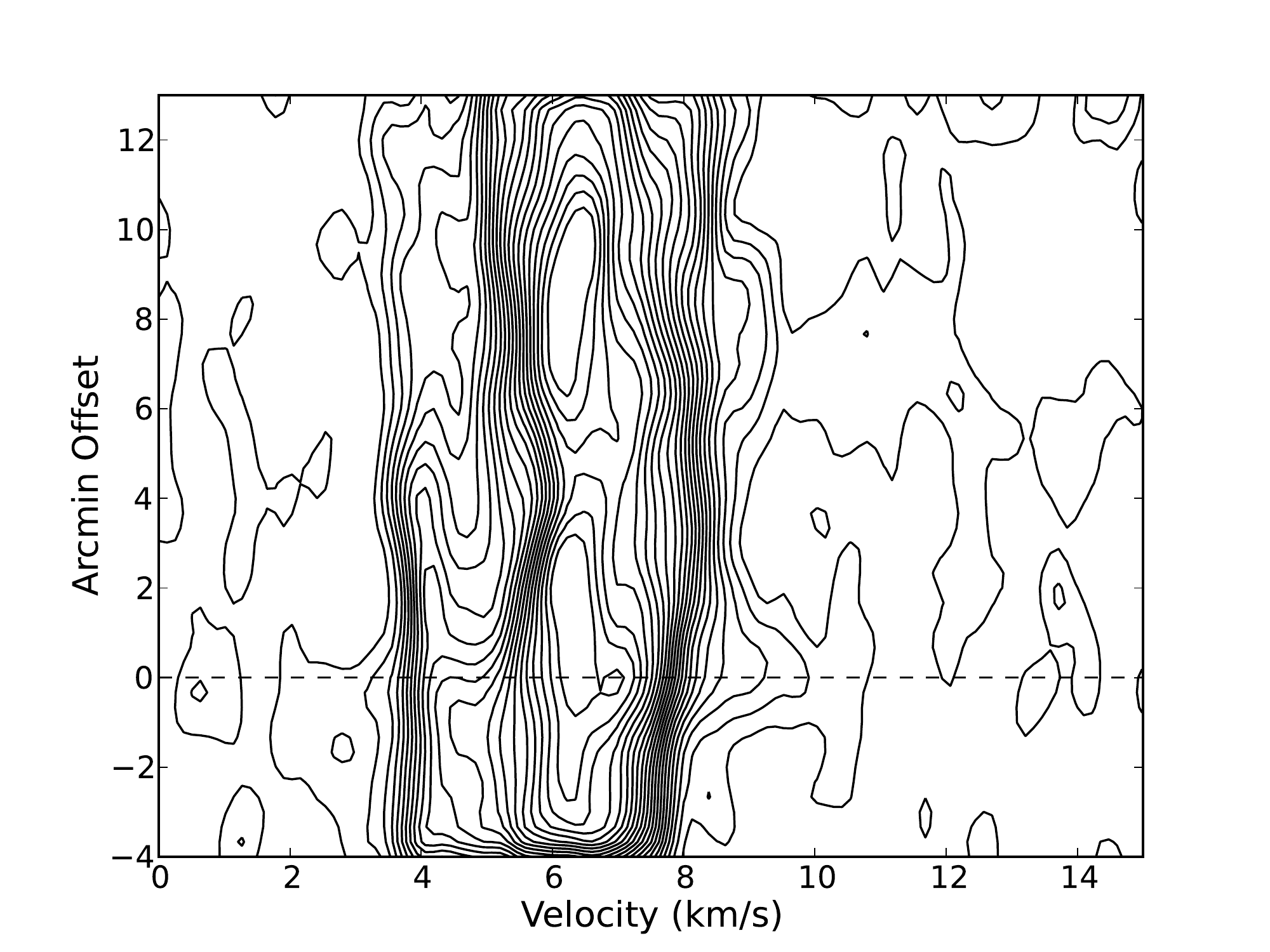}
\caption{Position velocity of \co\ emission towards the Haro~6-10
  region, through the slice at p.a. of $136^\circ$ shown in
  Figure~\ref{haro610_outflow}. The contour range is 0.1 to 4.3 K in
  steps of 0.3 K. Shown in dashed line is the position of the yellow
  circle shown in Fig~\ref{haro610_outflow}.
\label{haro610_posvel}}
\end{center}
\end{figure}

\begin{figure}%[hbp]
\begin{center}
\includegraphics[width=0.9\hsize]{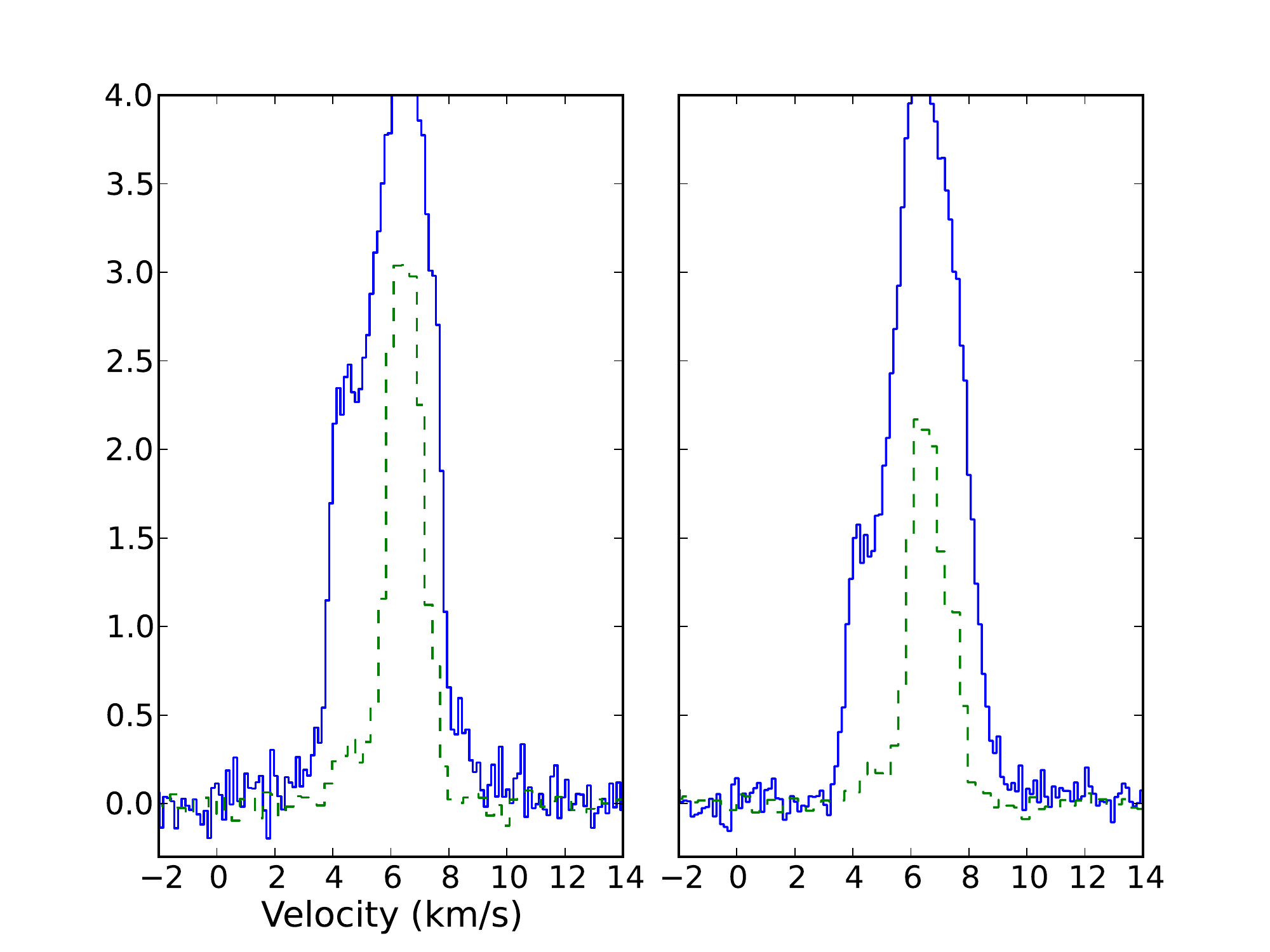}
\caption{Average Spectra of \co\ emission (blue solid lines) and
  \coa\ (green dashed lines) towards the blueshifted (left) and redshifted
  (right) lobes of the outflow towards the Haro~6-10 region shown in
  Fig~\ref{haro610_outflow}. The temperature scale is in
  T$_{A}^{*}$. 
\label{haro610_spectra}}
\end{center}
\end{figure}

Figure~\ref{haro610_outflow} shows the redshifted and blueshifted
emission in this region from our data.  The distribution of high
velocity emission is very similar to what was found by
\citet{stojimirovic2007}.  The position-velocity map along the cut
marked in Figure~\ref{haro610_outflow} is presented in
Figure~\ref{haro610_posvel} and shows weak redshifted and blue
shifted outflow emission.  The polygon averaged spectra, shown in
Figure~\ref{haro610_spectra}, also show evidence for weak high
velocity redshifted and blueshifted emission.

\subsubsection{IRAS 04278+2435 (ZZ Tau IRS)}

Extended high velocity redshifted emission was found in the region
around IRAS 04278+2435 by \citet{heyer1987} and they associated this
monopolar outflow with the pre-main sequence star ZZ Tau.  The catalogue
of YSOs of \citet{kenyon2008} lists both ZZ Tau AB and ZZ Tau IRS
(IRAS 04278+2435) that are separated by less than 1 arcminute, and
either, based on the morphology of the outflow, may be the origin of
this flow.  This outflow has a distinct redshifted velocity feature
similar to that seen in the L1551 IRS 5 outflow \citep{snell1980}, and
as in L1551 IRS 5 may be evidence for a swept-up shell.  Their map
revealed little evidence for any significant blueshifted emission.
\citet{gomez1997} detected a [SII] emission knot (HH 393) in this
region and argue that ZZ Tau IRS is the most likely driving source for
this outflow.

\begin{figure}%[hbp]
\begin{center}
\includegraphics[width=0.9\hsize]{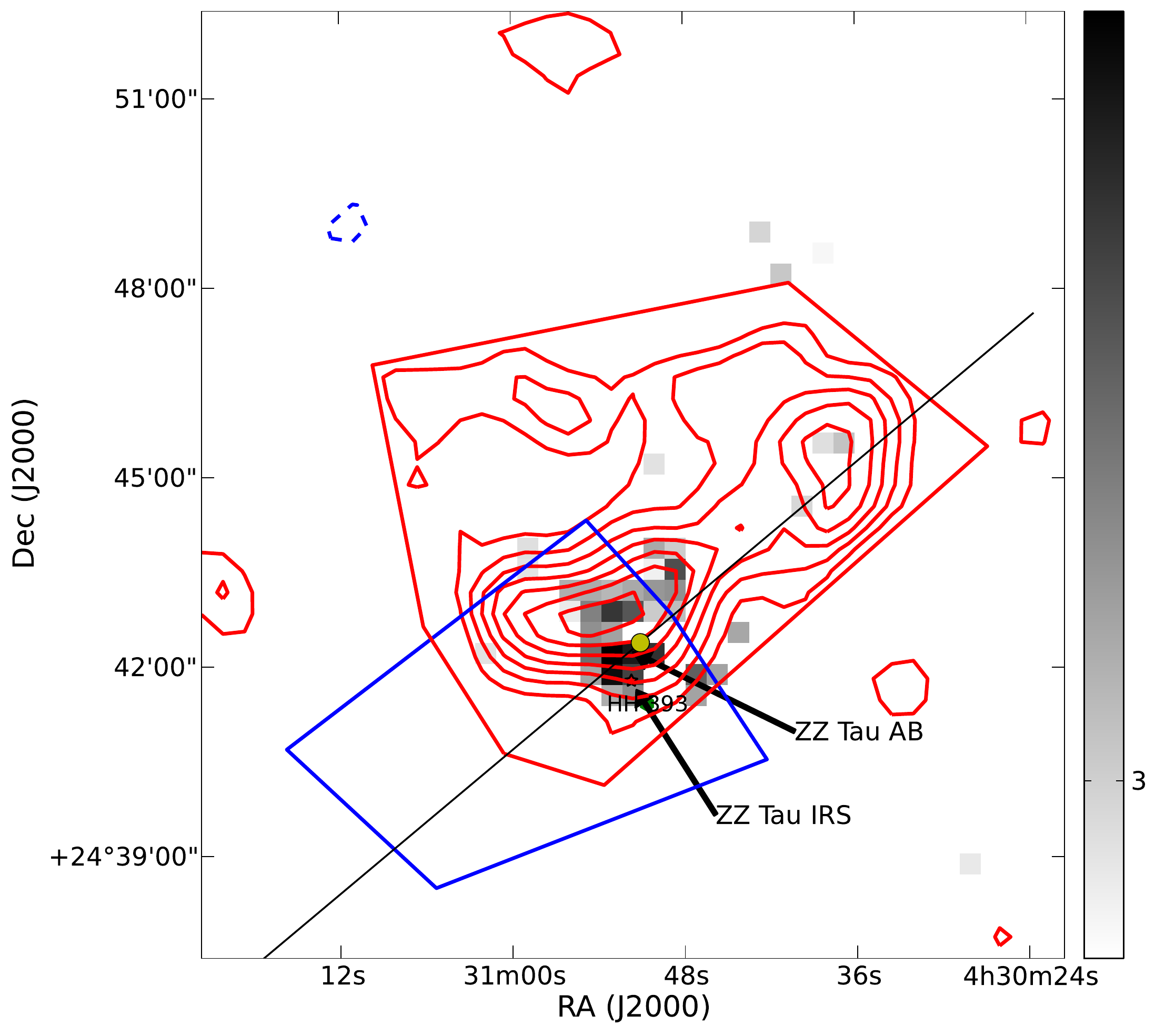}
\caption{Contour map of blueshifted and redshifted gas about a
  $15^\prime\times 15^\prime$ region near ZZ Tau. See
  Figure~\ref{041159_outflow} for details on symbols and markers.
  \co\ blueshifted and redshifted integrated intensity is for
  velocities of {\bf 0 to 3.5 \kms} and {\bf 8.5 to 12 \kms}
  respectively. Blueshifted contours range from 0.42 to 1.7 in steps
  of 0.075 \kkms, and redshifted contours range from 0.42 to 2.8 in
  steps of 0.075 \kkms. 
\label{zztau_outflow}}
\end{center}
\end{figure}

\begin{figure}%[hbp]
\begin{center}
\includegraphics[width=0.9\hsize]{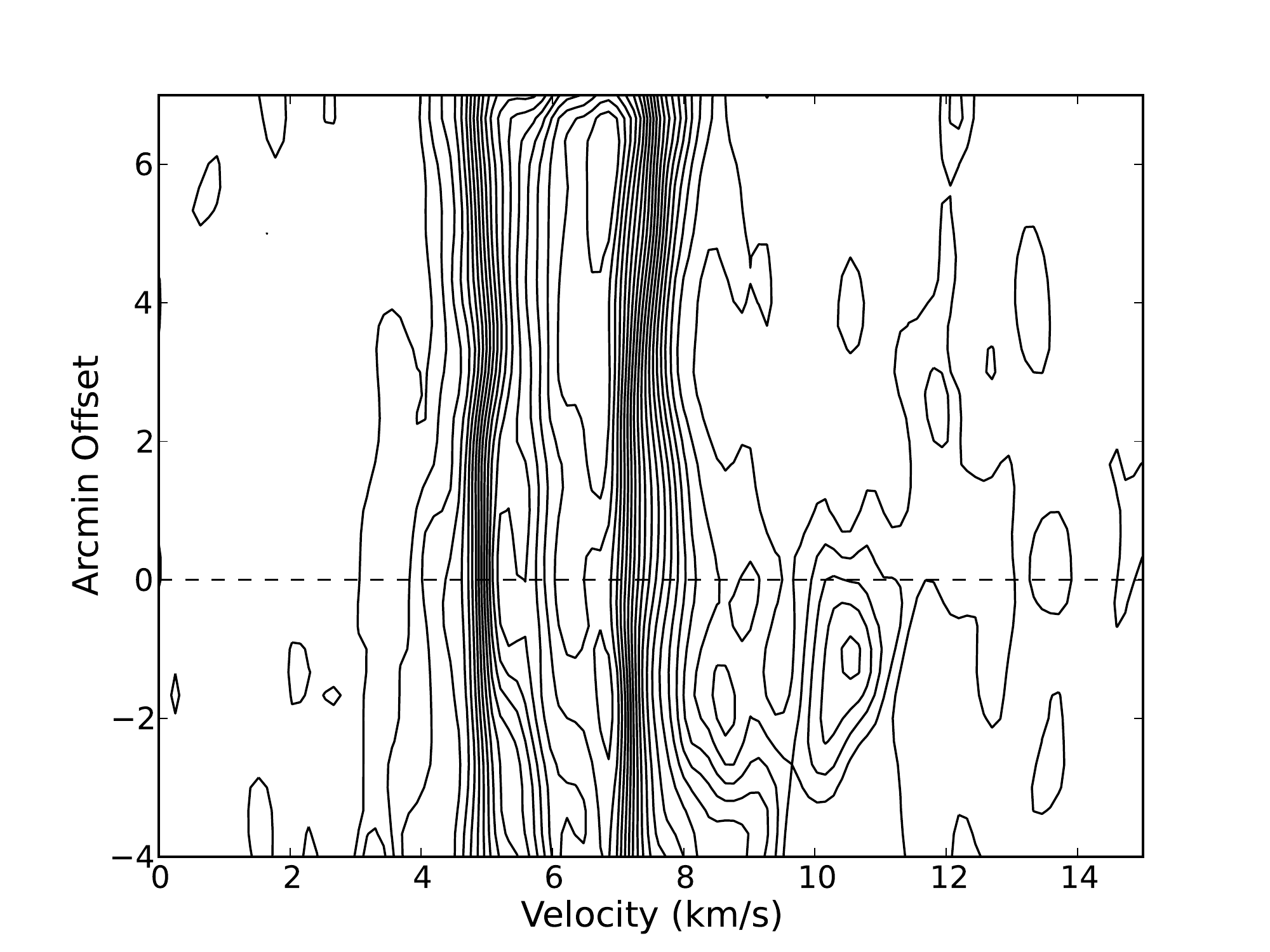}
\caption{Position velocity of \co\ emission towards the ZZ Tau region,
  through the slice at p.a. of $40^\circ$ shown in
  Figure~\ref{zztau_outflow}. The contour range is 0.13 to 3.93 K in
  steps of 0.3 K. Shown in dashed line is the position of the yellow
  circle shown in Fig~\ref{zztau_outflow}.
\label{zztau_posvel}}
\end{center}
\end{figure}

\begin{figure}%[hbp]
\begin{center}
\includegraphics[width=0.9\hsize]{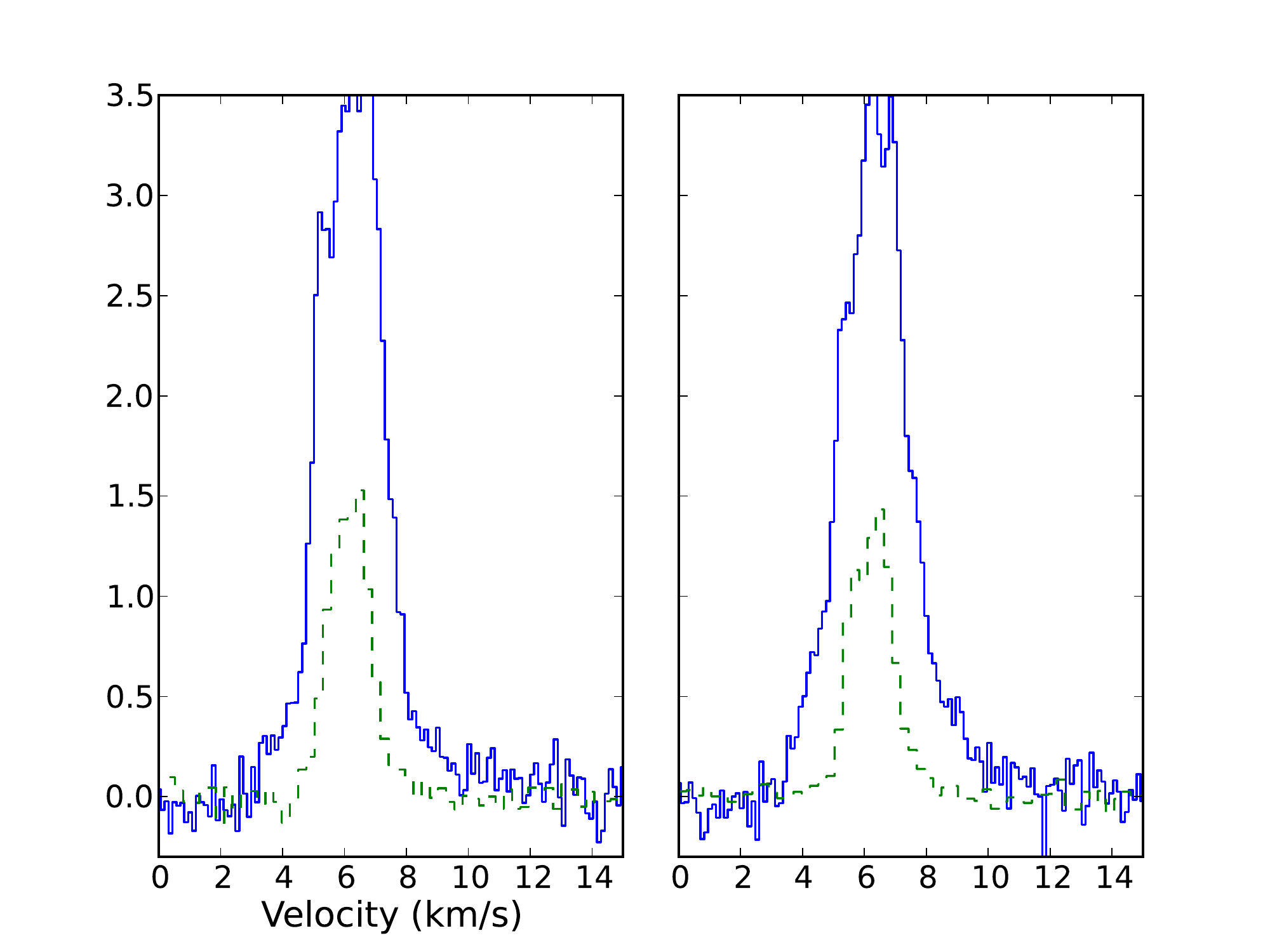}
\caption{Average Spectra of \co\ emission (blue solid lines) and
  \coa\ (green dashed lines) towards the presumed blueshifted (left) and
  redshifted (right) lobes of the outflow towards the ZZ Tau region
  shown in Fig~\ref{zztau_outflow}. The temperature scale is in
  T$_{A}^{*}$.
\label{zztau_spectra}}
\end{center}
\end{figure}

Our map of the high velocity emission is shown in
Figure~\ref{zztau_outflow}.  We detect only redshifted emission and
the morphology of this emission is similar to that found by
\citet{heyer1987}.  A position-velocity map along the line marked in
Figure~\ref{zztau_outflow} is shown in Figure~\ref{zztau_posvel} where
the distinct velocity feature at about 11 \kms\ can be readily
seen.  Averaged spectra are shown in Figure~\ref{zztau_spectra} which
shows clear high velocity redshifted emission.  A polygon was selected
where we might expect blueshifted emission to be present and the
averaged spectra shows some evidence for high velocity blueshifted
emission that may indicate that this is a asymmetrical bipolar
outflow.

\subsubsection{Haro 6-13}
\label{hktau}

TMC-2A is one of the Taurus cloud cores identified by
\citet{myers1983} and in the vicinity of this core are four IRAS
sources, three known to be associated with pre-main sequence stars
\citep{kenyon2008}: 04288+2417 (HK Tau), 04292+2422 (Haro 6-13) and
04294+2413 (FY/FZ Tau).  Both 04288+2417 and 04292+2422 were studied
by \citet{heyer1987}, \citet{myers1988} and
\citet{moriarty-schieven1994} and all found little evidence for high
velocity gas toward these sources.  Recently, \citet{Jiang2002} mapped
the CO J=3-2 emission in this region and found extended high velocity
emission.  The most prominent high velocity emission forms a bipolar
outflow roughly centred on IRAS 04292+2422 (Haro 6-13).  This outflow
was labeled TMC2A in the catalogue of \citet{wu2004}. The outflow emission
is complicated and \citet{Jiang2002} suggest that there may be multiple
outflows in the region.

\begin{figure}%[hbp]
\begin{center}
\includegraphics[width=0.9\hsize]{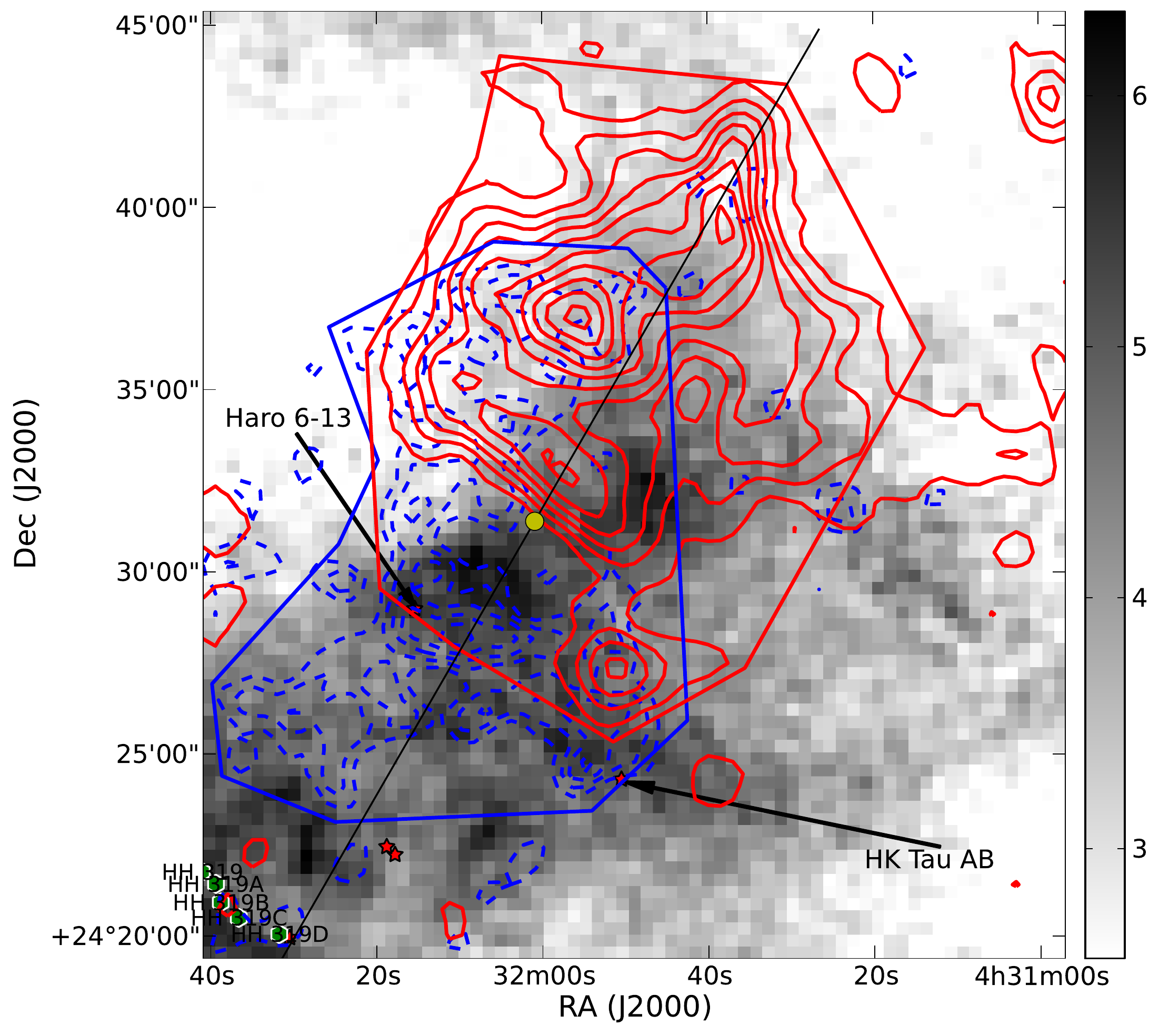}
\caption{Contour map of blueshifted and redshifted gas about a
  $25^\prime\times 25^\prime$ region near HK Tau. See
  Figure~\ref{041159_outflow} for details on symbols and
  markers. \co\ blueshifted and redshifted integrated intensity is for
  velocities of {\bf 0 to 4.0 \kms} and {\bf 8.5 to 12 \kms}
  respectively. Blueshifted contours range from 0.39 to 1.83 in steps
  of 0.075 \kkms, and redshifted contours range from 0.4 to 4.0 in
  steps of 0.075 \kkms. 
  \label{hktau_outflow}}
\end{center}
\end{figure}

\begin{figure}%[hbp]
\begin{center}
\includegraphics[width=0.9\hsize]{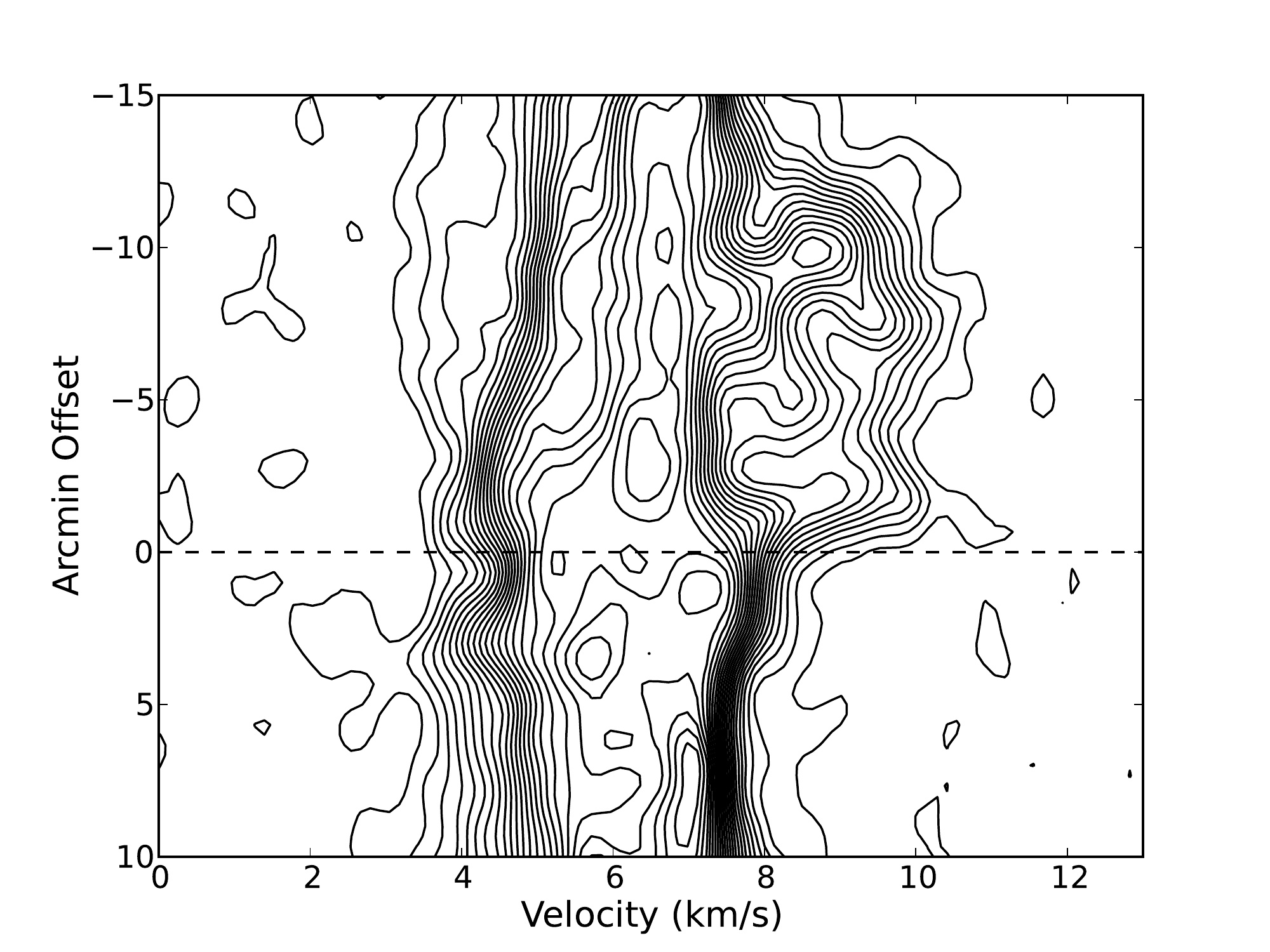}
\caption{Position velocity of \co\ emission towards the HK Tau region,
  through the slice at p.a. of $60^\circ$ shown in
  Figure~\ref{hktau_outflow}. The contour range is 0.13 to 3.93 K in
  steps of 0.3 K. Shown in dashed line is the position of the yellow
  circle shown in Fig~\ref{hktau_outflow}.
\label{hktau_posvel}}
\end{center}
\end{figure}

\begin{figure}%[hbp]
\begin{center}
\includegraphics[width=0.9\hsize]{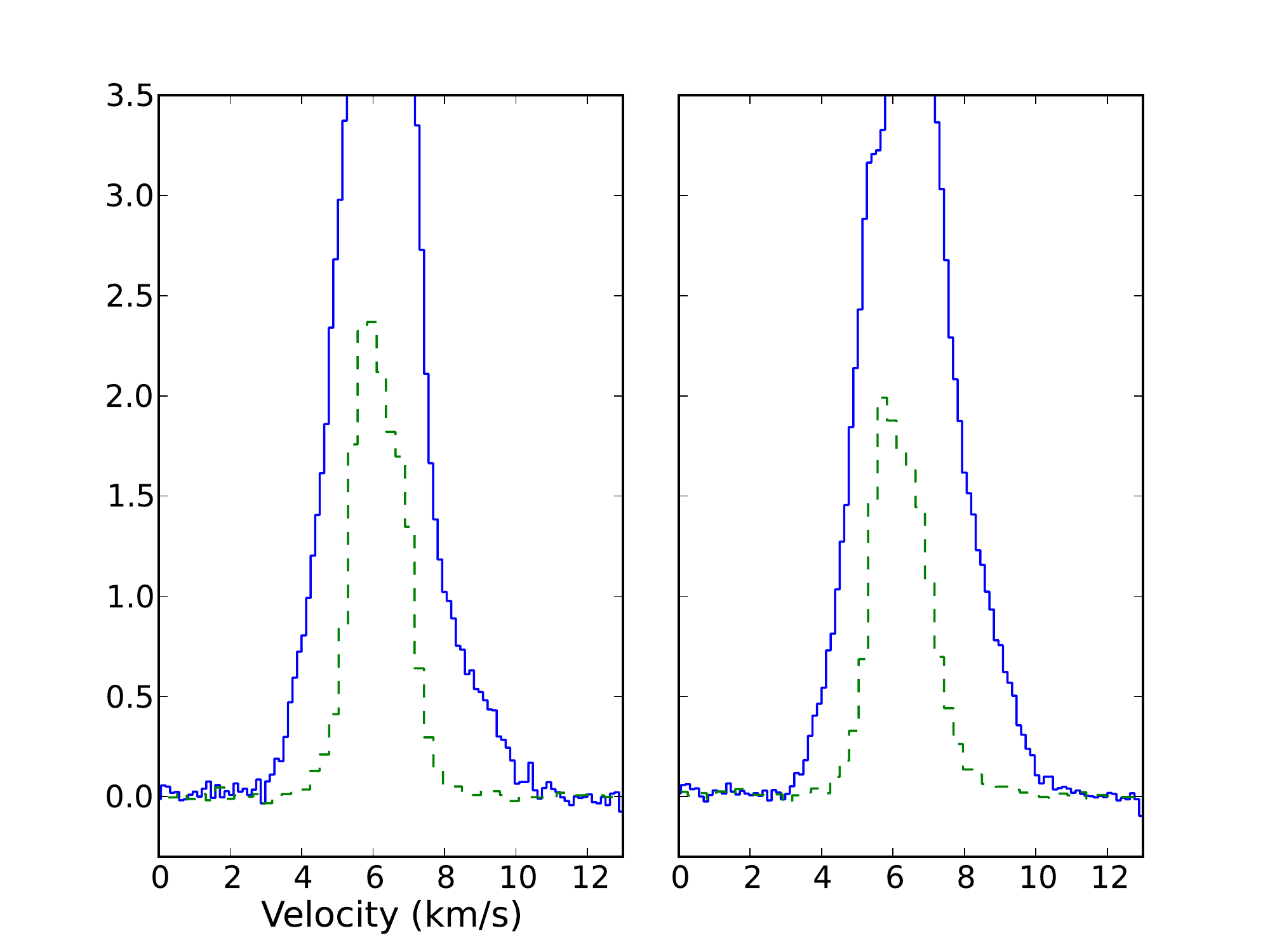}
\caption{Average Spectra of \co\ emission (blue solid lines) and
  \coa\ (green dashed lines) towards the blueshifted (left) and redshifted
  (right) lobes of the outflow towards the HK Tau region shown in
  Fig~\ref{hktau_outflow}. The temperature scale is in
  T$_{A}^{*}$. 
\label{hktau_spectra}}
\end{center}
\end{figure}

Our map of the high velocity red and blue emission is shown in
Figure~\ref{hktau_outflow} and the distribution of high velocity
emission is similar to that presented in \citet{Jiang2002}.  A large
bipolar outflow is clearly present and Haro 6-13, although slightly
offset to the south-east of the centroid of the outflow, is most
likely the driving source for this outflow.  The redshifted gas is
much more prominent than the blueshifted gas and that can be better
seen in the p-v diagram shown in Figure~\ref{hktau_posvel}.  Averaged
spectra are shown in Figure~\ref{hktau_spectra} which reveal
relatively low velocity outflow emission.

The redshifted and blueshifted features near the Class 0 object HK Tau
are suggestive of an outflow (see Figure~\ref{hktau_outflow}), but do
not meet our criteria for being an outflow.

\subsubsection{IRAS 04325+2402}

This outflow is located in the L1535 cloud and was first detected and
mapped in the CO J=1-0 line by \citet{heyer1987}.  Only redshifted
high velocity emission was detected in this outflow, however the
outflow is quite large with an angular size of about 15 arcminutes.
IRAS 04325+2402 is at the apex of this one-sided outflow.  This source
was in the survey of \citet{moriarty-schieven1992}, who found a total
line width of 13.9 \kms.

\begin{figure}%[hbp]
\begin{center}
\includegraphics[width=0.9\hsize]{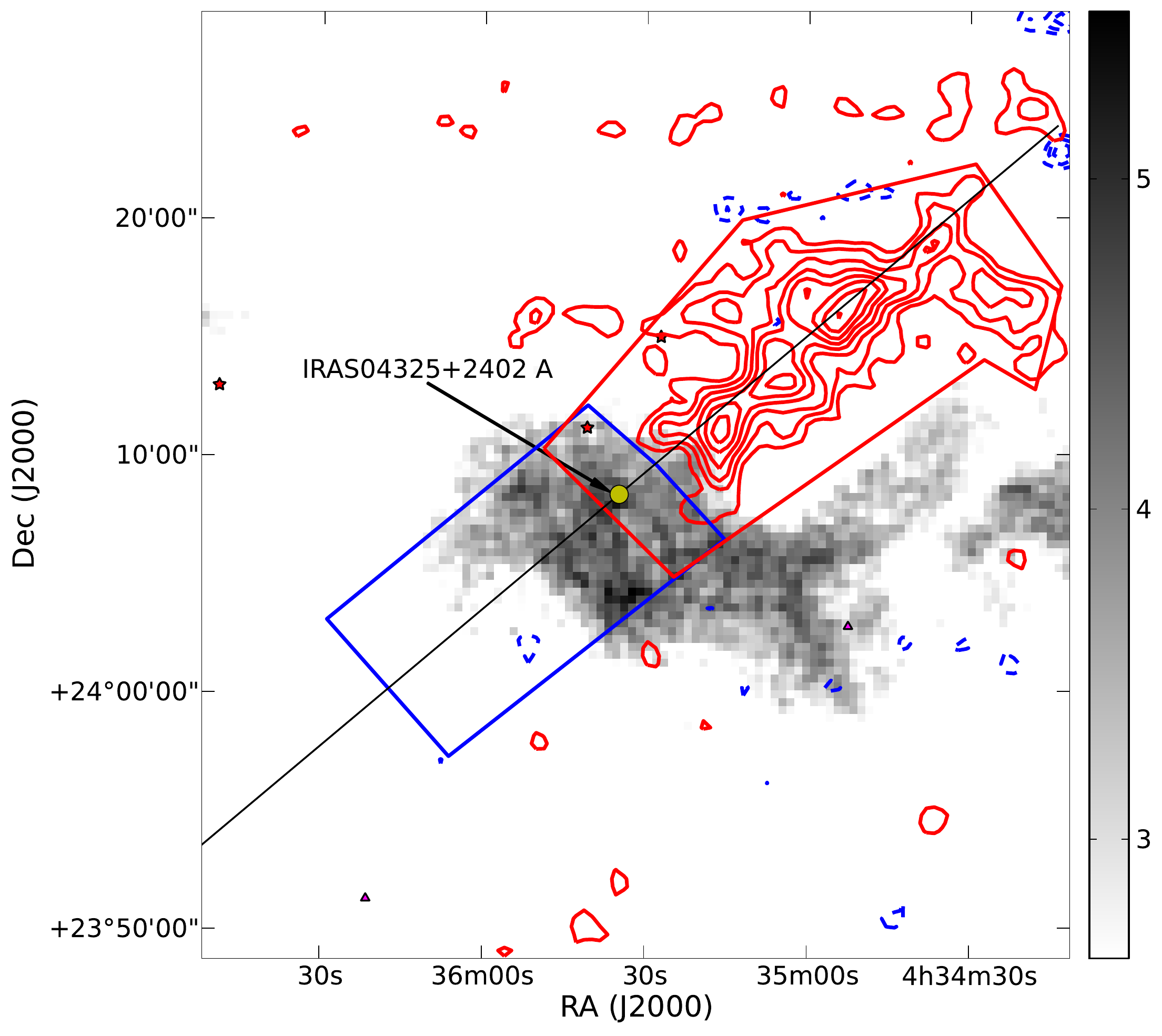}
\caption{Contour map of blueshifted and redshifted gas about a
  $25^\prime\times 25^\prime$ region near IRAS04325+2402. See
  Figure~\ref{041159_outflow} for details on symbols and markers.
  \co\ blueshifted and redshifted integrated intensity is for
  velocities of {\bf -1 to 3.8 \kms} and {\bf 8.5 to 12 \kms}
  respectively. Blueshifted contours range from 0.43 to 2.1 in steps
  of 0.075 \kkms, and redshifted contours range from 0.43 to 2.5 in
  steps of 0.075 \kkms. 
\label{l1535_outflow}}
\end{center}
\end{figure}

Our map of the high velocity emission is shown in
Figure~\ref{l1535_outflow}.  We detect only redshifted emission and
the morphology of the outflow is very similar to that found by
\citet{heyer1987}.  The protostellar source IRAS 04325+2402 is assumed
to be the origin of this outflow.  This protostellar source is complex
with at least two components and a complex bipolar scattered light
nebula \citep{hartmann1999}.  Sources A/B are located at the apex of
the bipolar nebula, however its orientation is not consistent with the
molecular outflow.  \citet{hartmann1999} suggest that the expected
outflow from a fainter component C may be better aligned with the
monopolar molecular outflow.

\begin{figure}%[hbp]
\begin{center}
\includegraphics[width=0.9\hsize]{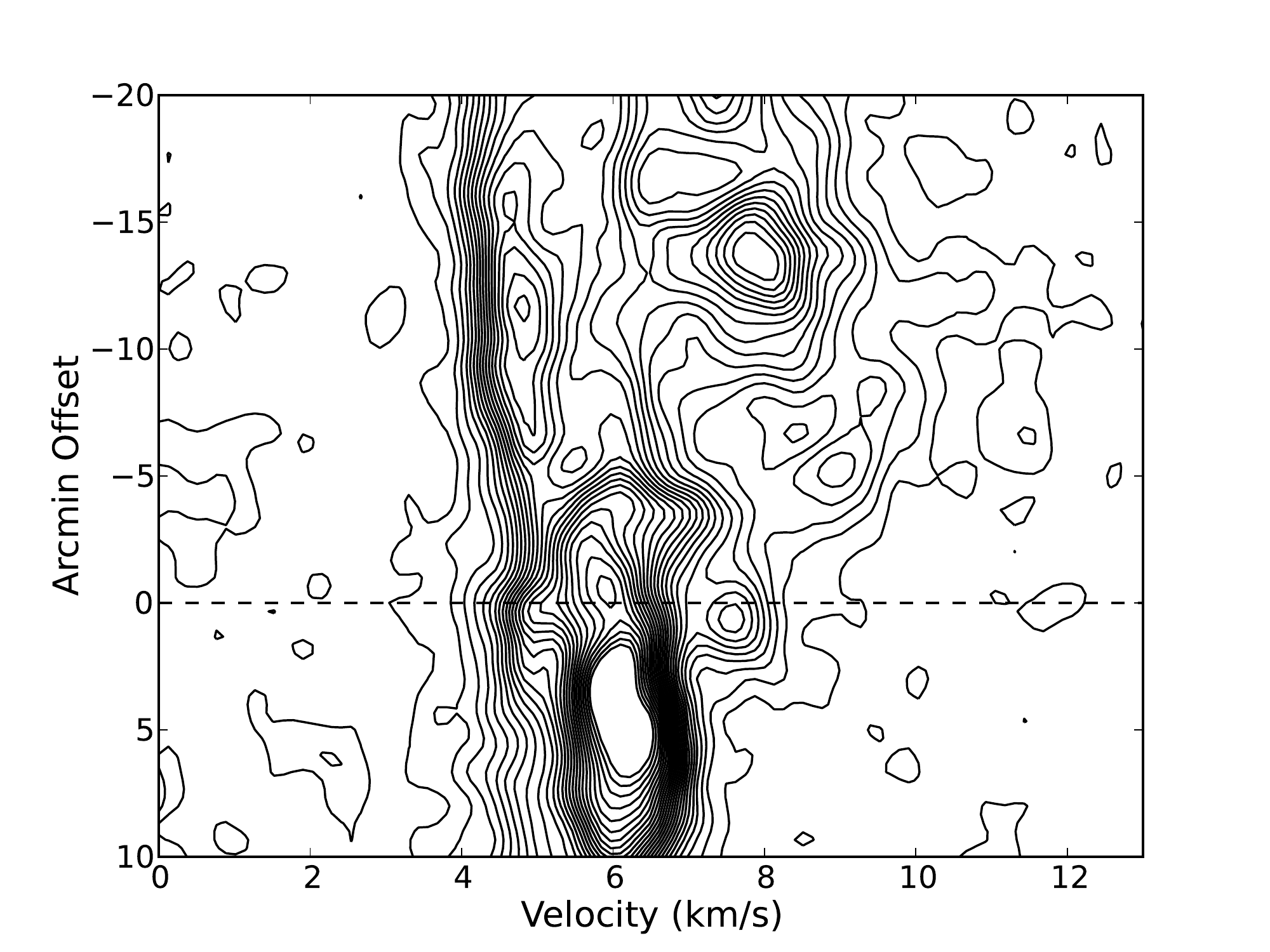}
\caption{Position velocity of \co\ emission towards the IRAS04325+2402
  region, through the slice at p.a. of $40^\circ$ shown in
  Figure~\ref{l1535_outflow}. The contour range is 0.11 to 5.1 K in
  steps of 0.3 K. Shown in dashed line is the position of the yellow
  circle shown in Fig~\ref{l1535_outflow}.
\label{l1535_posvel}}
\end{center}
\end{figure}

\begin{figure}%[hbp]
\begin{center}
\includegraphics[width=0.9\hsize]{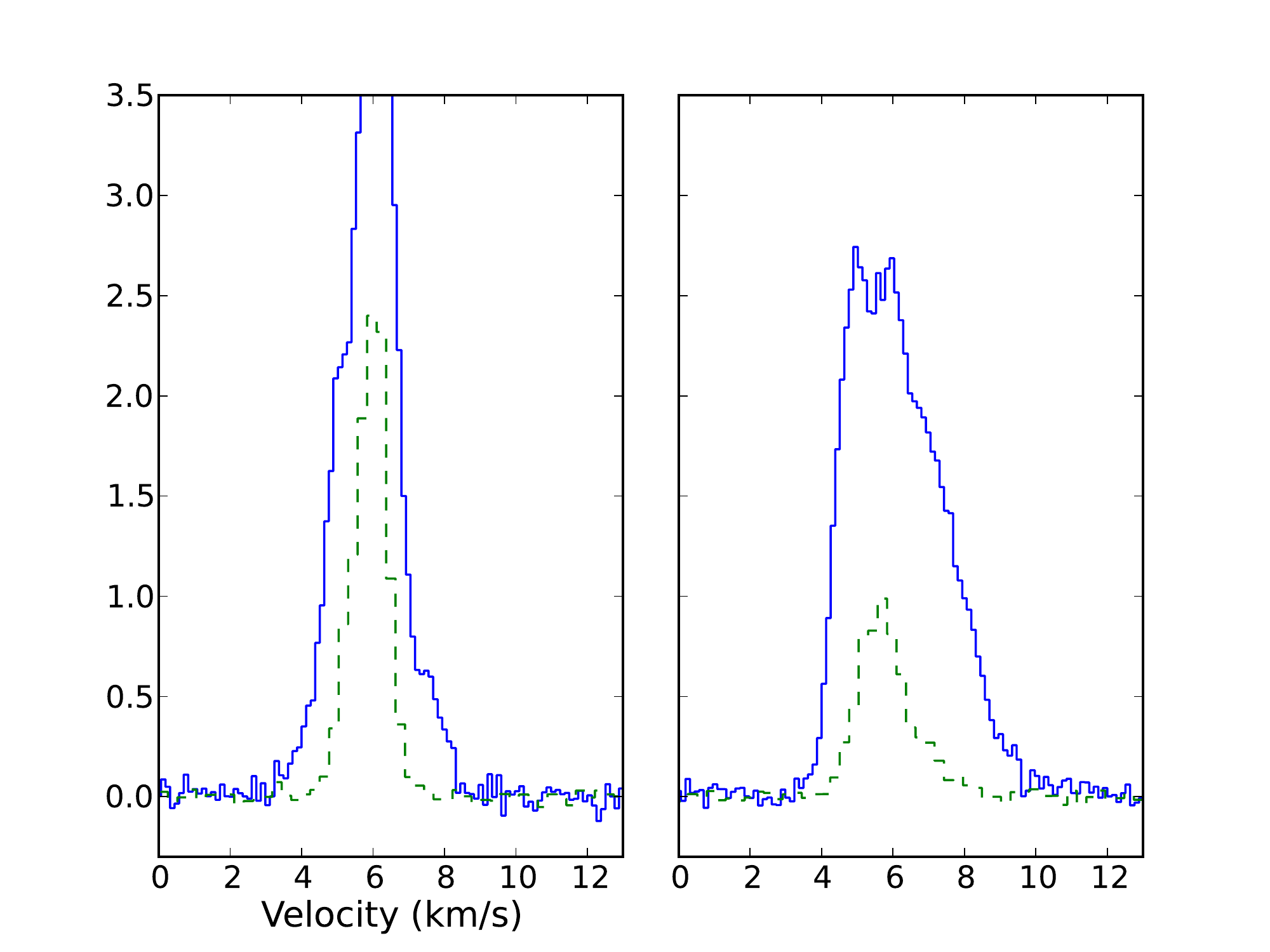}
\caption{Average Spectra of \co\ emission (blue solid lines) and
  \coa\ (green dashed lines) towards the blueshifted (left) and
  redshifted (right) lobes of the outflow seen towards IRAS04325+2402
  region shown in Fig~\ref{l1535_outflow}. The temperature scale is in
  T$_{A}^{*}$. \label{l1535_spectra}}
\end{center}
\end{figure}

In Figure~\ref{l1535_posvel} a position-velocity map along the line
marked in Figure~\ref{l1535_outflow} shows the prominent redshifted
outflow emission.  The redshifted outflow has a distinct secondary
velocity feature at 8 \kms\ located approximately 13 arcminutes
north-west of IRAS 04325+2402.  This feature may be part of shell-like
structure, similar to the ZZ Tau outflow.  Averaged spectra, shown in
Figure~\ref{l1535_spectra}, show clearly the high velocity redshifted
outflow and only very weak evidence for any blueshifted emission.

\subsubsection{HH 706 Outflow}

\begin{figure}%[hbp]
\begin{center}
\includegraphics[width=0.9\hsize]{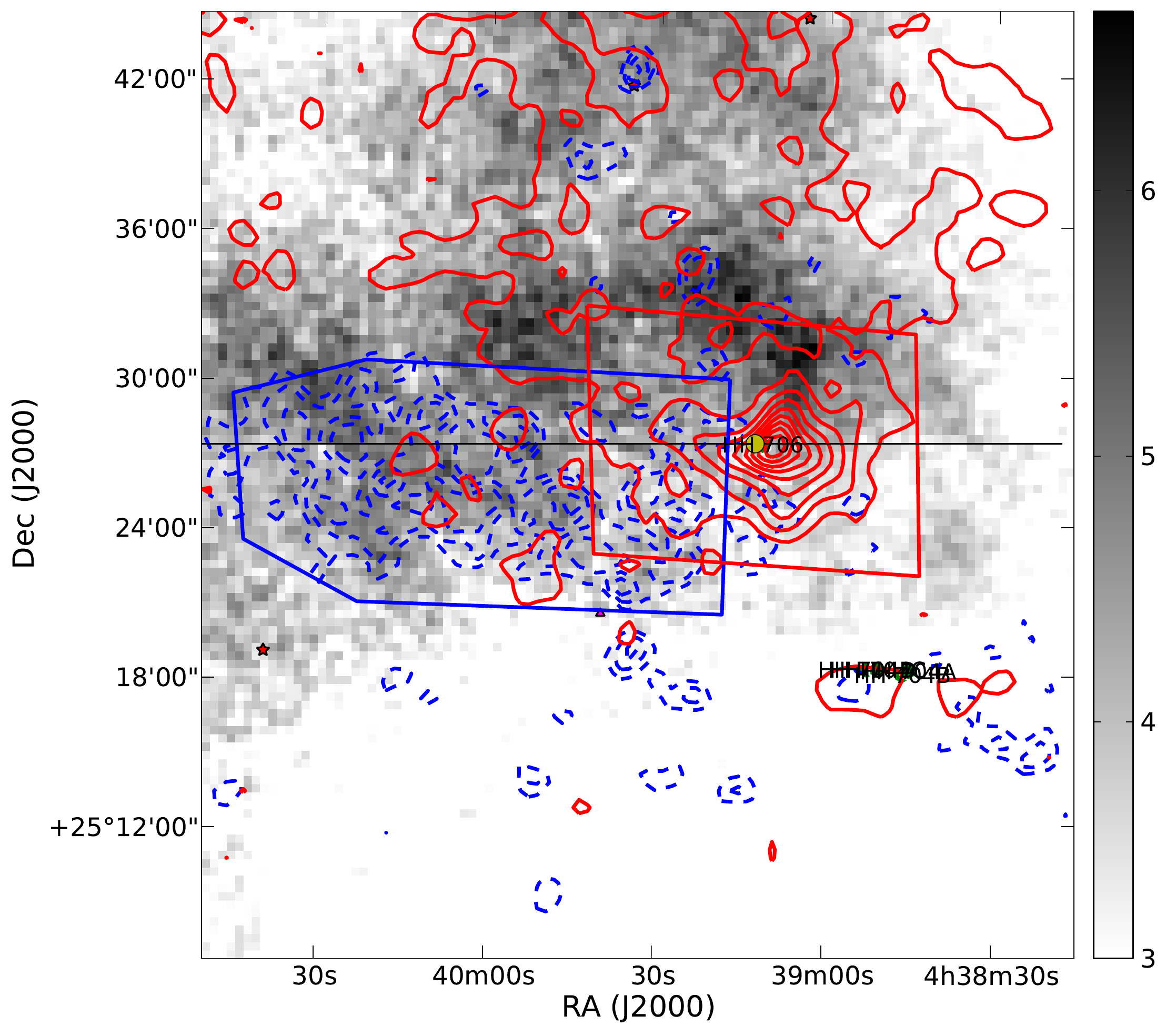}
\caption{Contour map of blueshifted and redshifted gas about a
  $35^\prime\times 45^\prime$ region centred at RA of 04:39:35 and
  Dec of 25:25:16. See Figure~\ref{041159_outflow} for details on symbols and
  markers.  \co\ blueshifted and redshifted 
  integrated intensity are for velocities of {\bf -1 to 4.0 \kms} and
  {\bf 8 to 13 \kms} respectively. Blueshifted contours
  range from 0.59 to 2.7 in steps of 0.075 \kkms, and redshifted
  contours range from 0.59 to 9.4 in steps of 0.075 \kkms. 
\label{nbs3_outflow}}
\end{center}
\end{figure}

\begin{figure}%[hbp]
\begin{center}
\includegraphics[width=0.9\hsize]{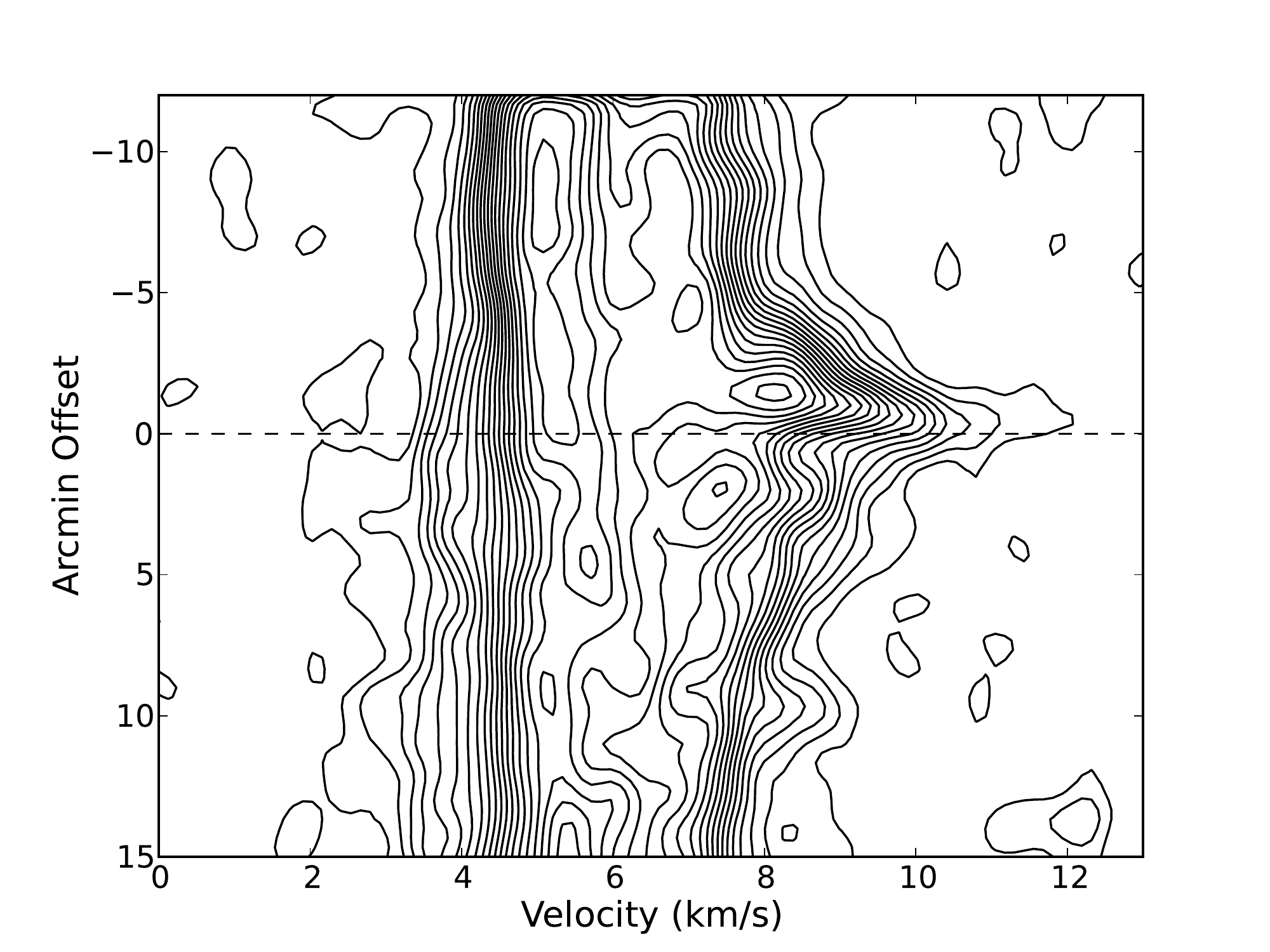}
\caption{Position velocity of \co\ emission towards the HH706 region,
  through the slice at p.a. of $0^\circ$ shown in
  Figure~\ref{nbs3_outflow}. The contour range is 0.1 to 5.3 K in
  steps of 0.2 K. Shown in dashed line is the position of the yellow
  circle shown in Fig~\ref{nbs3_outflow}.
\label{nbs3_posvel}}
\end{center}
\end{figure}

\begin{figure}%[hbp]
\begin{center}
\includegraphics[width=0.9\hsize]{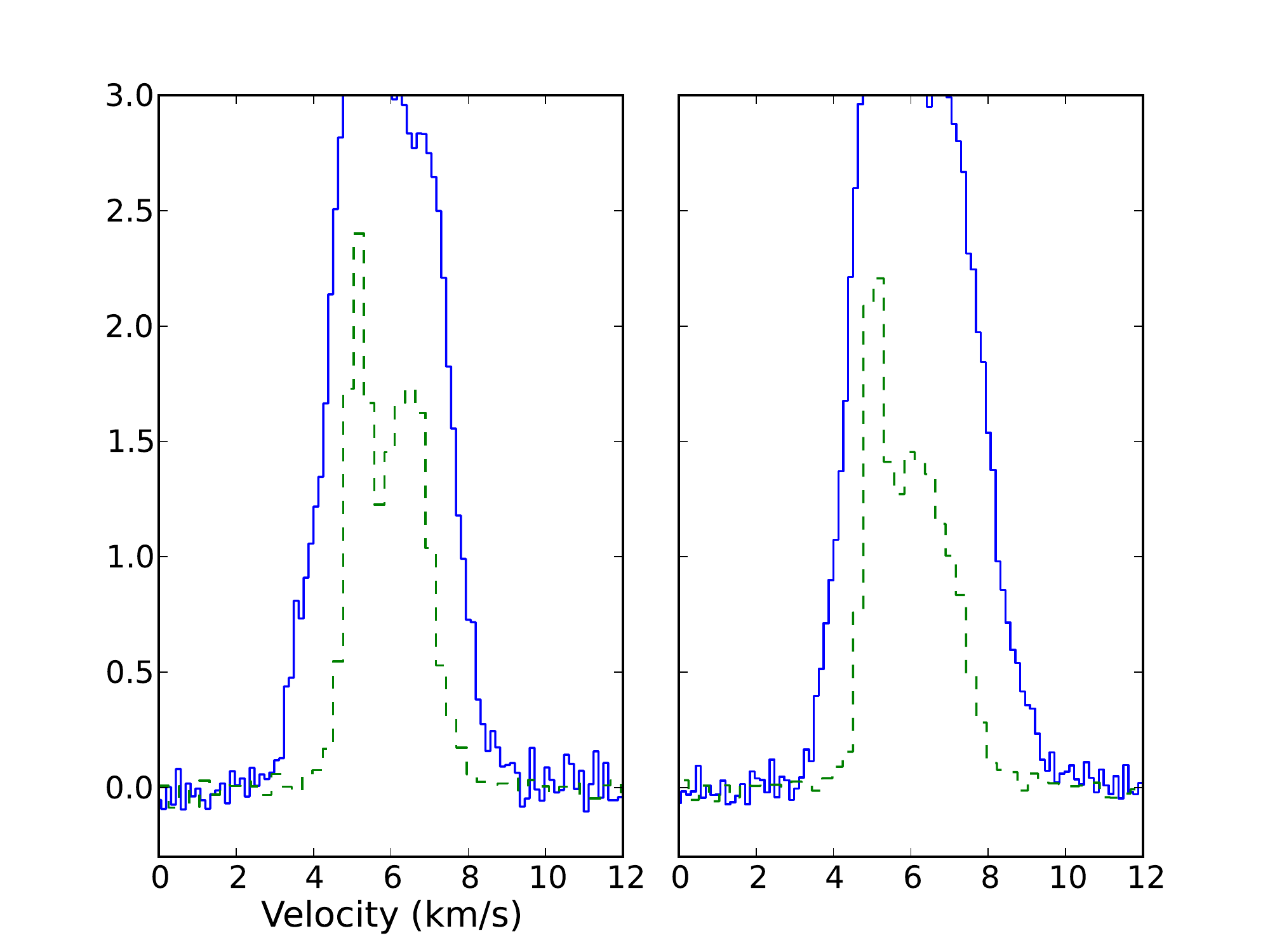}
\caption{Average Spectra of \co\ emission (blue solid lines) and
  \coa\ (green dashed lines) towards the blueshifted (left) and
  redshifted (right) lobes of the outflow seen towards HH706 region
  shown in Fig~\ref{nbs3_outflow}. The temperature scale is in
  T$_{A}^{*}$. \label{nbs3_spectra}}
\end{center}
\end{figure}

A newly discovered bipolar outflow is found associated with the
Herbig-Haro object HH 706 \citep{sun2003}.  We have labeled this
molecular outflow HH 706 Outflow, as we were unable to identify in the
Spitzer catalogue any source that might drive this outflow.  See the
overview figure of the Heiles Cloud 2 region in
Figure~\ref{ic2087_overview} for the location of the HH 706 flow in
this region. The distribution of redshifted and blueshifted gas toward
HH 706 is shown in Figure~\ref{nbs3_outflow}.  The blueshifted
emission is much more extended than the redshifted emission and HH 706
is located at the centre of the region of strong redshifted high
velocity emission.  A p-v plot along the axis shown in
Figure~\ref{nbs3_outflow} is presented in Figure~\ref{nbs3_posvel}.
The p-v plot shows prominent redshifted emission, while the
blueshifted outflow has lower velocity and is less well delineated.
The averaged spectra shown in Figure~\ref{nbs3_spectra} show weak high
velocity redshifted and blueshifted emission.

\subsubsection{HH 705 Outflow}

\begin{figure}%[hbp]
\begin{center}
\includegraphics[width=0.9\hsize]{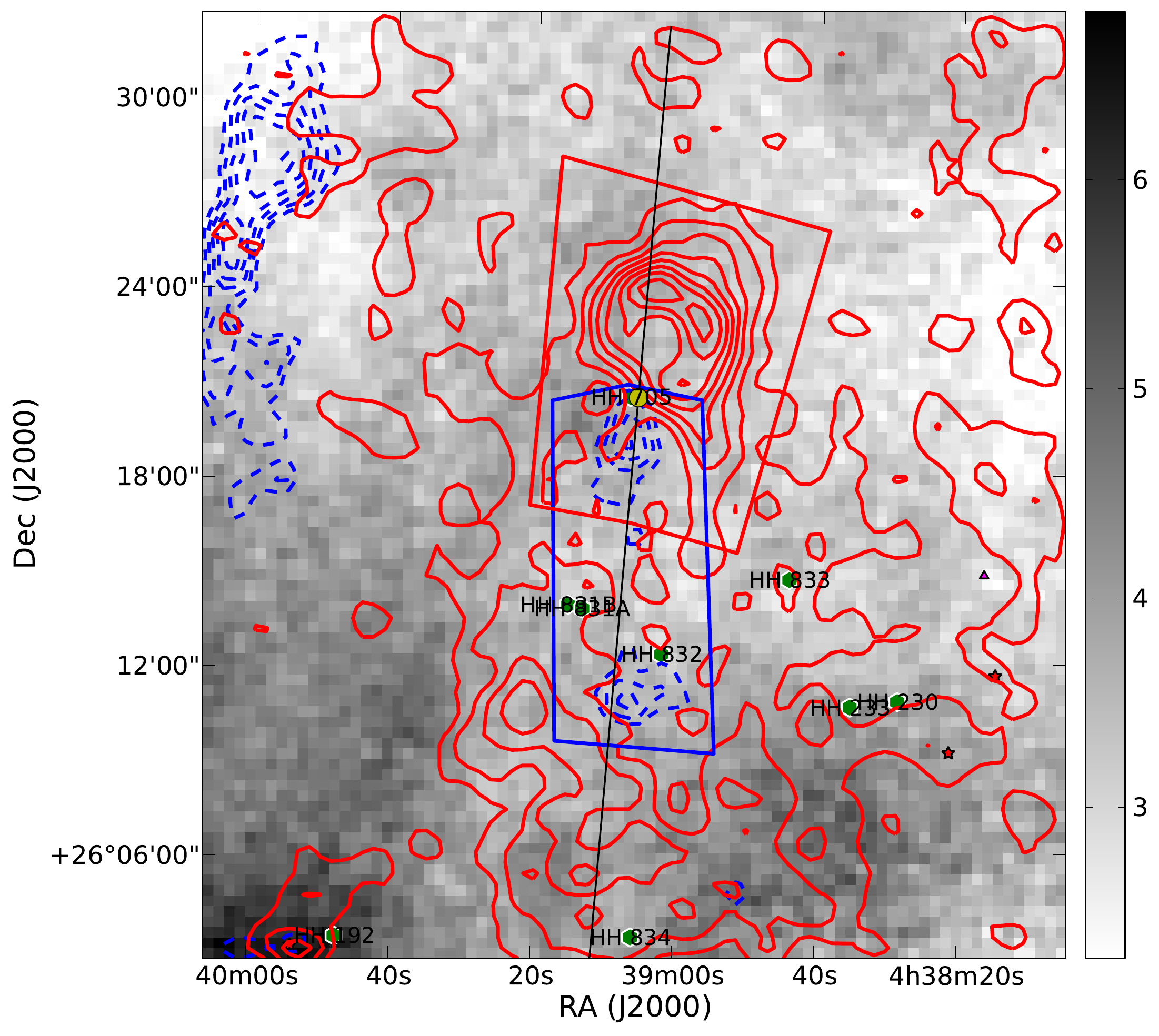}
\caption{Contour map of blueshifted and redshifted gas about a
  $35^\prime\times 45^\prime$ region centred at RA of 04:39:07 and
  Dec of 26:17:20. See Figure~\ref{041159_outflow} for details on symbols and
  markers. \co\ blueshifted and redshifted 
  integrated intensity are for velocities of {\bf -1 to 3.9 \kms} and
  {\bf 8.3 to 13 \kms} respectively. Blueshifted
  contours range from 0.57 to 2.3 in steps of 0.075 \kkms, and
  redshifted contours range from 0.57 to 3.5 in steps of 0.075
  \kkms. 
\label{nbs4_outflow}}
\end{center}
\end{figure}

The bipolar molecular outflow associated with HH 705 \citep{sun2003}
is newly discovered.  The distribution of redshifted and blueshifted
emission is shown in Figure~\ref{nbs4_outflow}.  The Herbig-Haro
object HH 705 is located at the centroid of this bipolar outflow.
\citet{mcgroarty2004} suggested that HH 705 could be associated with HV
Tau C, however more recent proper motion measurements
\citep{mcgroarty2007} seem to rule out this possibility.  The proper
motion measurements show that the four knots that compose HH 705 are
moving to the south and southwest with velocities varying from 100 to
300 \kms.  There is also [SII] and H$\alpha$ emission that extends
nearly 2 arcminutes south of HH 705 \citet{mcgroarty2004}.  Further
south along the bipolar outflow axis lies HH 831 and HH 832
\citep{mcgroarty2004}, however the proper motion vectors for HH 831A
are directly westward, while those for HH 831B are directed southeast
\citep{mcgroarty2007}.  It is unclear whether these Herbig-Haro objects
are related to this large bipolar outflow.  The driving source for the
large molecular outflow and HH 705 is unknown.  We have labeled this
molecular outflow HH 705 Outflow.

\begin{figure}%[hbp]
\begin{center}
\includegraphics[width=0.9\hsize]{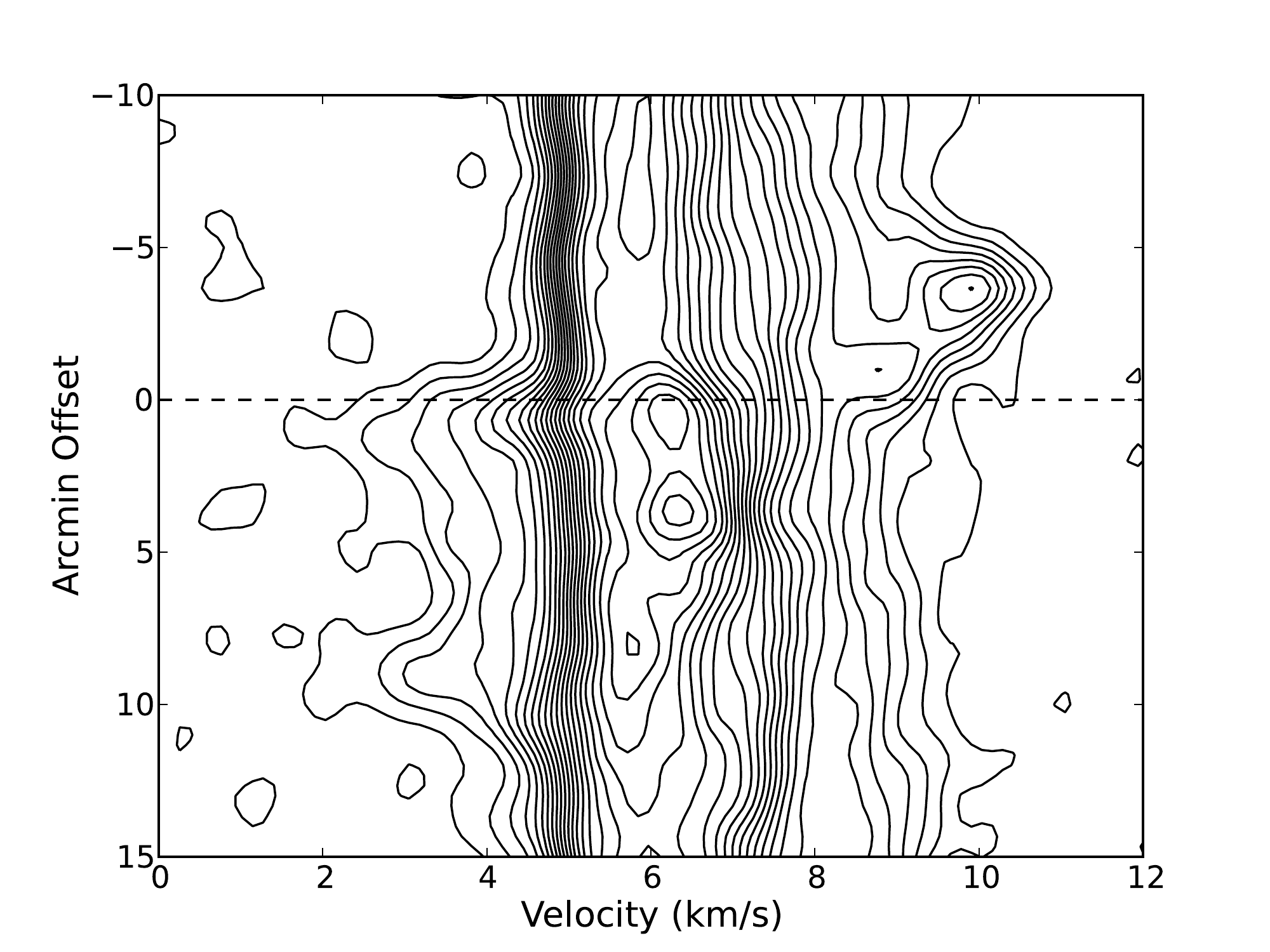}
\caption{Position velocity of \co\ emission towards the HH705 region,
  through the slice at p.a. of $85^\circ$ shown in
  Figure~\ref{nbs4_outflow}. The contour range is 0.1 to 5.3 K in
  steps of 0.2 K. Shown in dashed line is the position of the yellow
  circle shown in Fig~\ref{nbs4_outflow}.
\label{nbs4_posvel}}
\end{center}
\end{figure}

\begin{figure}%[hbp]
\begin{center}
\includegraphics[width=0.9\hsize]{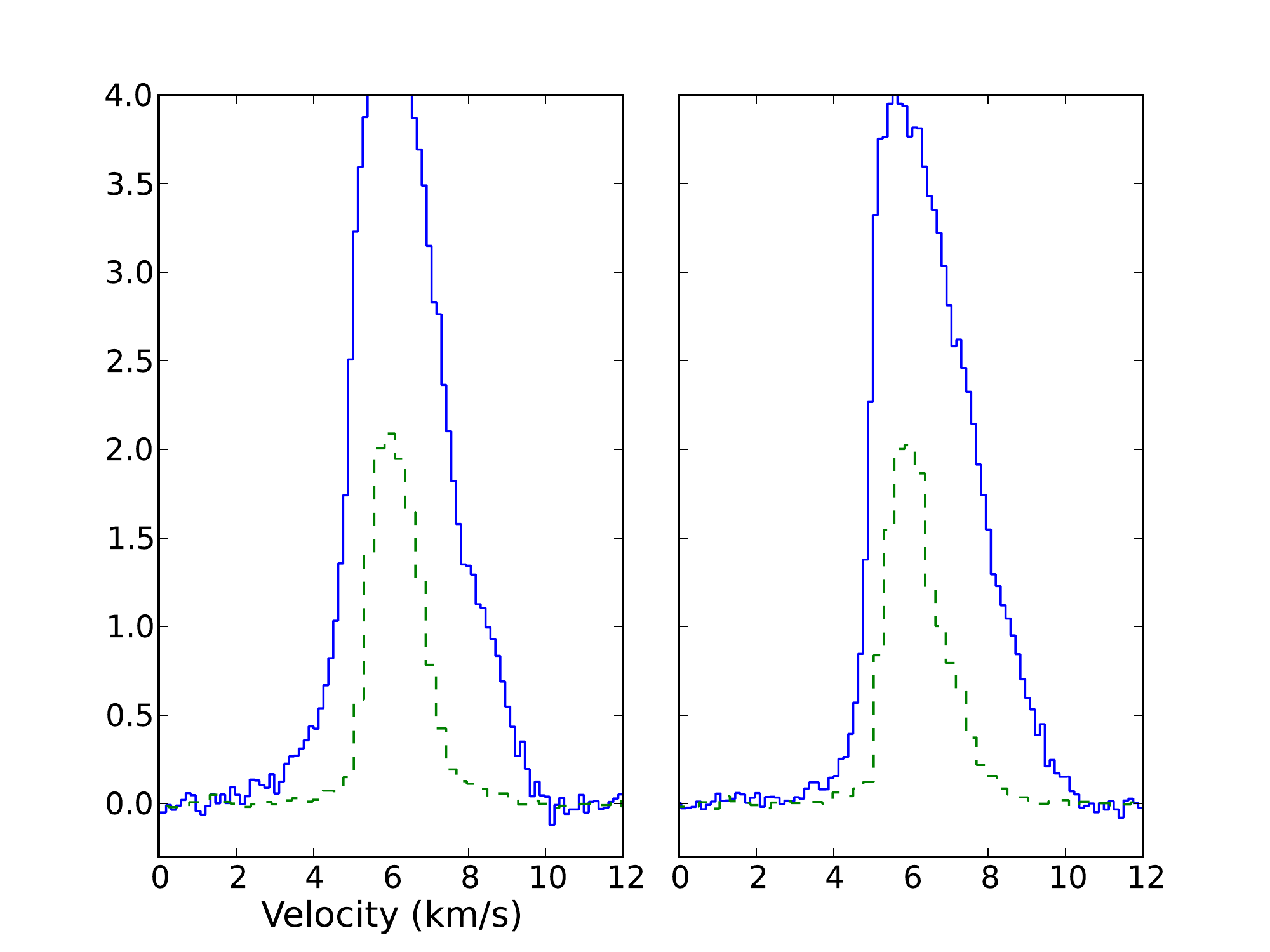}
\caption{Average Spectra of \co\ emission (blue solid lines) and
  \coa\ (green dashed lines) towards the blueshifted (left) and
  redshifted (right) lobes of the outflow seen towards HH705 region
  shown in Fig~\ref{nbs4_outflow}. The temperature scale is in
  T$_{A}^{*}$. \label{nbs4_spectra}}
\end{center}
\end{figure}

The p-v plot along the axis defined in Figure~\ref{nbs4_outflow}
is shown in Figure~\ref{nbs4_posvel}.  The bipolar nature of this outflow
can be clearly seen as well as its large extend of nearly 20 arcminutes.  
The averaged spectra in the two polygons are shown in 
Figure~\ref{nbs4_spectra} and also show clearly the high velocity blueshifted
and redshifted gas associated with this outflow.

\subsubsection{TMR-1 (IRAS 04361+2547)}

Based on their small interferometer map of the CO J=1-0 emission,
\citet{terebey1990} were first to detect an outflow associated IRAS
04361+2547, which they labeled TMR-1. In the J=3-2 CO line toward this
source, \citet{moriarty-schieven1992} measured a total line-width of
14.7 km s$^{-1}$.  A region larger than that observed by
\citet{terebey1990} was mapped by \citet{bontemps1996} in the CO J=2-1
line and \citet{hogerheijde1998} in the CO J=3-2 line, and these
studies revealed that this outflow was extended on angular scales of
at least 2 arcminutes, however no clear bipolar morphology was seen.
\citet{terebey1998} resolved TMR-1 into a binary (TMR-1AB) that is
surrounded by extended nebulosity.  Near infrared images of this
region \citep{petr-gotzens2010} reveal nebulosity extended south-east
and north-west of IMR-1.  Near infrared spectra of the emission to the
south-east shows emission lines of molecular hydrogen which
\citet{petr-gotzens2010} interpret as shocked excited.  Thus there is
strong evidence for a extended shocked gas emission directed to the
southeast of TMR-1AB.

\begin{figure}%[hbp]
\begin{center}
\includegraphics[width=0.9\hsize]{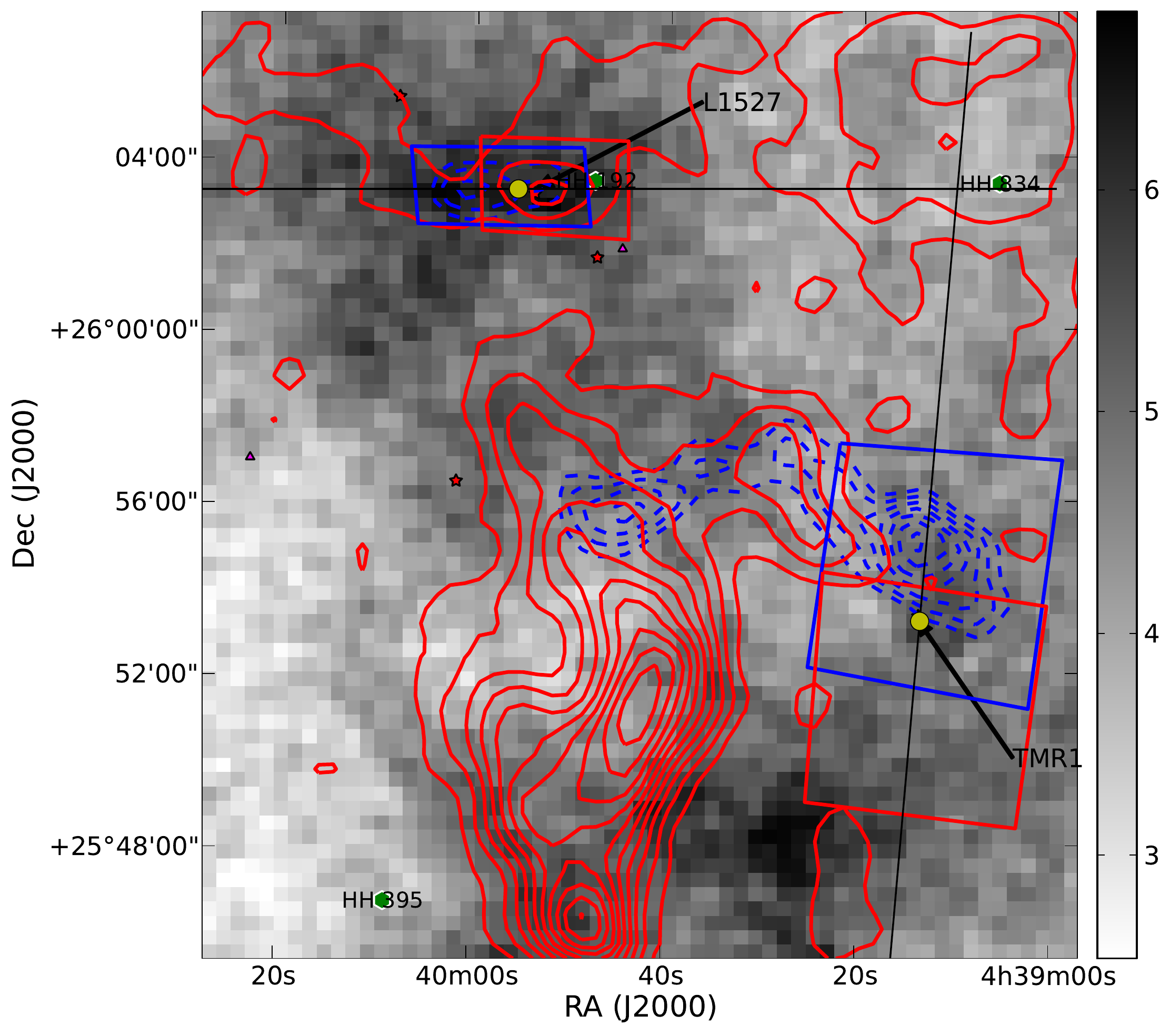}
\caption{Contour map of blueshifted and redshifted gas about a
  $20^\prime\times 20^\prime$ region centred at RA of 04:39:44 and Dec
  of 25:56:00. See Figure~\ref{041159_outflow} for details on symbols
  and markers.  \co\ blueshifted and redshifted integrated intensity
  are for velocities of {\bf -1 to 4.1 \kms} and {\bf 8.1 to 13 \kms}
  respectively. Blueshifted contours range from 0.63
  to 3.0 in steps of 0.075 \kkms, and redshifted contours range from
  0.63 to 5.6 in steps of 0.075 \kkms. Three outflows are seen in the
  figure, those due to TMR1, L1527 and IC2087 (the latter being the
  central big redshifted lobe). 
\label{04361_outflow}}
\end{center}
\end{figure}

\begin{figure}%[hbp]
\begin{center}
\includegraphics[width=0.9\hsize]{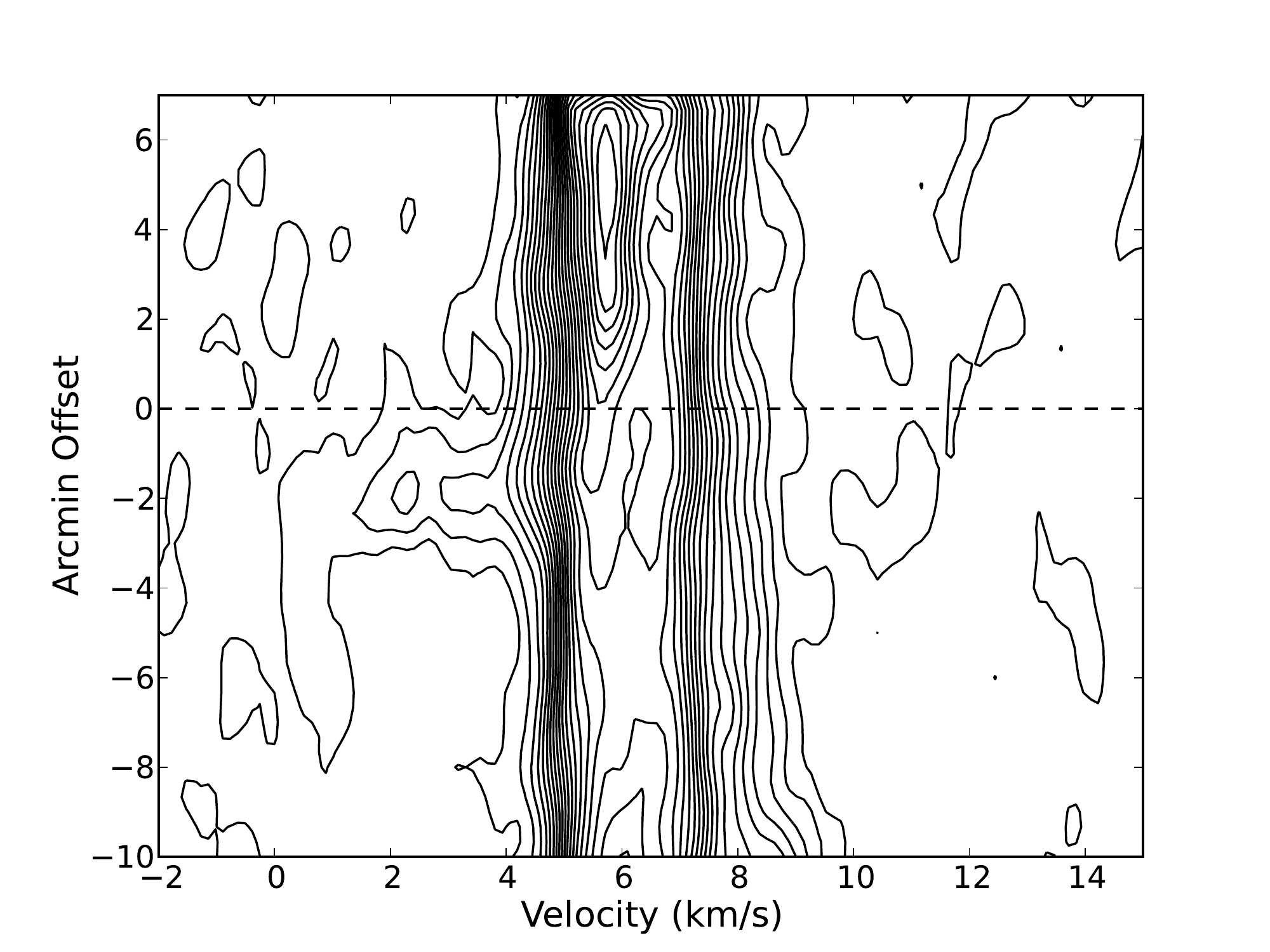}
\caption{Position velocity of \co\ emission towards the TMR1 region,
  through the slice at p.a. of $85^\circ$ shown in
  Figure~\ref{04361_outflow}. The contour range is 0.12 to 5.32 K in
  steps of 0.2 K. Shown in dashed line is the position of the yellow
  circle shown in Fig~\ref{04361_outflow}.
\label{04361_posvel}}
\end{center}
\end{figure}

Our map of the high velocity redshifted and blueshifted emission is
shown in Figure~\ref{04361_outflow}.  We detect primarily blueshifted
emission associated with this outflow, similar to the maps of
\citet{bontemps1996} and \citet{hogerheijde1998}, although the outflow
emission is somewhat more extended.  We also see weak blueshifted
emission extending to the east of TMR-1, however it is unclear whether
this is related to the TMR-1 outflow.  The p-v plot along the
direction marked in Figure~\ref{04361_outflow} is shown in
Figure~\ref{04361_posvel}.  Prominent blueshifted emission is seen,
but no evidence is seen for redshifted emission.  Likewise the
averaged spectra show blueshifted emission, but no sign of redshifted
emission.  The redshifted emission shown in the plots of
\citet{bontemps1996} is for a velocity range of 6 to 8.6 \kms\ that
may be strongly contaminated by the ambient cloud emission.

\begin{figure}%[hbp]
\begin{center}
\includegraphics[width=0.9\hsize]{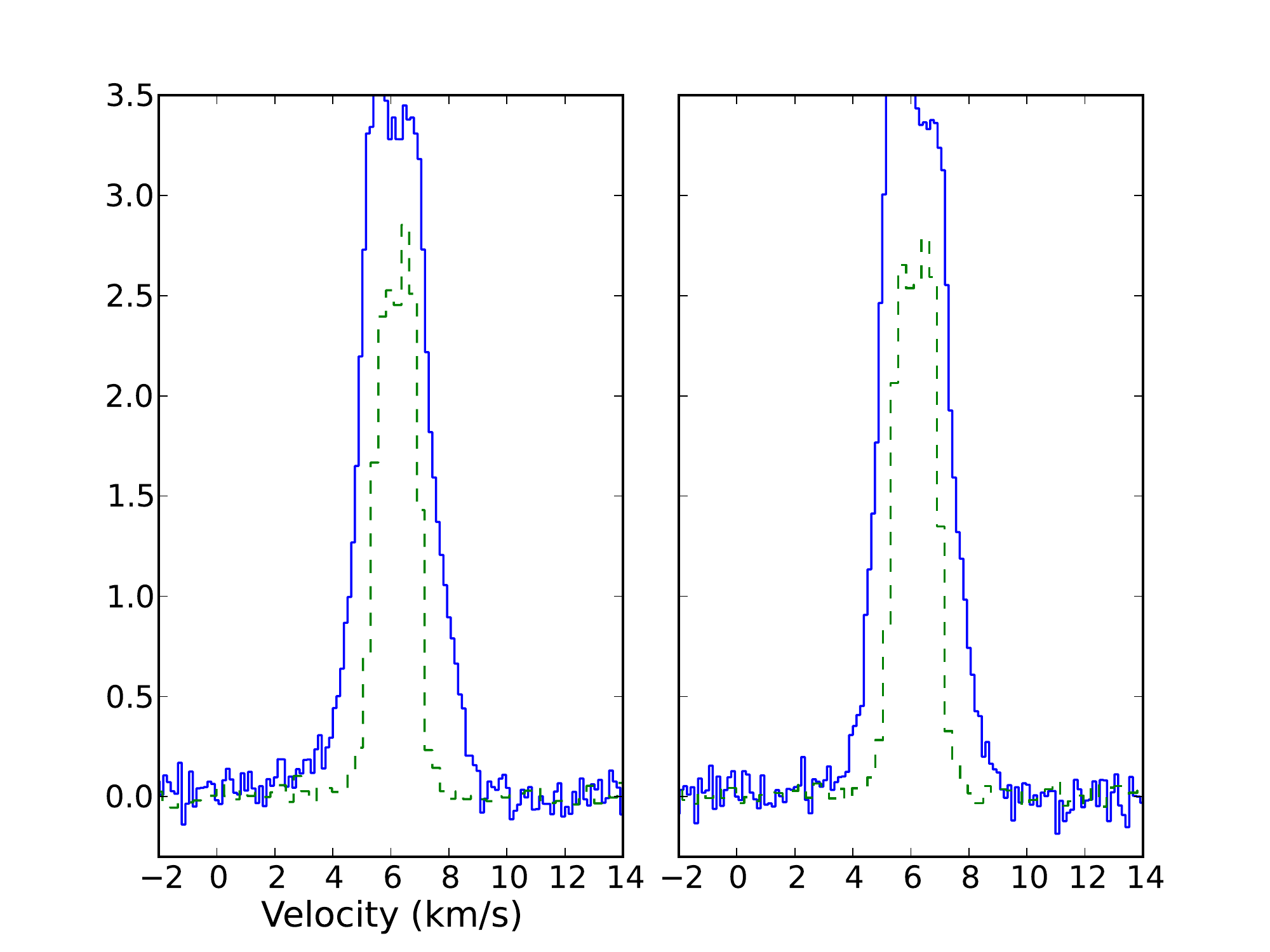}
\caption{Average Spectra of \co\ emission (blue solid lines) and
  \coa\ (green dashed lines) towards the blueshifted (left) and redshifted
  (right) lobes of the outflow seen towards the TMR1 region shown in
  Fig~\ref{04361_outflow}. The temperature scale is in
  T$_{A}^{*}$. 
  \label{04361_spectra}}
\end{center}
\end{figure}

\subsubsection{L1527 (IRAS 04368+2557)}

IRAS 04368+2557 is also called L1527 IRS \citep{kenyon2008}.  As in
many of these outflows, the first evidence for high velocity gas came
from the CO J=3-2 survey of \citet{moriarty-schieven1992} who found a
total line width of 19.4 km s$^{-1}$ towards this source.  Subsequent maps
by \citet{bontemps1996}, \citet{tamura1996} and \citet{zhou1996}
revealed a small bipolar outflow oriented east-west.  More extensive
mapping by \citet{hogerheijde1998} showed that the bipolar outflow had
an angular extent of approximately 4 arcminutes, much larger than was
suspected from the earlier observations.  Their outflow map showed
significant overlap between the redshifted and blueshifted high
velocity emission implying a large inclination angle as was first
suggested by \citet{tamura1996}.

There is also evidence for an optical jet from this source based on
the emission line nebulosity located east of L1527 IRS
\citet{eiroa1994}. A series of Herbig-Haro objects (HH 192 A,B,C) are
located on either side of L1527 IRS oriented east-west along the
molecular outflow axis \citep{gomez1997}.  L1527 IRS is also known to
have a flattened, nearly edge on disk seen in molecular emission
\citep{ohashi1997} and in scattered light imaging \citep{tobin2010}.

\begin{figure}%[hbp]
\begin{center}
\includegraphics[width=0.9\hsize]{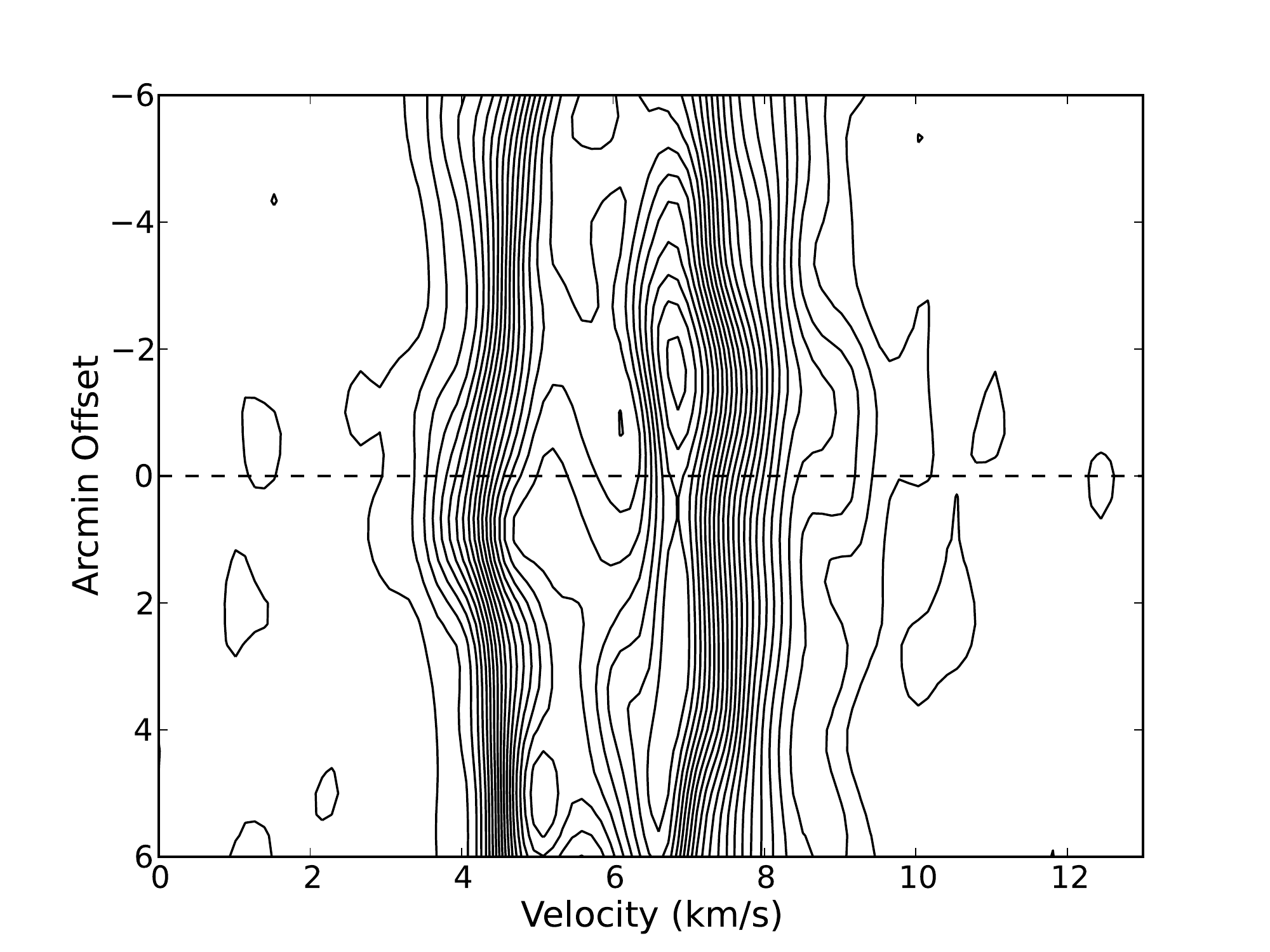}
\caption{Position velocity of \co\ emission towards the L1527 region,
  through the slice at p.a. of $0^\circ$ through L1527 shown in
  Figure~\ref{04361_outflow}. The contour range is 0.1 to 5.3 K in
  steps of 0.2 K. Shown in dashed line is the position of the yellow
  circle shown in Fig~\ref{04361_outflow}.
\label{l1527_posvel}}
\end{center}
\end{figure}

\begin{figure}%[hbp]
\begin{center}
\includegraphics[width=0.9\hsize]{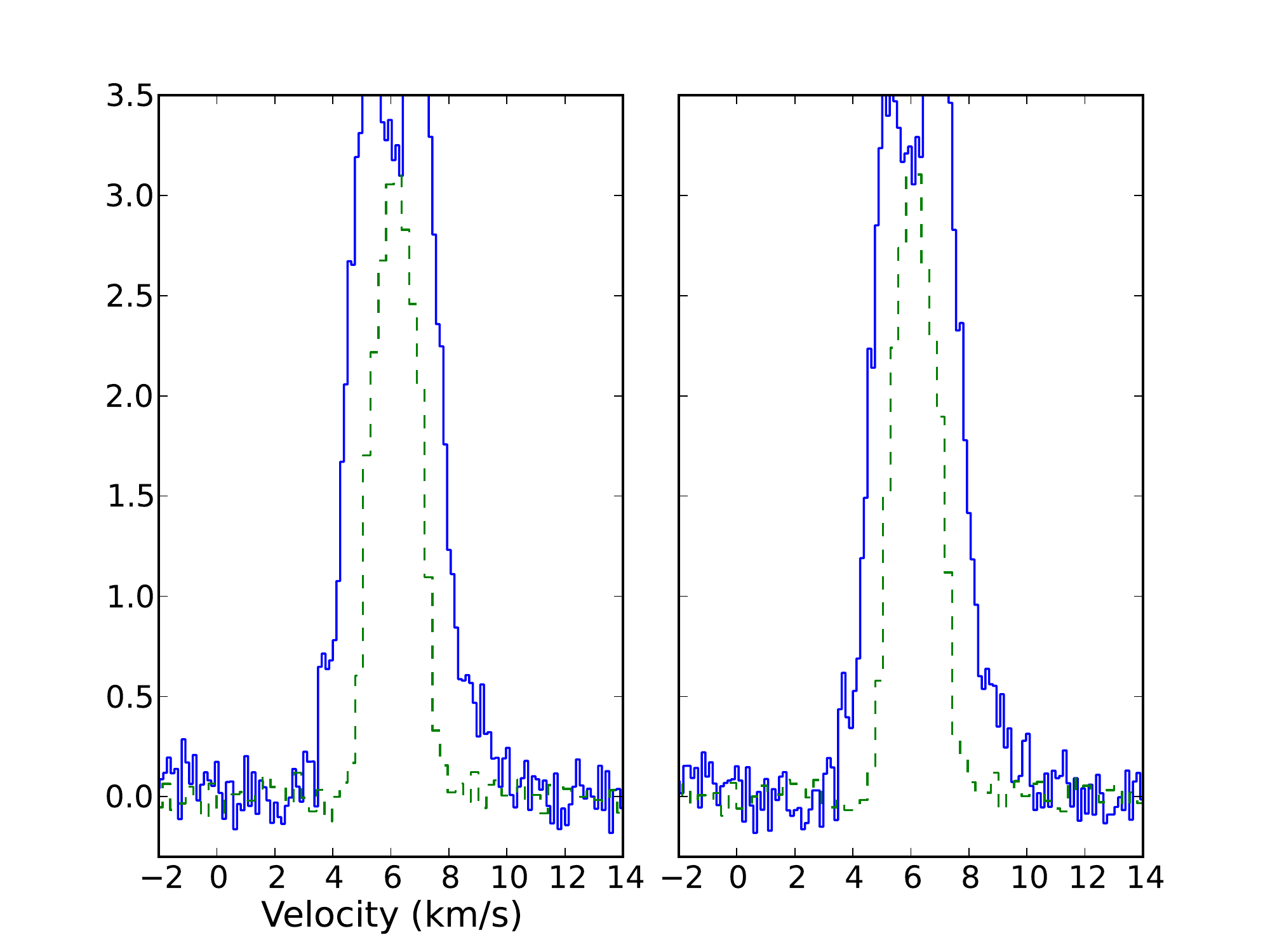}
\caption{Average Spectra of \co\ emission (blue solid lines) and
  \coa\ (green dashed lines) towards the blueshifted (left) and
  redshifted (right) lobes of the outflow seen towards the L1527
  region shown in Fig~\ref{04361_outflow}. The temperature scale is in
  T$_{A}^{*}$. \label{l1527_spectra}}
\end{center}
\end{figure}

Our map of the high velocity emission is shown in the top part of 
Figure~\ref{04361_outflow}.  We see a bipolar outflow
oriented east-west with a morphology and angular size similar to that
seen in \citet{hogerheijde1998}.  An east-west p-v plot is presented in
Figure~\ref{l1527_posvel} showing a clear bipolar signature.  Averaged
spectra are shown in Figure~\ref{l1527_spectra} and along with the p-v
plot reveal weak and relatively low velocity outflow emission.

\subsubsection{IRAS 04369+2539 (IC 2087 IR)}

\citet{heyer1987} first detected high velocity redshifted emission
associated with IRAS 04369+2539.  Their maps revealed a one-sided,
redshifted only outflow with an angular extent of 14 arcminutes.  An
extended reflection nebula, IC 2087, is associated with this source
and the YSO is often refered to as IC 2087 IR.  Also associated with
this outflow are the two Herbig-Haro objects HH 395A and HH 395B
\citep{gomez1997} located approximately 2 arcmin north-east of IC 2087
IR.  The relationship of these Herbig-Haro objects to the outflow is
unclear.

\begin{figure}%[hbp]
\begin{center}
\includegraphics[width=0.9\hsize]{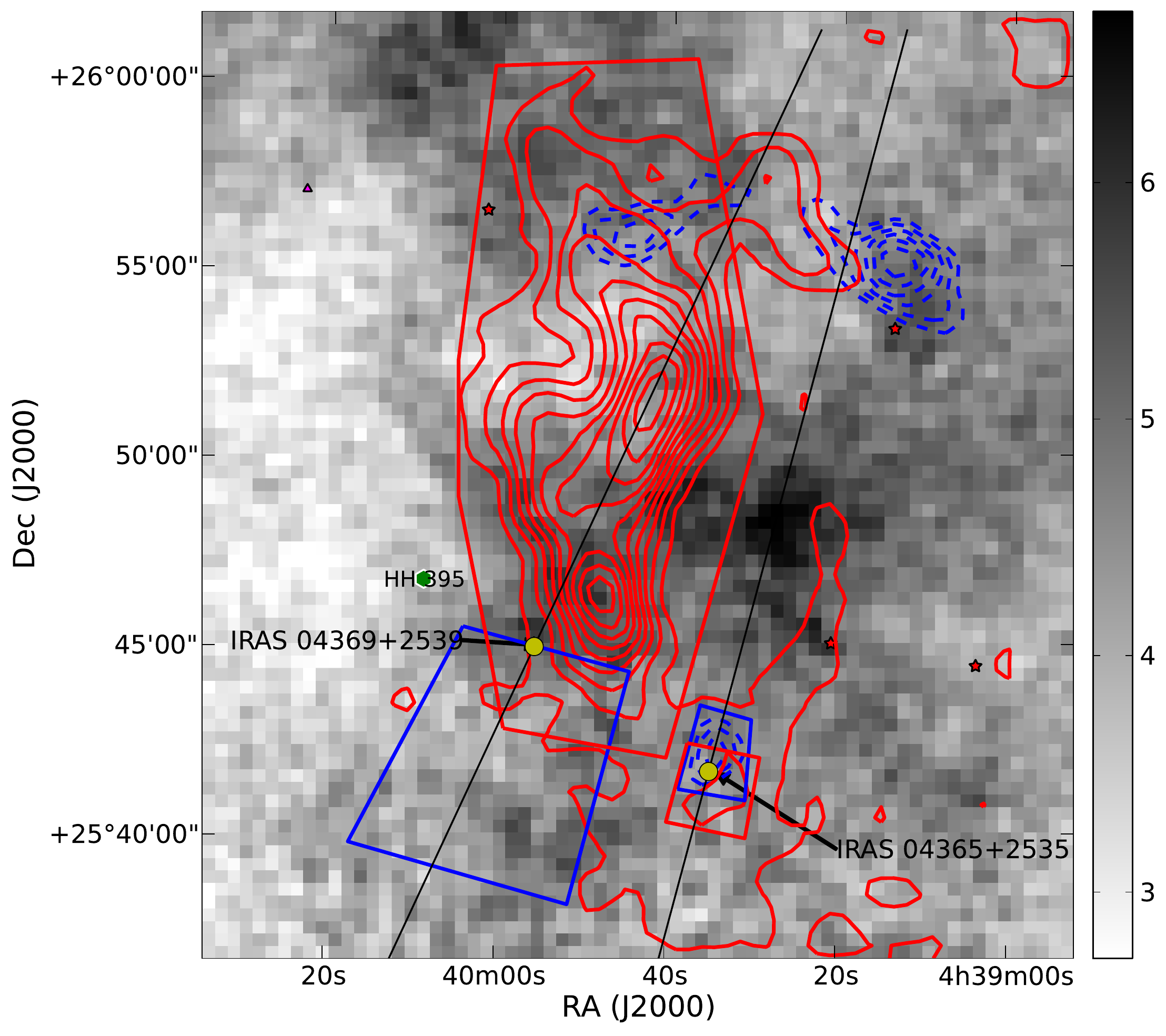}
\caption{Contour map of blueshifted and redshifted gas about a
  $25^\prime\times 25^\prime$ region centred at RA of 04:39:45 and
  Dec of 25:49:00. Two outflows, IC 2087 and TMC1A can be seen in the figure. See
  Figure~\ref{041159_outflow} for details on symbols and markers.
  \co\ blueshifted and redshifted integrated
  intensity are for velocities of {\bf -1 to 3.9 \kms} and {\bf 8.2 to
    13 \kms} respectively. Blueshifted contours range from 0.66 to 2.7
  in steps of 0.075 \kkms, and redshifted contours range from 0.66 to
  5.4 in steps of 0.075 \kkms.
\label{ic2087_outflow}}
\end{center}
\end{figure}

\begin{figure}%[hbp]
\begin{center}
\includegraphics[width=0.9\hsize]{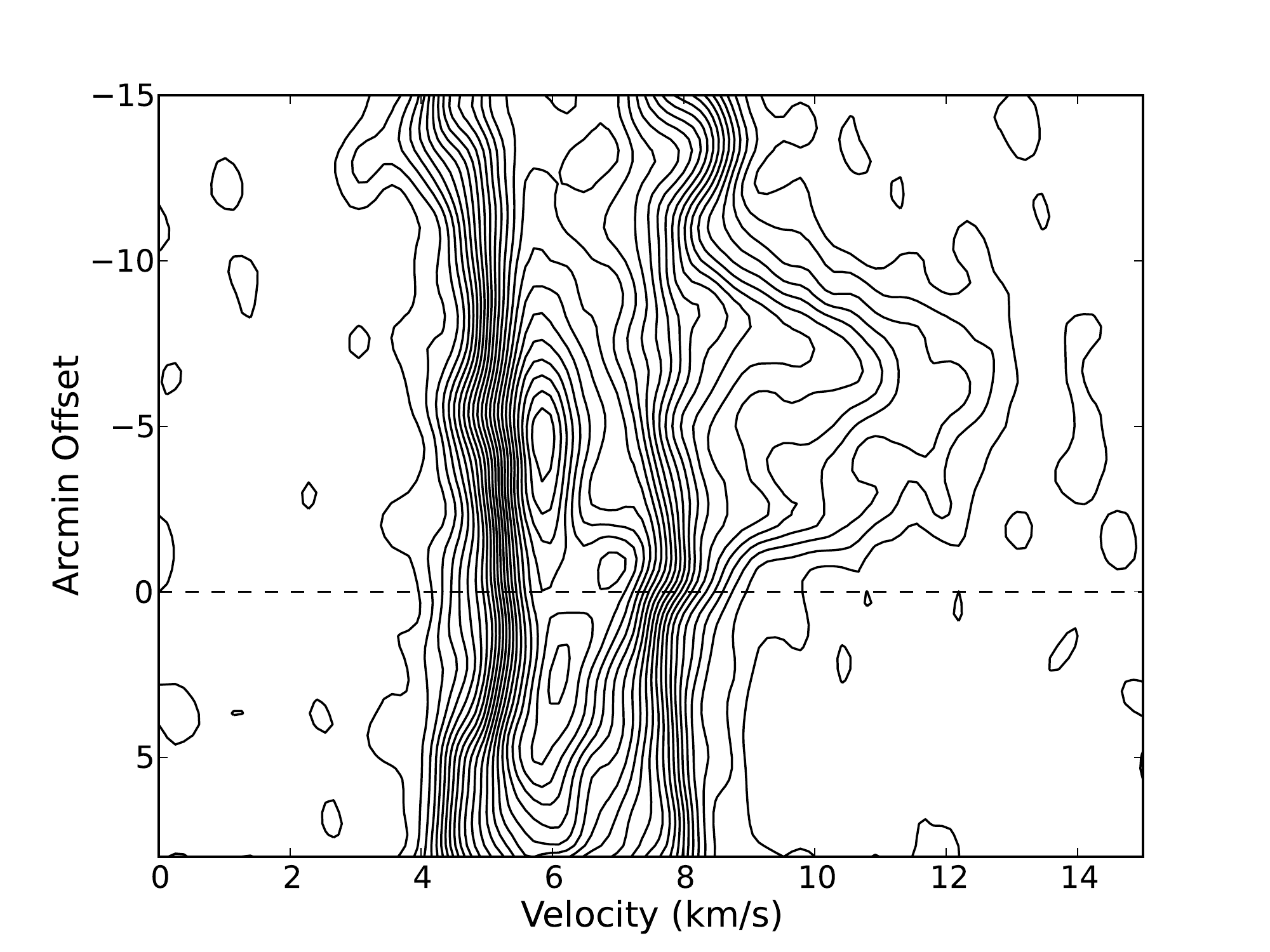}
\caption{Position velocity of \co\ emission towards the IC2087 region,
  through the slice at p.a. of $65^\circ$ through IRAS 04369+2539
  shown in Figure~\ref{04361_outflow}. The contour range is 0.12 to
  5.32 K in steps of 0.2 K. Shown in dashed line is the position of
  the yellow circle shown in Fig~\ref{ic2087_outflow}.
\label{ic2087_posvel}}
\end{center}
\end{figure}

\begin{figure}%[hbp]
\begin{center}
\includegraphics[width=0.9\hsize]{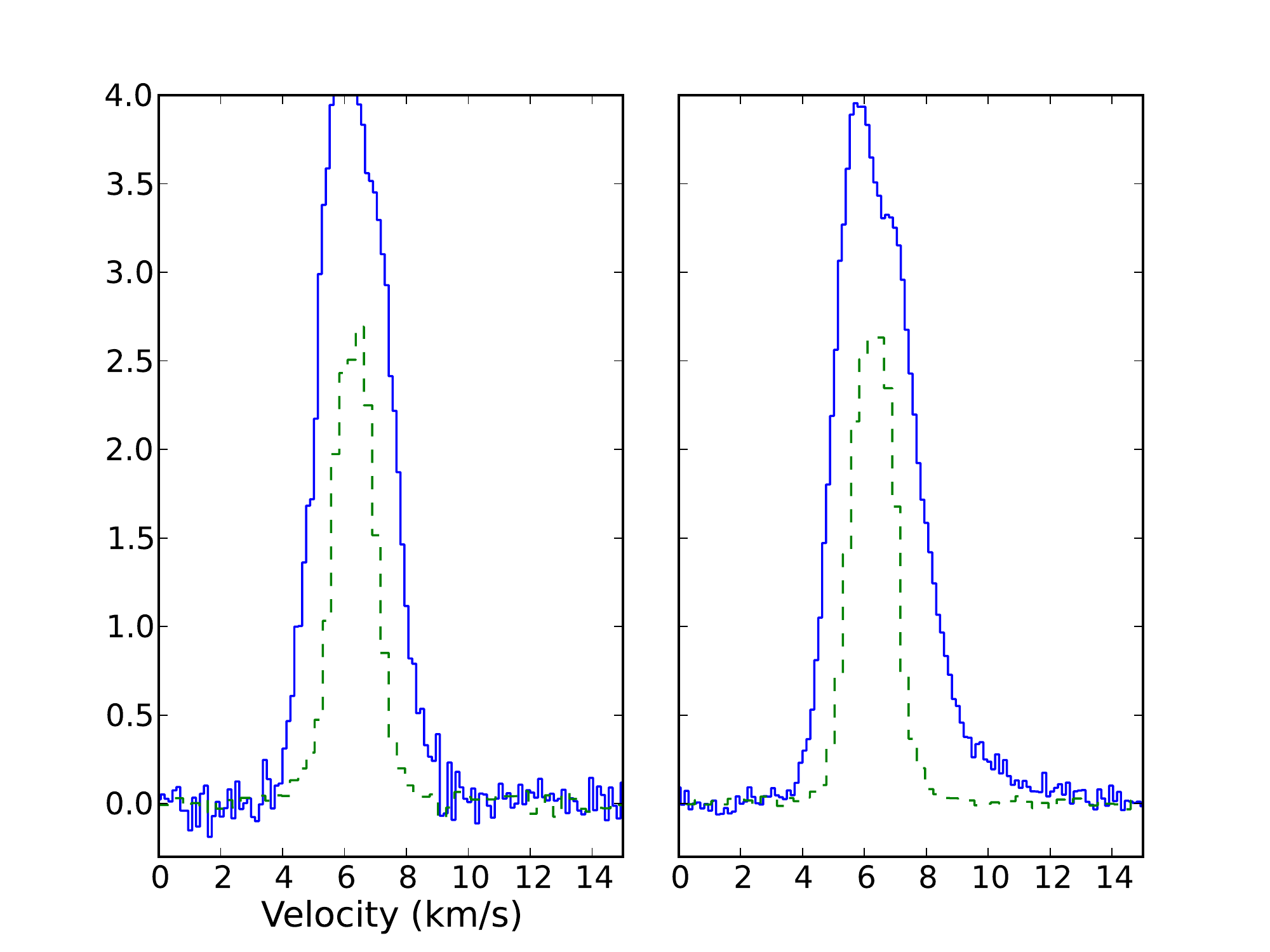}
\caption{Average Spectra of \co\ emission (blue solid lines) and
  \coa\ (green dashed lines) towards the blueshifted (left) and redshifted
  (right) lobes of the outflow seen towards the IC2087 region shown in
  Fig~\ref{ic2087_outflow}. The temperature scale is in
  T$_{A}^{*}$. \label{ic2087_spectra}}
\end{center}
\end{figure}

Our map of the high velocity redshifted and blueshifted emission is
shown in Figure~\ref{ic2087_outflow}.  The distribution of redshifted
outflow emission is very similar to that shown in \citet{heyer1987}.
A p-v plot along the axis through IC 2087 IR, marked in
Figure~\ref{ic2087_outflow} is shown in Figure~\ref{ic2087_posvel} and
shows prominent redshifted emission.  Spectra averaged within the
polygons shown in Figure~\ref{ic2087_outflow} are shown in
Figure~\ref{ic2087_spectra}, and like the p-v plot, show prominent
redshifted emission and little evidence for blueshifted high velocity
emission.

\subsubsection{IRAS 04365+2535 (TMC-1A)}

This outflow is located in the TMC 1A core, hence its name.  The first
evidence for high velocity gas was provided in the snap-shot
interferometer survey by \citet{terebey1989}.  This sources was
observed by \citet{moriarty-schieven1992}, who found a total line
width of 29.8 \kms\ in the CO J=3-2 line, however the line was
asymmetrical with much higher blueshifted velocities than redshifted.
\citet{tamura1996}, \citet{chandler1996}, \citet{bontemps1996} and
\citet{hogerheijde1998} all mapped this region revealing an outflow
centred on the IRAS source and extended on angular scales of at least
2 arcminutes.  The outflow is bipolar, however the redshifted emission
is very weak.

\begin{figure}%[hbp]
\begin{center}
\includegraphics[width=0.9\hsize]{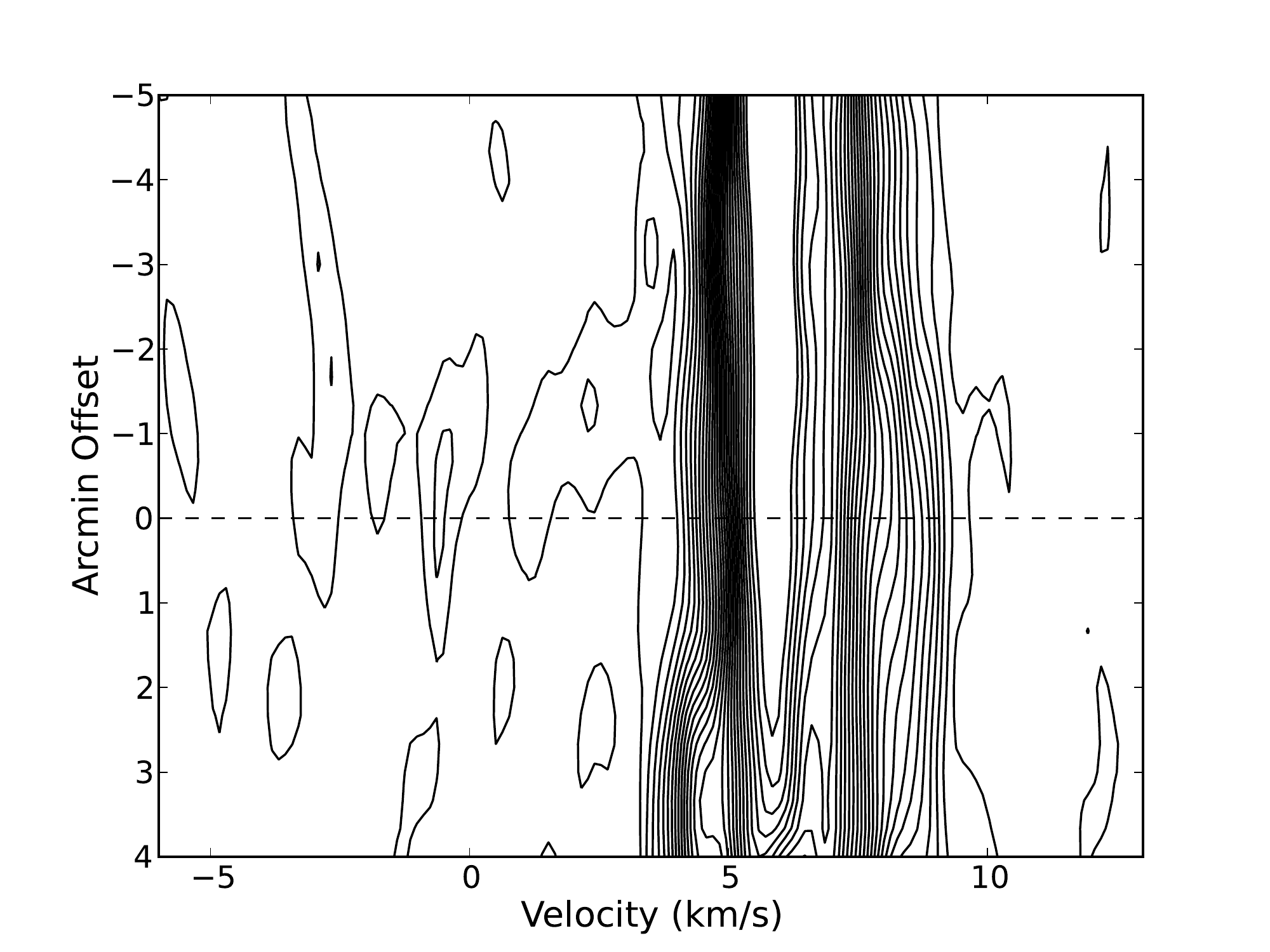}
\caption{Position velocity of \co\ emission towards the TMC1A region,
  through the slice at p.a. of $75^\circ$ through IRAS 04365+2535
  shown in Figure~\ref{ic2087_outflow}. The contour range is 0.12 to 4
  K in steps of 0.15 K. Shown in dashed line is the position of the
  yellow circle shown in Fig~\ref{ic2087_outflow}.
\label{tmc1a_posvel}}
\end{center}
\end{figure}

\begin{figure}%[hbp]
\begin{center}
\includegraphics[width=0.9\hsize]{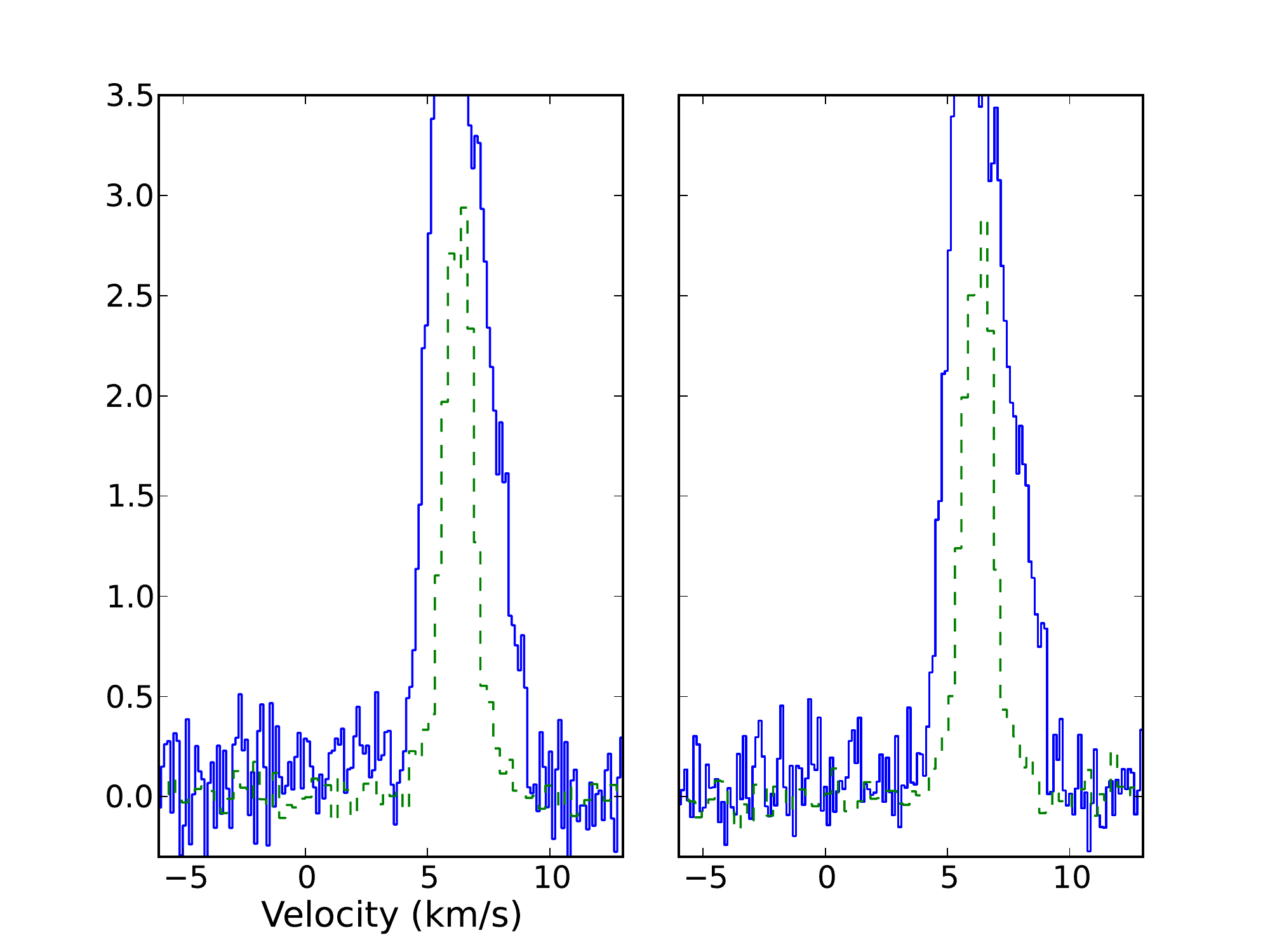}
\caption{Average Spectra of \co\ emission (blue solid lines) and
  \coa\ (green dashed lines) towards the blueshifted (left) and
  redshifted (right) lobes of the outflow seen towards the TMC1A
  region shown in Fig~\ref{ic2087_outflow}. The temperature scale is
  in T$_{A}^{*}$. \label{tmc1a_spectra}}
\end{center}
\end{figure}

Our map of the high velocity emission is shown in the same figure as
the IC 2087 outflow (see Figure~\ref{ic2087_outflow}, and is the small
outflow to the bottom right). We detect only blueshifted emission.
This outflow is very small with weak emission and is near our
threshold for outflow detection.  A p-v plot is shown in
Figure~\ref{tmc1a_posvel} and averaged spectra in
Figure~\ref{tmc1a_spectra}, and in both figures only weak blueshifted
emission is detected.

\subsubsection{TMC-1 North Outflow}

\begin{figure}%[hbp]
\begin{center}
\includegraphics[width=0.9\hsize]{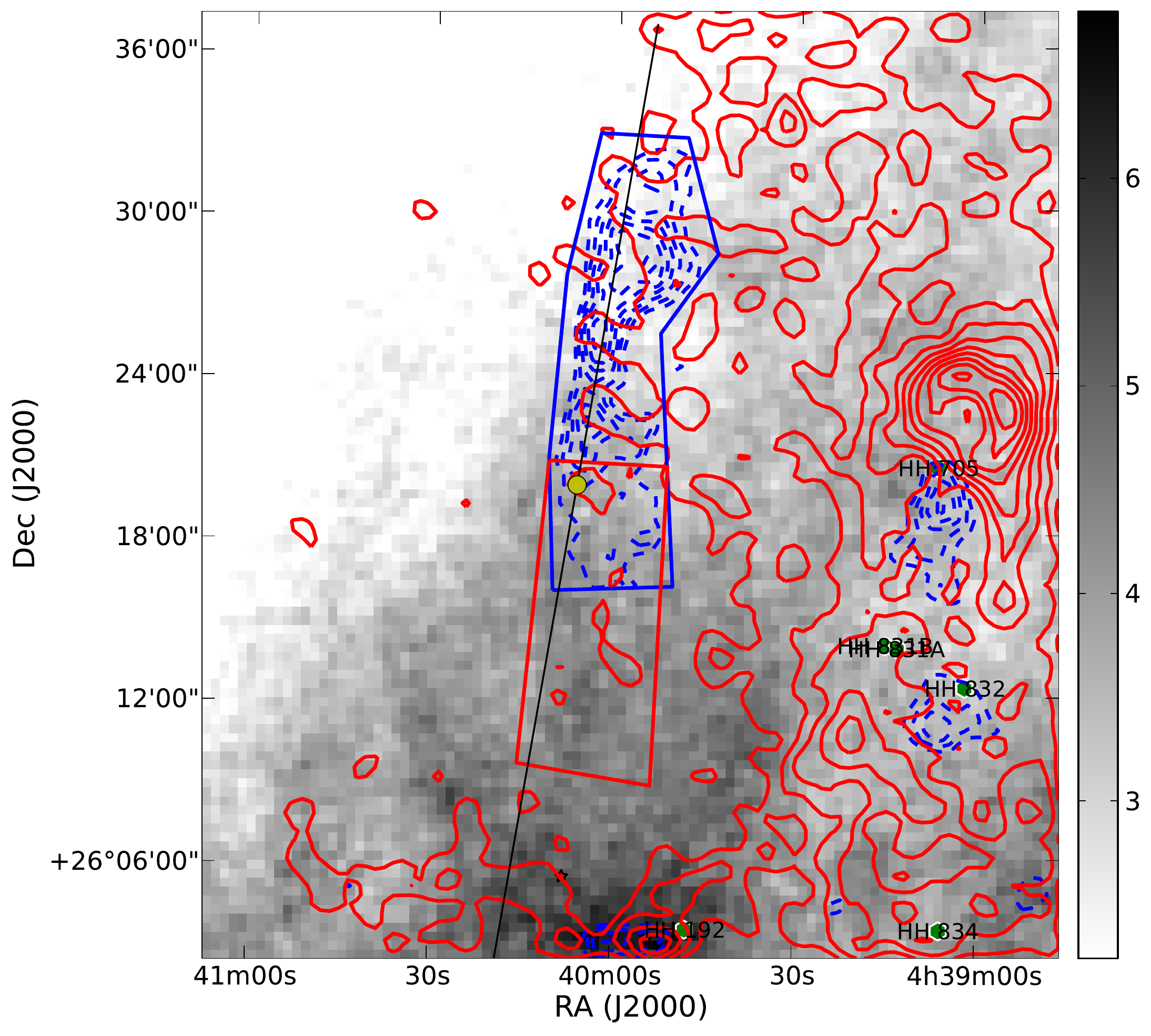}
\caption{Contour map of blueshifted and redshifted gas about a
  $35^\prime\times 35^\prime$ region in the region around TMC-1
  North. See Figure~\ref{041159_outflow} for details on symbols and
  markers.  \co\ blueshifted and integrated intensity are for
  velocities of {\bf -1 to 4.1 \kms} and {\bf 8.1 to 13 \kms}
  respectively. Blueshifted contours range from 0.6 to 2.5 in steps of
  0.075 \kkms, and redshifted contours range from 0.6 to 3.6 in steps
  of 0.075 \kkms.
\label{nbs5_outflow}}
\end{center}
\end{figure}

\begin{figure}%[hbp]
\begin{center}
\includegraphics[width=0.9\hsize]{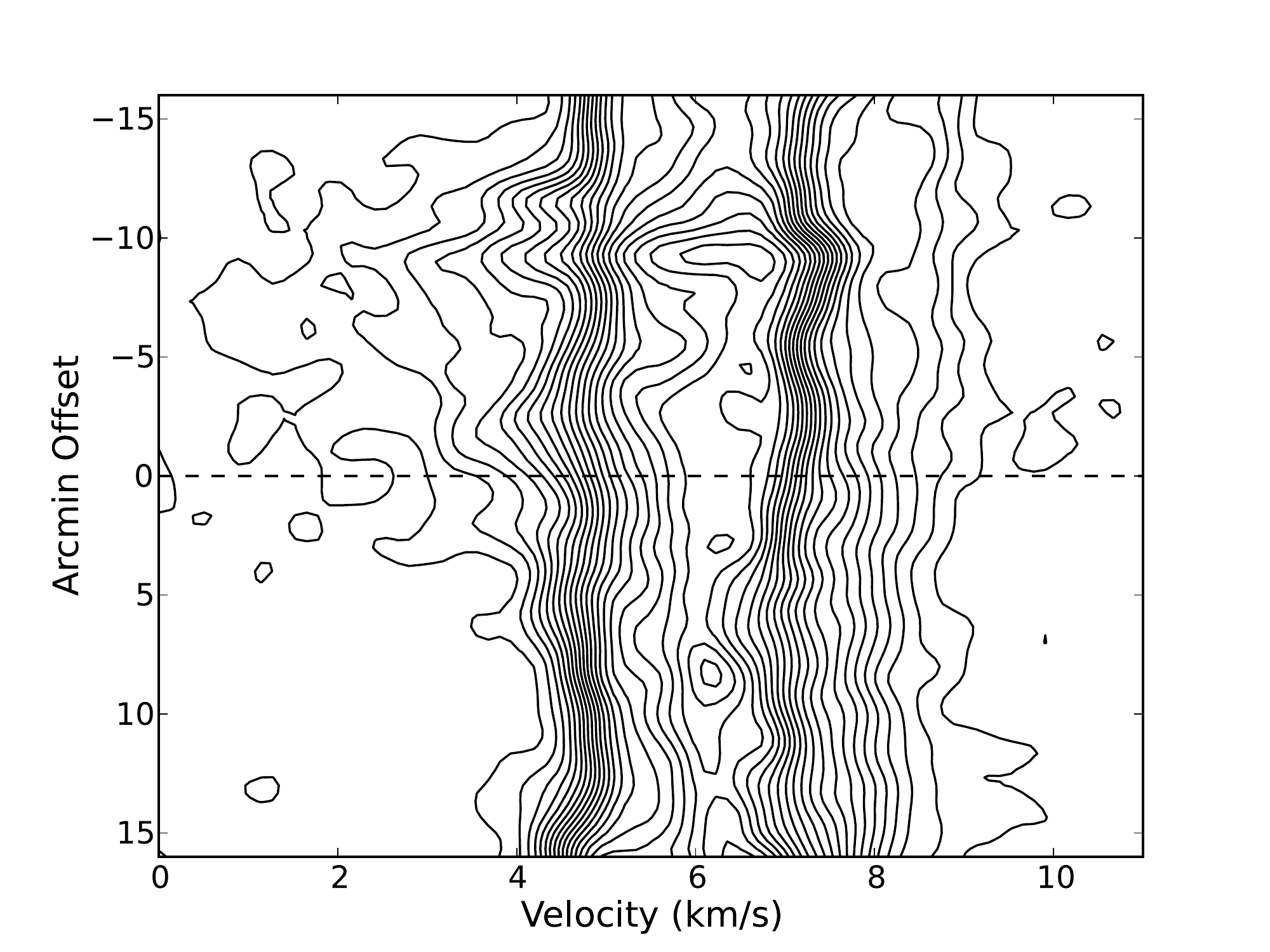}
\caption{Position velocity of \co\ emission towards the TMC-1 North
  region, through the slice at p.a. of $80^\circ$ shown in
  Figure~\ref{04361_outflow}. The contour range is 0.12 to 5.32 K in
  steps of 0.2 K. Shown in dashed line is the position of the yellow
  circle shown in Fig~\ref{nbs5_outflow}.
\label{nbs5_posvel}}
\end{center}
\end{figure}

\begin{figure}%[hbp]
\begin{center}
\includegraphics[width=0.9\hsize]{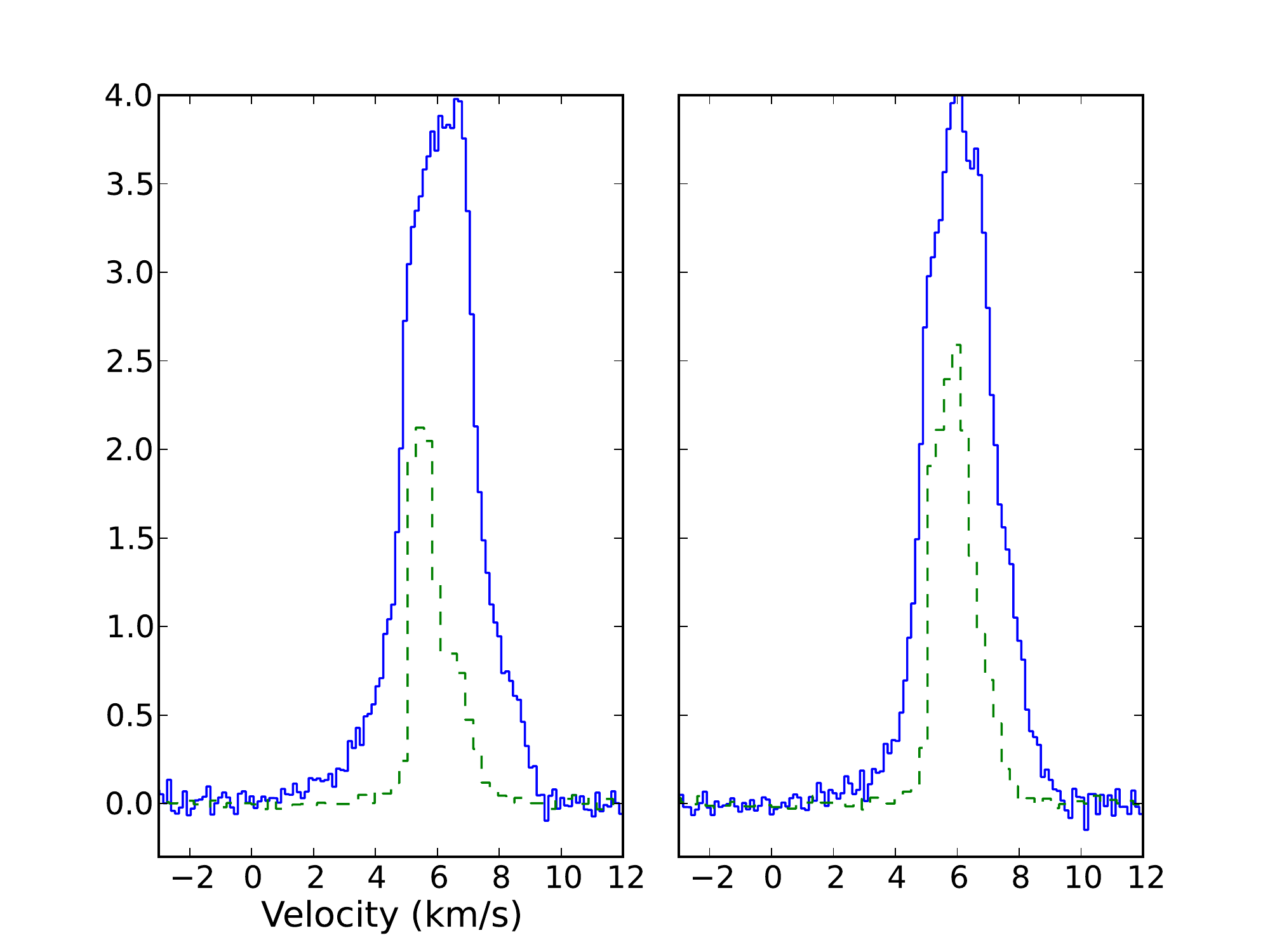}
\caption{Average Spectra of \co\ emission (blue solid lines) and
  \coa\ (green dashed lines) towards the blueshifted (left) and redshifted
  (right) lobes of the outflow seen towards the TMC-1 North region
  shown in Fig~\ref{nbs5_outflow}. The temperature scale is in
  T$_{A}^{*}$. \label{nbs5_spectra}}
\end{center}
\end{figure}

We have detected a new outflow with a striking blueshifted emission
north of the core TMC-1.  The outflow is located approximately 12
arcminutes east of HH 705 Outflow and 20 arcminutes north of IC 2087
IRS (see Figure~\ref{ic2087_overview}).  We label this outflow TMC-1
North Outflow.  A map of the high velocity emission is shown in
Figure~\ref{nbs5_outflow}.  This blueshifted only outflow is highly
collimated and has an angular extent of nearly 20 arcminutes.  There
is no known YSO in the vicinity, thus the source of this outflow is
unknown.  We have chosen a position at the south end of the outflow as
the outflow centroid, and this position is given in
Table~\ref{outflowlist}.  This position is close to one of the high
extinction regions from the data of \citet{padoan2002} and shown in
\citet{toth2004}. We show a p-v plot along the path marked in
Figure~\ref{nbs5_outflow} in Figure~\ref{nbs5_posvel} and this plot
shows clear evidence for high velocity blueshifted emission.  The
averaged spectra are shown in Figure~\ref{nbs5_spectra}, again clear
evidence is seen for high velocity blueshifted gas, however in the
polygon south of the chosen outflow centroid, no redshifted emission
is detected.

\subsubsection{IRAS 04381+2540 (TMC 1)}

IRAS 04381+2540 is located in the TMC 1 core.  Evidence for high
velocity gas was first found by \citet{moriarty-schieven1992} who
measured a total line width of 36.6 km s$^{-1}$ in the CO J=3-2 line
toward this source.  This region was subsequently mapped by
\citet{bontemps1996}.  \citet{chandler1996} and
\citet{hogerheijde1998} who all found a small bipolar outflow centred
on this IRAS source with an angular extent of about 1.5
arcminutes. Infrared imaging \citep{apai2005, terebey2006} reveal a
narrow jet and a wide-angle conical outflow cavity.  The jet, first
seen by \citet{gomez1997}, is directed nearly due north of the YSO.

\begin{figure}%[hbp]
\begin{center}
\includegraphics[width=0.9\hsize]{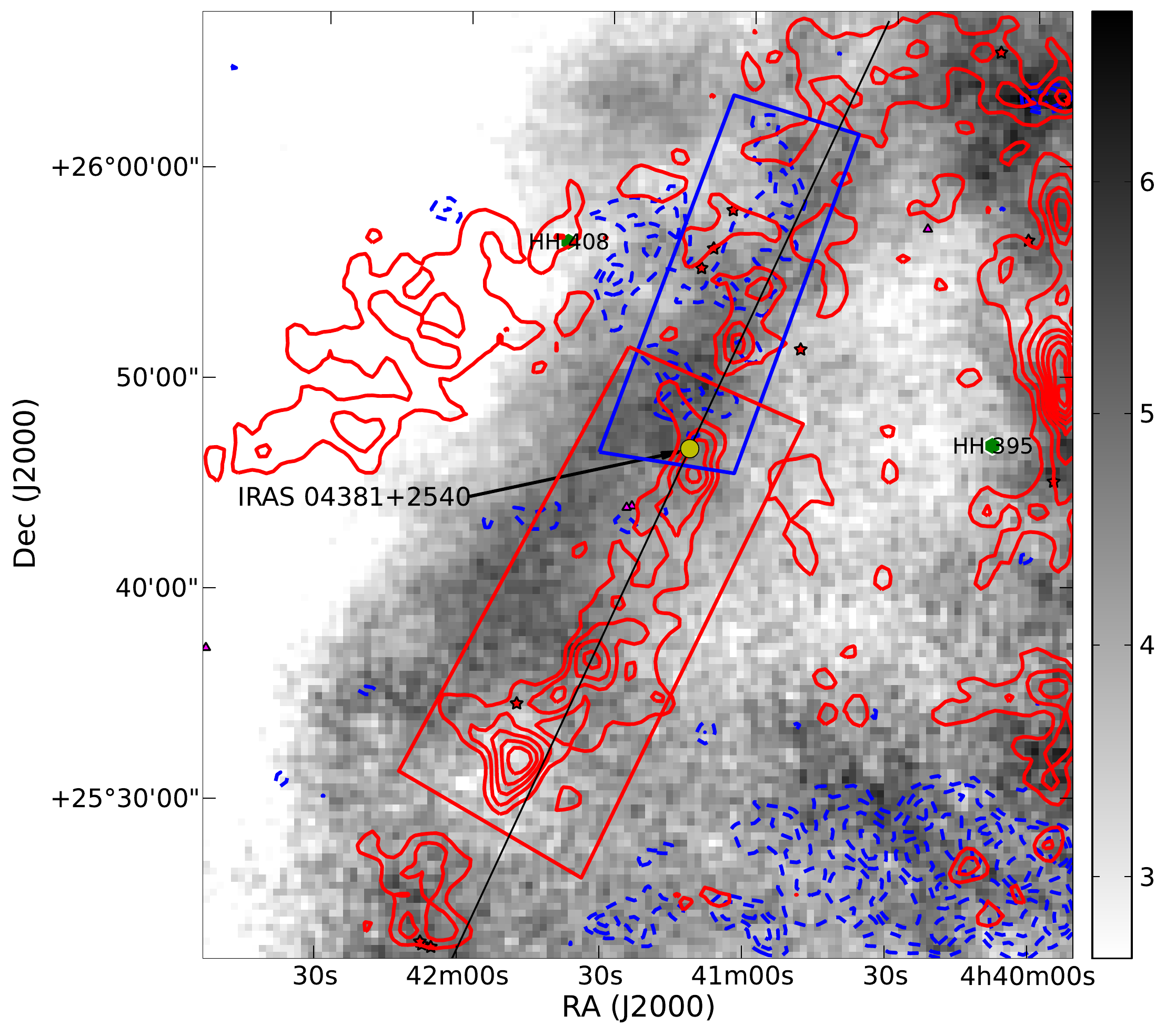}
\caption{Contour map of blueshifted and redshifted gas about a
  $45^\prime\times 45^\prime$ region centred on TMC1. See
  Figure~\ref{041159_outflow} for details on symbols and markers.
  \co\ blueshifted and redshifted integrated intensity are for
  velocities of {\bf -1 to 4.1 \kms} and {\bf 8. to 13 \kms}
  respectively. Blueshifted contours range from 0.61
  to 2.8 in steps of 0.075 \kkms, and redshifted contours range from
  0.61 to 4.4 in steps of 0.075 \kkms. 
\label{tmc1_outflow}}
\end{center}
\end{figure}

\begin{figure}%[hbp]
\begin{center}
\includegraphics[width=0.9\hsize]{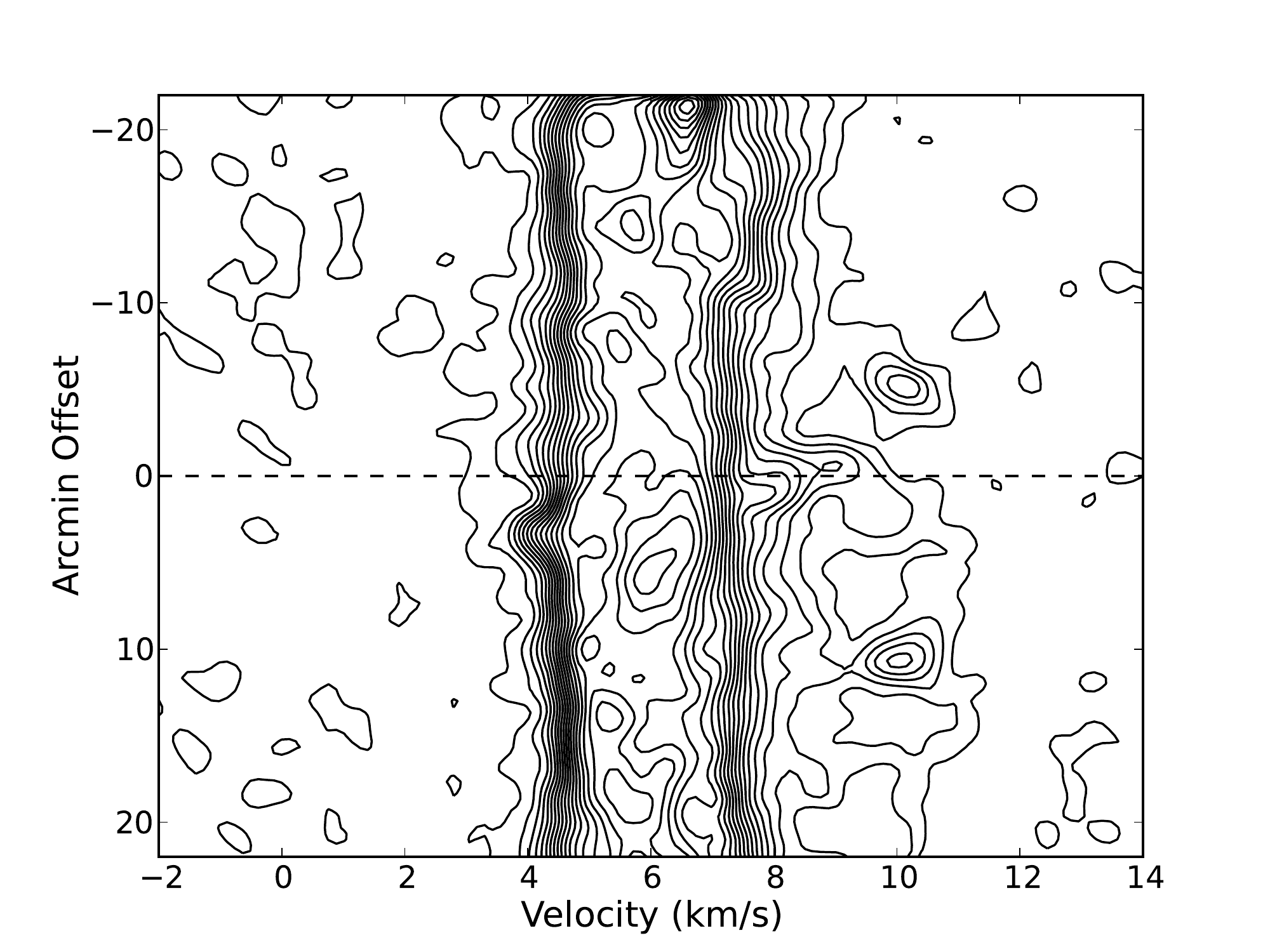}
\caption{Position velocity of \co\ emission towards the TMC1 region,
  through the slice at p.a. of $65^\circ$ shown in
  Figure~\ref{04361_outflow}. The contour range is 0.12 to 5.32 K in
  steps of 0.2 K. Shown in dashed line is the position of the yellow
  circle shown in Fig~\ref{tmc1_outflow}.
\label{tmc1_posvel}}
\end{center}
\end{figure}

\begin{figure}%[hbp]
\begin{center}
\includegraphics[width=0.9\hsize]{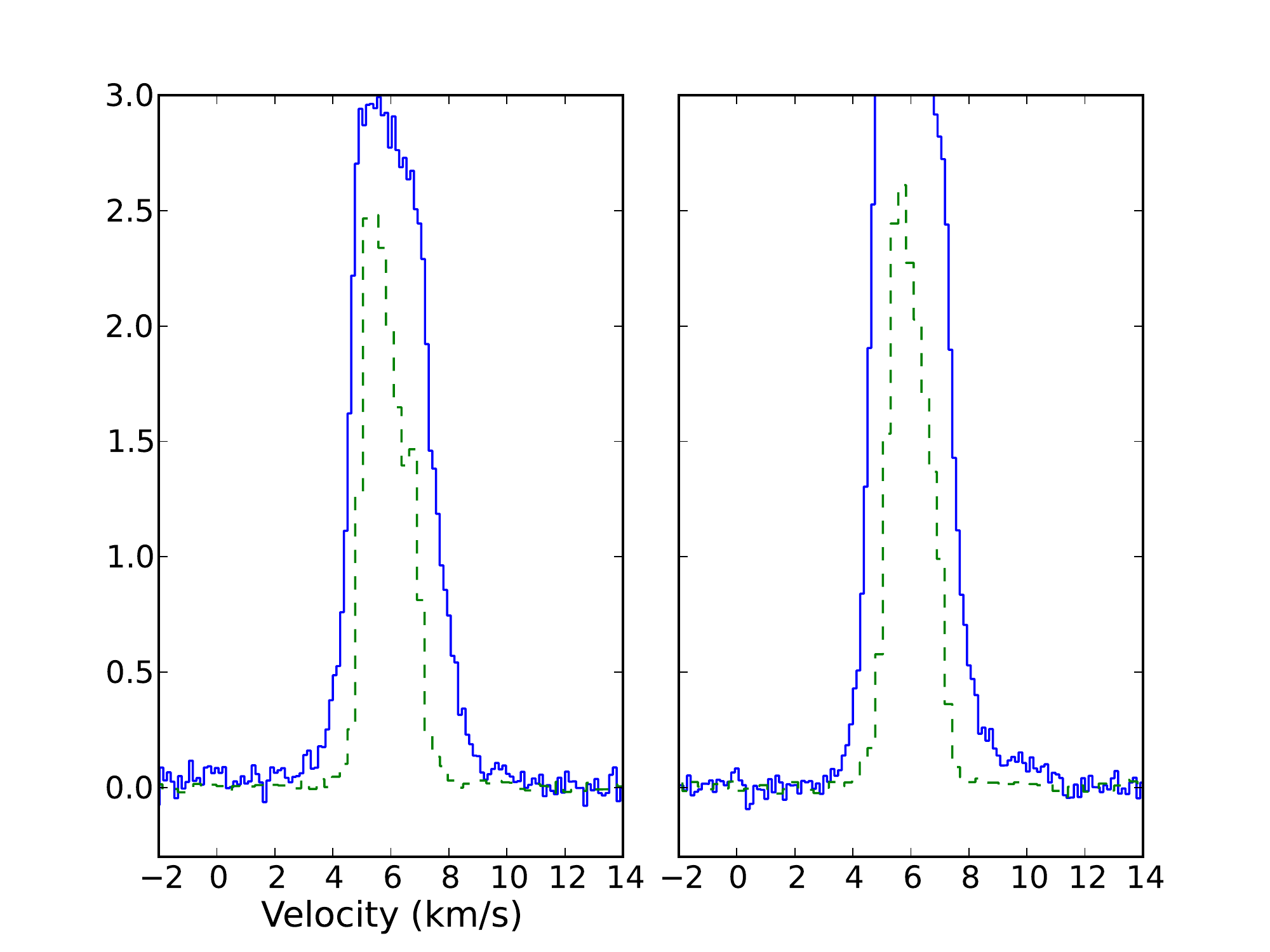}
\caption{Average Spectra of \co\ emission (blue solid lines) and
  \coa\ (green dashed lines) towards the blueshifted (left) and redshifted
  (right) lobes of the outflow seen towards the TMC1 region shown in
  Fig~\ref{tmc1_outflow}. The temperature scale is in
  T$_{A}^{*}$. \label{tmc1_spectra}}
\end{center}
\end{figure}

Our maps of the high velocity emission are shown in
Figure~\ref{tmc1_outflow} and reveal an outflow much more spatially
extended than previous maps.  The outflow near the central source is
nearly north-south, but the redshifted emission is extended nearly 20
arcminutes from the central source and curves to the south-east.  The
blueshifted emission is much weaker and the morphology of this part of
the outflow poorly determined.  The p-v plot along the line marked in
Figure~\ref{tmc1_outflow} is shown in Figure~\ref{tmc1_posvel}. The
p-v plot shows the extended nature of the redshifted lobe of the
outflow, however it misses some of the more intense regions of
redshifted and blueshifted emission.  The averaged spectra shown in
Figure~\ref{tmc1_spectra} shows clearly the redshifted outflow
emission, but the blueshifted emission is much weaker.

\subsubsection{Source 045312+265655}

\begin{figure}%[hbp]
\begin{center}
\includegraphics[width=0.9\hsize]{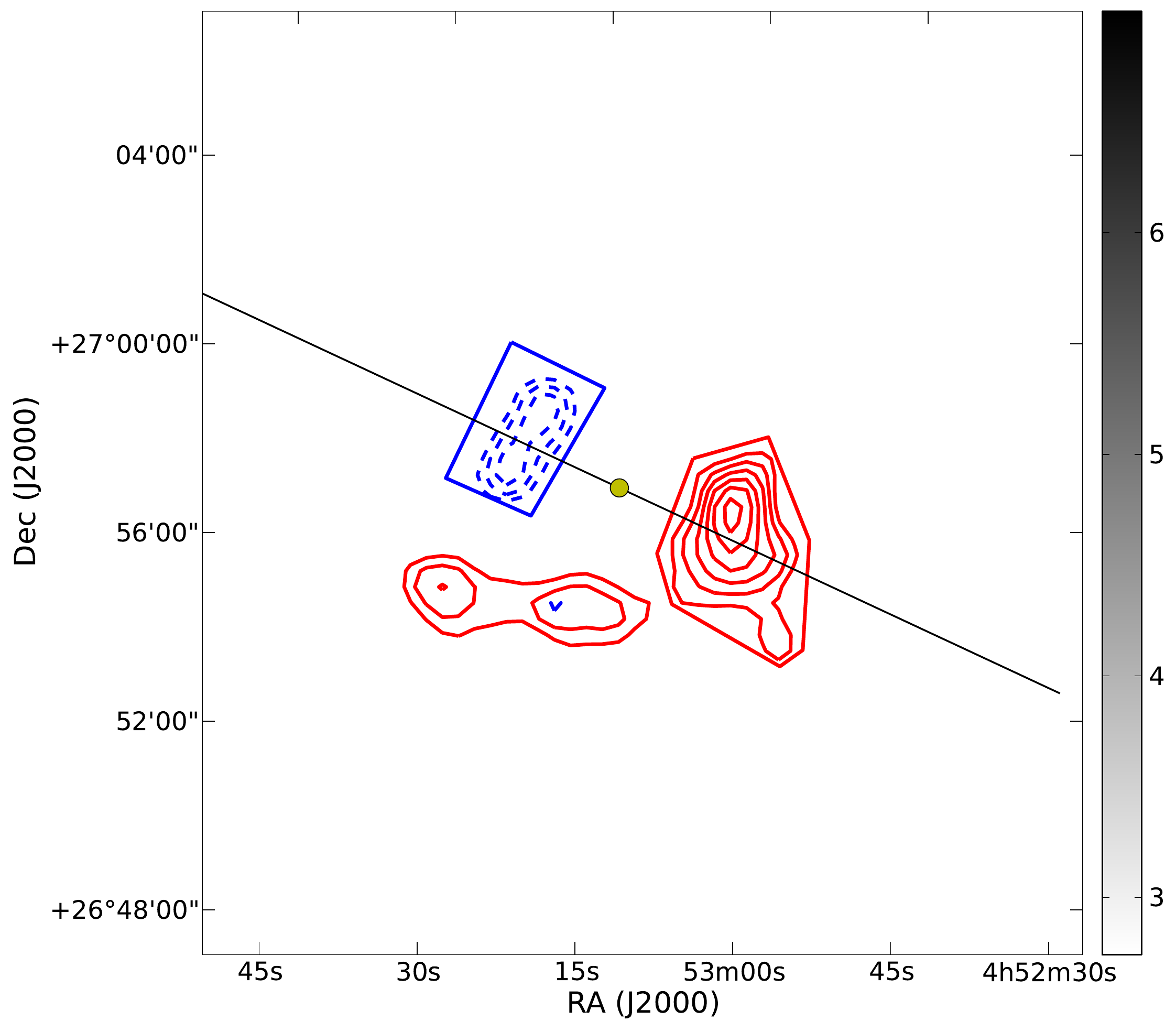}
\caption{Contour map of blueshifted and redshifted gas about a
  $20^\prime\times 20^\prime$ region centred on RA 04:53:12 and Dec
  26:56:55. No detectable \coa\ emission greater than 10 sigma is
  detected in this region. See Figure~\ref{041159_outflow} for details
  on symbols and markers.  \co\ blueshifted and redshifted integrated
  intensity are for velocities of {\bf -1 to 4.3 \kms} and {\bf 7.8 to
    13 \kms} respectively. Blueshifted contours range from 0.6 to
  2. in steps of 0.075 \kkms, and redshifted contours range from 0.6
  to 2.4 in steps of 0.075 \kkms. 
\label{nbs6_outflow}}
\end{center}
\end{figure}

\begin{figure}%[hbp]
\begin{center}
\includegraphics[width=0.9\hsize]{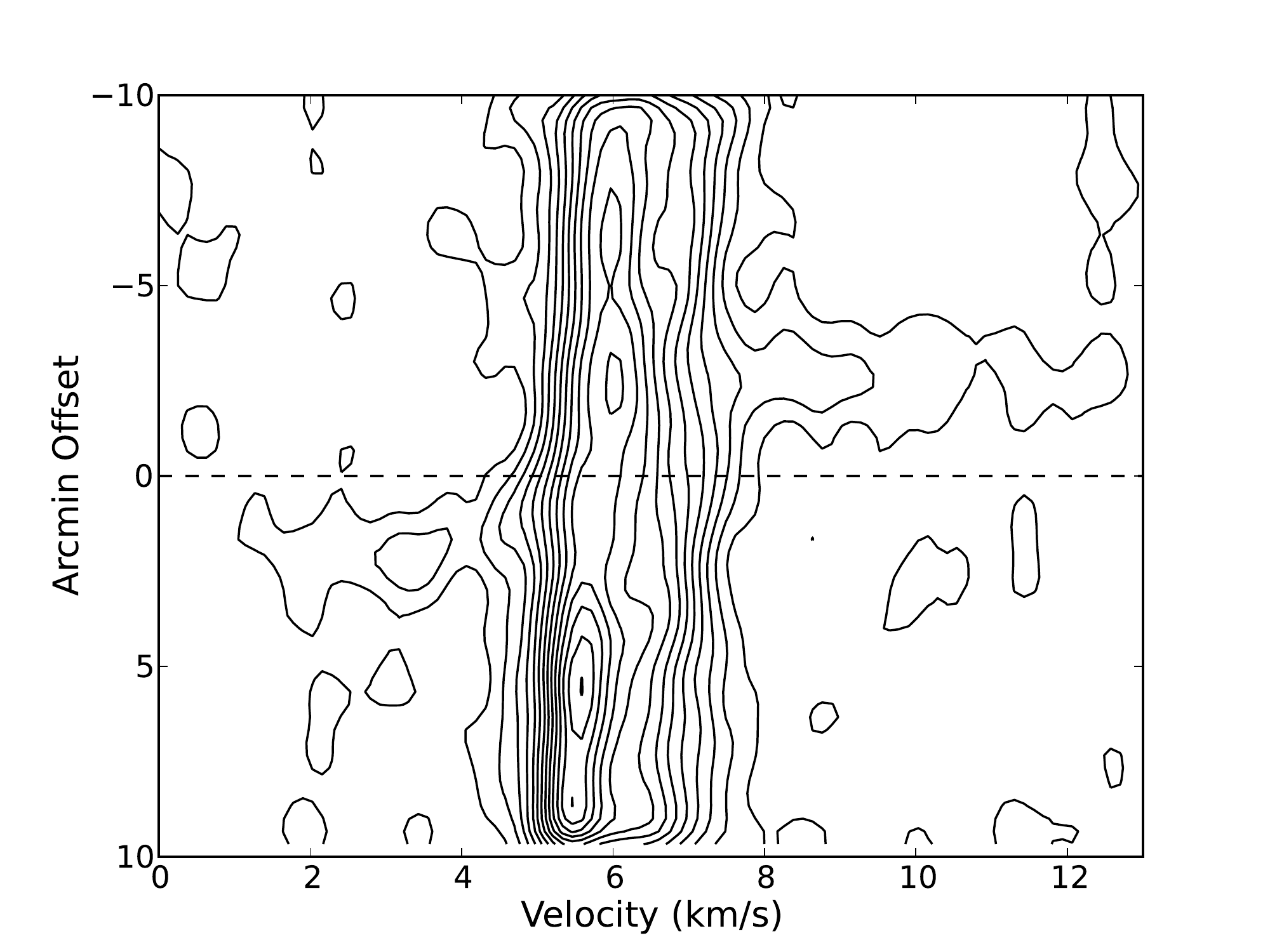}
\caption{Position velocity of \co\ emission towards the region shown
  in Figure~\ref{nbs6_outflow}, through the slice at p.a. of
  $155^\circ$ shown in that figure. The contour range is 0.12 to 5.32
  K in steps of 0.2 K. Shown in dashed line is the position of the
  yellow circle shown in Fig~\ref{nbs6_outflow}.
\label{nbs6_posvel}}
\end{center}
\end{figure}

\begin{figure}%[hbp] 
\begin{center}
\includegraphics[width=0.9\hsize]{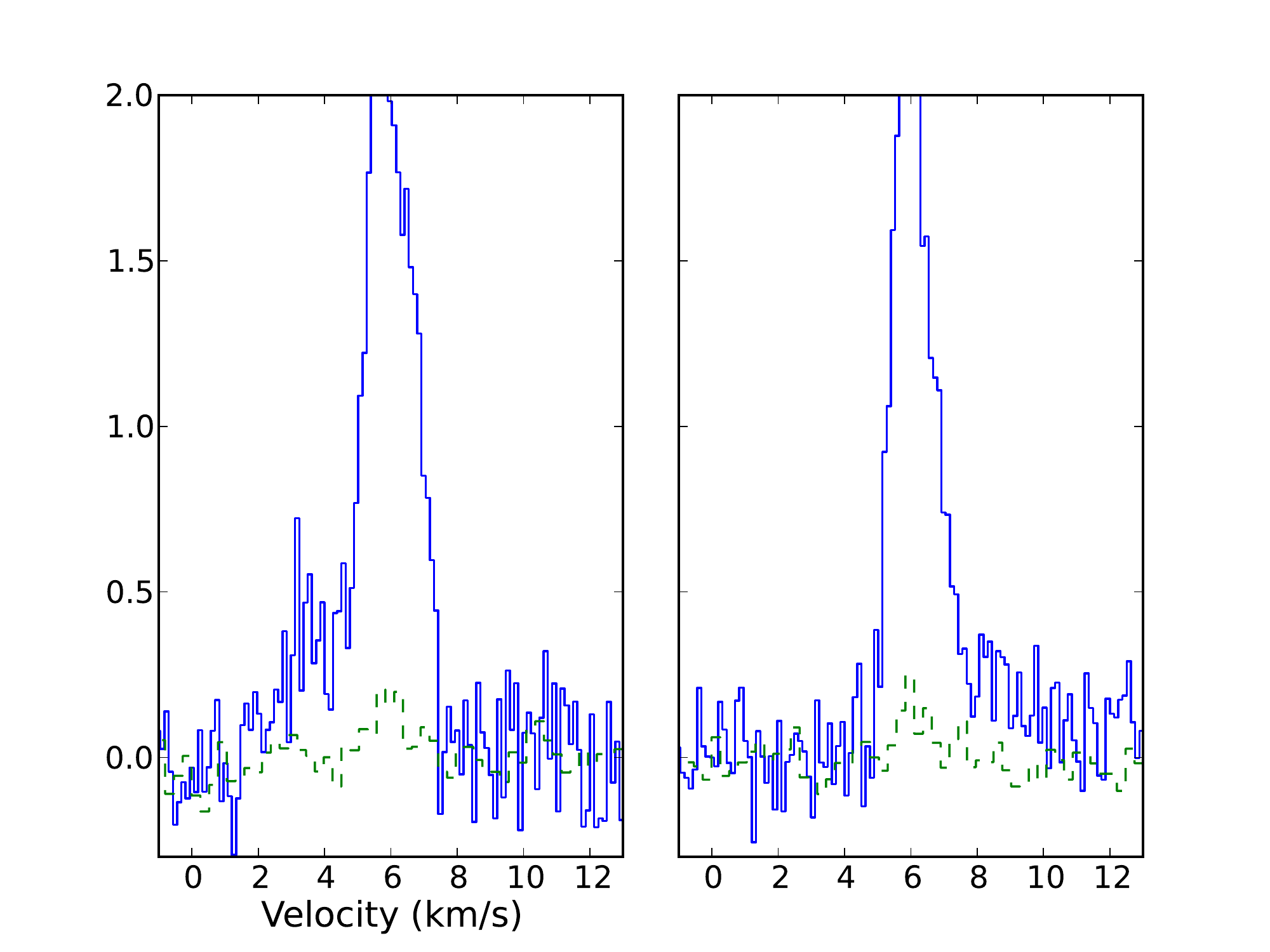}
\caption{Average Spectra of \co\ emission (blue solid lines) and
  \coa\ (green dashed lines) towards the blueshifted (left) and redshifted
  (right) lobes of the outflow seen towards the region shown in
  Fig~\ref{nbs6_outflow}. The temperature scale is in
  T$_{A}^{*}$. \label{nbs6_spectra}}
\end{center}
\end{figure}

Our outflow search process yielded a surprising detection of a
possible outflow at RA of 04:43:12 and Dec of 26:56:55 (J2000). We
call this outflow Source 045312+265655. Figure~\ref{nbs6_outflow}
shows a map of the high velocity emission towards this region. There
are no known YSOs in this area, although it is very likely
that the Spitzer survey did not cover this particular region. 
%There is almost no detectable \coa\ emission. 
This outflow is curious in a variety of ways. There is little ambient
gas around to sweep up as there is almost no detectable
\coa\ emission. The position velocity diagram shown in
Figure~\ref{nbs6_posvel} reveals a canonical outflow signature. The
averaged spectra in the blue and redshifted polygonal regions
(Figure~\ref{nbs6_spectra}) show high velocity wings in blue and
redshifted gas. So it is no wonder that our outflow detection
algorithm picked up this source. But it is very atypical in that we
cannot attribute any known driving source for this outflow. Moreover,
it is hard to conceive how a YSO could even form in such low column
density regions.

This region in the north-east section of the Taurus Molecular Cloud is
the same region studied by \citet{heyer2008}, where they found the low
column density substrate of gas with striations of elevated \co\ gas
that is well aligned with local magnetic field direction.

\subsection{Outflows Not Detected}
\label{non_detections}

There are a number of previously identified outflows that were not
detected in our survey.  From the outflows in
Table~\ref{known_outflow_table}, we were unable to detect four of
these even though they were within the region surveyed (IRAS
04181+2655, IRAS 04240+2559, L1529 and IRAS 04302+2247).

The outflow, IRAS 04181+2655 (CFHT-19), is a bipolar outflow first
detected by \citet{bontemps1996}. The recent observations by
\citet{davis2010} confirm the presence of a bipolar outflow in this
region, which in Table 2 of their paper is labeled J04210795+2702204.
This outflow is small and has relatively weak emission.  The
detections by \citet{bontemps1996} and \citet{davis2010} were made in
the CO J=2-1 and CO J=3-2 lines respectively.  For warm, optically
thin emission, as may be expected for outflows, the emission in
these higher rotational transitions of CO should be stronger than in the
CO J=1-0 line used in our survey.  It is likely that this outflow is
below our detection threshold.

Similarly, the redshifted only molecular outflow IRAS 04240+2559,
detected by \citet{mitchell1994}, and associated with DG Tau, has very
weak emission even in the CO J=3-2 line used in their study. We note
that the CO J=3-2 spectrum obtained toward this source by
\citet{moriarty-schieven1992} had very broad wings and had a total
velocity extent of 29.9 km s$^{-1}$, so there is little doubt that an
outflow is present. As in IRAS 04181+2655, this outflow must be below
our detection threshold.

The L1529 outflow was detected by \citet{lichten1982} in the CO
J=1-0 emission.  He found high velocity emission toward two positions
with full velocity extents as large as 30 km s$^{-1}$.  The
highest velocity emission was found in a region about 8 arcminutes
south-east of Haro 6-13.  This region was subsequently studied
by \citet{goldsmith1984} and their observations of the CO J=1-0 line
did not confirm the presence of high velocity emission, although
their sensitivity was sufficient to readily detect the high velocity
emission reported by \citet{lichten1982}.  We also found no evidence
for an outflow in our survey, so this outflow has never been confirmed
and it's existence is doubtful.

Finally, IRAS 04302+2247, which was mapped by \citet{bontemps1996} in
the CO J=2-1 line, was not detected in our survey.  Their small map
revealed a bipolar outflow with an irregular geometry.  As in previous
outflows, the emission is very weak and likely below our detection
limit.

In addition to the outflows in Table~\ref{known_outflow_table}, a number
of additional outflow candidates were found by \citet{davis2010} in
their CO J=3-2 map of the L1495 region.  The  outflows E-CO-R1,
E-CO-R2, SE-CO-R1, SE-CO-R2, SE-CO-flow1, CFHT-21 and SE-CO-B1
were not identified as outflows in our survey.

\subsection{Summary of Masses, Energy, etc.}
\label{masses}

To understand the effects of a molecular outflow in its immediate
environment, it is important to calculate the mass, momentum, and
mechanical energy output of the outflows as accurately as
possible. The availability of both \co\ and \coa\ data allow us to
mitigate some issues that typically can cause rather large
uncertainties in the estimation of the physical properties of
outflows. One major issue is the confusion of the slow-moving parts of
the outflow with the ``ambient'' gas at low velocities that are not
part of the outflow. By judging the velocity extent of \coa\ emission
(which because of its lower abundance is predominantly seen only in
the higher column density ambient gas), we try to avoid the ambient
gas contamination by simply eliminating any emission at ``low
velocities''. That means that our mass and energetics are only lower
limits, since we are probably missing outflowing gas projected to
lower velocities with respect to the cloud's LSR. Another major source
of uncertainty in determining physical properties is the unknown
inclination of the outflow axis to the plane of the sky. This problem
is much harder to deal with. There are many methods of inclination
correction in the outflow literature. The most robust method is to use
proper motion studies of HH objects in the flow. Since most of our
flows do not have corresponding HH objects, and proper motion studies
do not exist for many of these flows, we chose to present the physical
data of outflows uncorrected for inclination effects.

Table~\ref{tbl:mass} lists the mass, momentum, energy, length and
dynamical timescale of blue and redshifted lobes of the detected
outflows. Our algorithm for the calculation of column densities is the
same as that presented in \citet{goldsmith2008}. In the polygonal
regions of the outflow contour figures presented above, we derive
column density using both \co\ and \coa\ for each pixel, and derive
total blue-shifted and red-shifted column densities (after correcting
for the relative antenna main beam efficiencies at each frequency, 0.5
and 0.45 respectively for \coa\ and \co). We use an excitation
temperature of 25~K. Our assumption of excitation temperature (25 K)
could be lower or higher than the real value. For example, in
\citet{stojimirovic2006}, using the ratio of J=3-2 to J=1-0 of \co,
they estimate an excitation temperature of 16.5 K for outflowing
gas. Using this lower temperature will decrease the estimates of mass,
momentum and energy by a factor of 1.3. However, \citet{hirano2001}
show that it is not uncommon for outflowing molecular gas to have
excitation temperatures in the 50--100 K range. Using these higher
values would increase our estimate of the outflow mass, momentum and
energy by a factor of 1.7 to 3.1 for excitation temperature of 50 and
100 K respectively.

We also assume a \co\ to \coa\ abundance ratio of 65, and an abundance
of H$_2$ to \co\ of $1.1\times 10^4$.  Column densities are then
converted to mass using the distance to Taurus of 140~pc
\citep{elias1978}. Momentum and kinetic energy are estimated from
average velocities in each lobe. The length of the outflow lobe is
derived from the presumed driving source and the furthest extent of
the contours shown in the figures. The dynamical timescale in
Table~\ref{tbl:mass} is derived simply from the length and the average
velocity of the gas in each lobe.

It should be noted that the mass, momentum and energy listed in
Table~\ref{tbl:mass} are strictly lower limits, as there are several
effects that have not been taken into account, which if properly
estimated, would increase these quantities. The main effects that make
our estimates lower limits can be listed as follows, and is similar to
the arguments presented in \citet{arce2010}. We probably are missing a
rather large fraction of low-velocity outflowing gas in our effort to
avoid contamination with ambient cloud emission. We assign a factor of
2 for this unaccounted outflow emission. We are not correcting for
angle of inclination effects. But if we assume that the average
outflow is tilted by 45$^\circ$ to the plane of the sky, then the
momentum and energy are to be scaled up by factors of 1.4 and 2
respectively.
  %A similar approach is explained in detail by
  %\citet{arce2010} for a similar dataset and analysis for their
  %COMPLETE data in Perseus.  
Combining the factor of 2 due to missing
  outflowing gas at lower velocities and the factor of 2 due to
  inclination, the total scaling factor for outflow lobe momentum and
  energy will be as large as a factor of 2.8 and 4 respectively.  So
  it may be fair to multiply the momentum and energy values listed in
  Table~\ref{tbl:mass} by a factor of 2.8 and 4 respectively (however,
  the uncorrected numbers are listed in the Table). Assuming the
  average outflow is tilted by 45$^\circ$, we can ignore any
  correction for the calculation of the dynamic age, t$_{\rm dyn}$,
  since the correction for distance in the numerator and velocity in
  the denominator in estimating t$_{\rm dyn}$ cancel out.

\begin{table*}
\centering

\begin{minipage}{180mm}
\caption{Outflow Mass, Velocity and
Energy estimates\label{tbl:mass}} 
\begin{tabular}{lllccccccc}
\hline
{S.No} & {Name} & {Lobe} & {$v_{\rm avg}$} &
{Mass} & {Momentum} &
{Energy} & {Length} & {$t_{\rm dyn}$} &
{L$_{\rm flow}$}\\
{} & {} & {} & {km s$^{-1}$} & 
{M$_{\odot}$} & {M$_{\odot}$km s$^{-1}$} & 
{$\times 10^{43}$ (ergs)} & {pc} & {$10^5$ yrs}
& {$\times 10^{30}$ (ergs/s)}\\
\hline
1. & 041149+294226 & Blue & 5.3 & 0.077 & 0.41 & 2.15 & 1.0 &
1.9 & 3.6\\
   &                & Red & 5.0 & 0.046 & 0.23 & 1.13 & 0.55 & 1.1 & 3.3\\

2. & IRAS 04113+2758 & Blue & - & - & - & - & - & - & - \\
   &                 & Red &  4.3 & 0.018 & 0.08 & 0.32 & 0.25 & 0.6
& 1.7\\

3. & IRAS 04166+2706 & Blue & 5.0 & 0.044 & 0.22 & 1.10 & 0.67 & 1.3
& 2.7\\
   &                 & Red  & 4.5 & 0.027 & 0.12 & 0.54 & 0.50 & 1.1
& 1.6\\

4. & IRAS 04169+2702 & Blue & 4.9 & 0.039 & 0.19 & 0.94 & 0.35 & 0.7
& 4.3\\
   &                 & Red  & 4.6 & 0.012 & 0.06 & 0.26 & 0.23 & 0.5
& 1.7\\

5. & FSTAU B         & Blue & 5.0 & 0.016 & 0.08 & 0.39 & 0.55 & 1.1
& 1.1\\
   &                 & Red  & 3.4 & 0.005 & 0.02 & 0.06 & 0.17 & 0.5
& 0.4\\

6. & IRAS 04239+2436 & Blue & -   &  -    & -   &  -   & - & - & - \\
   &                 & Red  & 3.3 & 0.094 & 0.31 & 1.02 & 1.24 & 3.7
& 0.9\\

7. & IRAS 04248+2612 & Blue & 5.4 & 0.016 & 0.08 & 0.45 & 0.64 & 1.2
& 1.2\\
   &                 & Red  & 3.4 & 0.14 & 0.47 & 1.58 & 0.54 & 1.6 &
3.1\\

8. & Haro 6-10       & Blue & 4.4 & 0.005 & 0.02 & 0.10 & 0.17 & 0.4
& 0.8\\
   &                 & Red  & 4.2 & 0.02  & 0.09 & 0.35 & 0.44 & 1.0
& 1.1\\

9. & ZZ Tau IRS      & Blue & -  & - & - & - & - & - & -\\
   &                 & Red  & 4.4 & 0.023 & 0.10 & 0.44 & 0.30 & 0.7
& 2.0\\

10.& Haro 6-13       & Blue & 3.7 & 0.050 & 0.18 & 0.67 & 0.71 & 1.9
& 1.1\\
   &                 & Red  & 4.6 & 0.163 & 0.74 & 3.36 & 0.55 & 1.2
& 8.9\\

11.& IRAS04325+2402 & Blue & -  & - & - & - & - & - & -\\
   &                 & Red  & 4.5 & 0.077 & 0.35 & 1.56 & 0.83 & 1.8
& 2.8\\

12.& HH 706 flow     & Blue & 4.4 & 0.087 & 0.38 & 1.67 & 0.88 & 2.0
& 2.7\\
   &                 & Red  & 4.6 & 0.112 & 0.52 & 2.36 & 0.48 & 1.0
& 7.5\\

13.& HH 705 flow     & Blue & 4.5 & 0.014 & 0.06 & 0.27 & 0.46 & 1.0
& 0.9\\
   &                 & Red  & 4.8 & 0.062 & 0.29 & 1.39 & 0.43 & 0.9
& 4.9\\

14.& IRAS04361+2547  & Blue & 4.4 & 0.012 & 0.05 & 0.23 & 0.20 & 0.5
& 1.5\\
   &                 & Red  & 4.7 & 0.006 & 0.03 & 0.12 & - & - & -\\

15.& TMC1A           & Blue & 4.5 & 0.002 & 0.01 & 0.04 & 0.07 & 0.2
& 0.6\\
   &                 & Red  & 4.7 & 0.004 & 0.02 & 0.09 & 0.07 & 0.1
& 2.9\\

16.& L1527           & Blue & 4.4 & 0.004 & 0.02 & 0.07 & 0.12 & 0.3
& 0.7\\
   &                 & Red  & 4.7 & 0.006 & 0.03 & 0.13 & 0.15 & 0.3
& 1.4\\

17.& IC2087 IR       & Blue & 4.5 & 0.001 & 0.01 & 0.02 & -   & - & -\\
   &                 & Red  & 4.7 & 0.135 & 0.63 & 2.94 & 0.65 & 1.4
& 6.7\\

18.& TMC-1 North     & Blue & 4.4 & 0.042 & 0.19 & 0.81 & 0.56 & 1.3
& 2.0\\
   &                 & Red  & 4.7 & 0.017 & 0.08 & 0.36 & -  & - & -\\

19.& TMC1            & Blue & 4.4 & 0.038 & 0.17 & 0.72 & 0.78 & 1.3
& 1.8\\
   &                 & Red  & 4.6 & 0.094 & 0.43 & 1.98 & 1.0 & 2.1 &
3.0\\

20.& 045312+265655   & Blue & 4.3  & 0.003 & 0.01 & 0.05  & 0.14 &
0.3 & 0.5 \\
   &                 & Red  & 4.5 & 0.007 & 0.03 & 0.13 & 0.21 & 0.5
& 0.8\\
\hline
\end{tabular}
\end{minipage}
\end{table*}

\section{Discussion}
\label{discussion}

\subsection{Parsec Scale Outflows in Taurus}
\label{parsec_scale}

A somewhat surprising result in the study of HH flows from young stars
using larger format CCD cameras in the 1990s was that many of these
flows could extend to many parsecs from the driving source
\citep[e.g.][]{bally1994, reipurth1998}. When large-scale millimetre
wavelength molecular line mapping is performed, many molecular
outflows were shown to extend to parsec scales, as well
\citep[e.g.][]{bence1996, wolfchase2000, arce2010}. However, such
millimetre wavelength mapping studies have been hampered by the large
amounts of observational time required to adequately complete the
projects. Hitherto, in most studies of molecular outflows from YSOs,
even for the non parsec-scale ones, mapping of the flow has been done
primarily in the main isotope of \co, with some opacity correction
applied using pointed observations of \coa.

With the advent of large-format heterodyne focal plane arrays like
SEQUOIA \citep{erickson1999}, it is now possible to make sensitive,
large spatial extent, high spatial and velocity resolution maps at
millimetre wavelengths that were hitherto not possible with finite
amounts of observing time \citep{ridge2006, narayanan2008}. With the
100 square degree area covered by the Taurus Molecular Cloud Survey,
it becomes possible to study the true extent of molecular outflows in
this region. Knowledge of the true extent of outflows will in turn
allow a more accurate assessment of the impact of outflows from YSOs
on feedback mechanisms in molecular clouds, the role that outflows
play in pumping and maintaining turbulence in these clouds.

Of the 20 outflows listed in Table~\ref{tbl:mass}, 8 outflows (40\%)
have a combined length of redshifted and blueshifted lobes that are
greater than 1 pc in length. These 8 parsec-scale outflows have an
average length of 1.37 pc. Four of these parsec-scale flows,
041159+294236, IRAS04248+2612, Haro 6-13, and HH 706 flow are new
outflow detections. Even when the outflow is previously known and studied,
our results indicate that they are considerably longer than previously
suspected.

For example, in IRAS 04166+2706 (see Figure~\ref{04166_outflow}), the
entire extent of the outflow previously studied in great detail by
\citet{santiagogarcia2009} is only a few arcminutes in length, and
their maps are confined to regions close to the driving source. In
this latter study, using high angular resolution data derived from
interferometer and single-dish data, the authors are able to
distinguish the detailed kinematic information for the oppositely
directed winds and the swept-up shells. But what is being missed in
the study by \citet{santiagogarcia2009} is the mammoth scale of the
04166 outflow much beyond the region that they studied.

Our results suggest that a more careful census of outflowing gas
using large-scale mapping studies such as done here will elucidate the
true scales and reaches of molecular outflows and their impacts on
GMCs.

\subsection{Statistics of Outflows and Young Stars in Taurus}

We have detected 20 outflow sources in the Taurus Molecular Cloud of
which 8 are new identifications. Given that our sensitivity limits
prevent the identification of at least 3 other sources (see
\S\ref{non_detections}), we can postulate that there are at least 23
outflow sources in Taurus. The Spitzer map of Taurus
\citep{rebull2010}, while not covering all of the mapped Taurus
Molecular Cloud, has 215 YSOs in the previously known candidate list,
and 148 new YSOs as newly identified candidates. In the
\citet{rebull2010} catalogue, YSOs are classified as Class I, Flat,
Class II or Class III based on the slope of the spectral energy
distribution (SED). Since Spitzer observations cannot distinguish
Class 0 objects from Class I (and the most embedded Class 0 objects
may not be detected by Spitzer), we can assume that the Class I
and Flat spectrum sources represent the most embedded and youngest
population of YSOs, with the Class I objects forming the younger
subset. Of the Spitzer detected YSOs, there are 48 Class I objects
and 33 Flat spectrum objects. There are thus 81 total embedded objects
in a list of 363 YSOs ($\sim 22$\%) in the Spitzer catalogue.

Of the 23 outflow sources in Taurus, 18 can be associated with known
Spitzer identifications, and their spectral classification can be
derived. Table~\ref{outflowlist} lists this SED class in the YSO class
column. The remaining 5 outflow sources are all new detections from
this study, and all but one (045312+265655), are in the regions
covered by Spitzer, so the Spitzer non-detection of the driving
sources might imply that these are Class 0 sources. The three
non-detected outflows of this study (see \S\ref{non_detections}) are
all classified as Class I sources. Of the 18 outflows with identified
driving sources, 14 are Class I objects, and 4 are Flat spectrum
objects. We conclude from this that $\sim 30$\% of Class I sources and
$\sim 12$\% of Flat spectrum sources in Taurus have
outflows. Recently, a comprehensive list of known Class 0 protostars
was compiled \citep{froebrich2005}, and the list is being actively
maintained as a Class 0 database
online\footnote{http://astro.kent.ac.uk/protostars/}. This list
contains five Class 0 protostars in the Taurus region surveyed: B213
(what they refer to as PS041943.00+271333.7, which is also
IRAS04166+2706), L1521-F IRS, IRAS 04248+2417 (HK Tau), IRAS
04325+2402 (L1535) and IRAS 04368+2557 (L1527). Of these five sources,
three of the objects, IRAS04166+2706, L1535 and L1527 are outflows
detected in this study. In the Spitzer classification of
\citet{rebull2010}, HK Tau (IRAS 04248+2417) is actually classified as
a Class II object, which is also confirmed by recent Akari
observations \citep{aikawa2012}. \citet{stapelfeldt1998} using Hubble
Space Telescope observations found HK Tau to be an edge-on disk, which
may explain its mis-classification as a Class 0 object in the online
database. So we drop HK Tau from the list of Class 0 objects in
Taurus. L1521-F IRS is known to be a very low-luminosity source which
is probably a very young Class 0 object \citep{terebey2009}. We detect
no outflow towards L1521-F. 
%Depending on whether or not we include the
%low-luminosity L1521-F IRS in the list of Taurus Class 0
%objects
We conclude that 75\% of known Class 0 objects in
Taurus have outflows.
%% If we just count the Class I sources, there are 48 such sources in
%% Taurus, so with the 23 outflows detected in Taurus hitherto, we have a
%% detection rate of $\sim 48$\% for outflows in embedded YSOs. It can be
%% seen from Table~\ref{outflowlist} that all of the detected outflows
%% (when the driving sources are associated as Spitzer candidates) are
%% Class I or Flat spectrum sources. From this, we conclude that outflows
%% are seen only in the earliest embedded phases of star-formation in
%% Taurus. 

But what could explain the non-detection of outflows towards 63 other
Spitzer-detected embedded Class I and Flat spectrum sources in Taurus?
It is possible that our sensitivity limits and our relatively coarse
angular resolution prevents the identification of small-scale, low
intensity flows. But it is more likely that these 63 embedded sources
without outflows represent an older population, where the swept-up
molecular outflows have slowed down to ambient cloud velocities, and
are hence not detectable in high-velocity wings. The mean dynamic age
of the outflows in our study is $1.1\times 10^5$ years, with the
maximum being $3.7\times 10^5$ years. When compared against the
average lifetime of about $5\times 10^5$ years for the Class I
protostellar phase \citep{evans2009}, the non-detection of outflows in
a large number of Class I and Flat spectrum sources in Taurus could
mean that molecular outflows are a short-lived phenomenon marking the
youngest phase of protostellar life.

\subsection{Turbulence in Molecular Cloud and Outflows as Injection
  Mechanism}
\label{turbulence}

%% The supersonic velocity dispersion seen ubiquitously in molecular
%% clouds has been shown to be due to turbulent motions using different
%% lines of evidence, the chief amongst which is the size-linewidth
%% relationship first described by \citet{larson1981}. While these first
%% studies applied to size-linewidth relations between different GMCs,
%% more recent studies \citep[e.g.][]{heyer2004} show the same
%% relationship holds even within individual GMCs. 
The sources of turbulent motions in molecular clouds have been
intensely debated over the past three decades \citep[see for e.g. from
][]{larson1981, heyer2004}. At larger scales, kinetic energy injection
from supernovae and galactic differential rotation can provide sources
of turbulent energy. At smaller scales, stellar feedback in the form
of HII regions, radiatively driven winds, and accretion-driven
outflows are believed to be the sources of turbulent energy. 
%% If
%% self-gravity is included, gravitational contraction from large scales
%% can itself produce velocity dispersions consistent with Larson's laws
%% \citep{field2008,field2011}. This represents an alternative way to
%% drive turbulent-like motions in molecular clouds. 
The question of what powers the turbulence is important. However,
questions on the so-called injection scale of turbulence, and whether
there is in fact more than one injection scale where energy is
deposited into the turbulent spectrum are equally important. Another
unanswered question is what sustains the turbulence in the parsec and
sub-parsec scale clumps. It is clear that the turbulence needs to be
driven continually over timescales longer than the crossing time
either internally using stellar feedback processes, or externally
using some cascade down process from the ISM.

A very different question from the origin of turbulent motions in
star-forming clouds is the role that turbulence provides during the
star-formation process. In the absence of turbulent support, most of
the mass within a given structure, be it a GMC, cloud or core, would
collapse into stars within one free-fall time with an efficiency per
free fall time, \eff\ approaching 1 \citep[see][for definition of
  \eff]{krumholz2005, mckee2007}. However, direct observations provide
very low values of \eff\ ranging from 0.01 to 0.1 in GMCs and
substructures \citep{krumholz2005, krumholz2007}. This low value of
\eff\ may be due to internal kinetic support provided by turbulent
motions against collapse. So outflows can play an important role by
providing {\em local} turbulent feedback in regulating star forming
efficiency, quite apart from the question of whether outflows are an
important contributor of the global energy budget of turbulence at the
molecular cloud level \citep{arce2010}.

\citet{arce2010} used the COMPLETE data obtained towards Perseus in
the \co\ and \coa\ transitions with the same angular resolution as
this study to gauge the effect of outflows on cloud turbulence,
feedback, star-formation efficiency, and the role
outflows play in the disruption of the parent clouds. This latter study
on Perseus can be gainfully compared against the outflow study
presented here in Taurus, and contrast the effect of outflows in these
two nearby star-forming clouds.

In order to estimate the cloud-wide contribution of outflows in Taurus
to turbulence, we can compare the total outflow energy to an estimate
of the Taurus Molecular Cloud's turbulent energy. Given that
star-forming clouds exhibit evidence for turbulence being maintained
in some way, a better method to assess the importance of outflows in
driving turbulence is to compare the total outflow energy {\em rate}
into the cloud, i.e. the total outflow luminosity, with the energy
{\em rate} needed to maintain the turbulence in the gas.  We can estimate
the outflow luminosity by dividing the outflow energy with its
corresponding dynamical timescale, $t_{\rm dyn}$ (from
Table~\ref{tbl:mass}). There is considerable uncertainty in the
determination of this dynamical timescale. To derive the dynamical
timescale of outflows, we need a good identification of the driving
source, and a measurement of the velocities of the shocks associated
with the outflow. Another method to determine dynamical ages, as used
in \citet{arce2010} is to simply use an average value between median
jet dynamical timescales of $3\times 10^3$ yrs and the average
lifetime of a typical Class I protostar stage of $\sim 0.5$ Myr
\citep{evans2009}. In our analysis, we choose to use the direct
estimate of the dynamical timescale derived from the molecular
outflow, recognising that there could be an uncertainty of order 2 or
so when accounting for the inclination correction.

In order to estimate the turbulent energy dissipation rate, we need an
estimate of the timescale for the dissipation of magnetohydrodynamic
(MHD) turbulence. This latter number has been theoretically estimated
by several authors, and is given in the numerical study of
\citet{maclow1999} as: $t_{\rm diss} \sim ({{3.9\kappa}\over {M_{\rm
      rms}}})t_{\rm ff}$, where $\kappa = \lambda_d/\lambda_J$, the
ratio of driving wavelength to the Jean's length of the region, and
M$_{\rm rms}$ is the Mach number of the turbulence, i.e. the ratio of
turbulence velocity dispersion to the sound speed. For the same
reasons specified in \citet{arce2010}, we assume $\kappa=1$ and
M$_{\rm rms} = 10$. Using the known expression for $t_{\rm ff}$, we get:
%The free-fall timescale $t_{\rm ff} =
%\sqrt{{3\pi}\over{32G\rho}}$, where $\rho$ is mass density. Using
%normal expressions for density, we derive $t_{\rm ff} =
%\pi\sqrt{{R_{\rm reg}^3}\over{8GM_{\rm reg}}}$, where M$_{\rm reg}$
%and R$_{\rm reg}$ are the mass and radius of the region under
%consideration. Substituting, we get 
\begin{equation}
t_{\rm diss} = 0.39\pi\sqrt{{R_{\rm reg}^3}\over{8GM_{\rm reg}}},
\end{equation}
as an expression of the dissipation timescale, where M$_{\rm reg}$ and
R$_{\rm reg}$ are the mass and radius of the region under
consideration respectively.

The H$_{\rm 2}$ mass of the entire Taurus region mapped in
\citet{narayanan2008} was estimated by \citet{pineda2010} to be
$1.5\times 10^4$ \Msun.  We adopt a linewidth of 2 \kms\ based on the
average \coa\ FWHM of typical regions of Taurus. The turbulent energy
of the cloud can then estimated using $E_{\rm turb} = (3/(16\ {\rm
  ln\ 2}))M_{\rm cloud}\Delta v^2$, which with the above values gives
$3.2 \times 10^{47}$ ergs. Summing up the energy from the detected
outflows in Table~\ref{tbl:mass} yields $3\times 10^{44}$ ergs. Even
if we scale up the energy by a factor of 4 (see explanation in
\S\ref{masses}), we see that the Taurus Molecular Cloud has $\sim 270$
times more turbulent energy than the kinetic energy of all outflows in
Taurus. Using an effective radius $R\sim 13.8$ pc for the entire 100
square region of Taurus, and using equation 1, the dissipation rate
for turbulence for the entire Taurus Molecular Cloud is $2.7\times
10^6$ years. The rate of turbulent dissipation is $L_{\rm turb} =
3.8\times 10^{33}$ ergs~s$^{-1}$. Summing up the outflow luminosities
from Table~\ref{tbl:mass} yields a net outflow luminosity, L$_{\rm
  flow}$ of $8\times 10^{31}$ ergs~s$^{-1}$. Multiplying the outflow
luminosity again by a factor of 4, we see that the turbulent energy
dissipation rate is a factor of 12 greater than the net luminosity of
all outflows in Taurus. This indicates that outflows by themselves
cannot account for and sustain all the turbulence in Taurus.

Given that several previously known outflows have not been detected in
this study (see \S\ref{non_detections}), it is worth asking if our
survey could be missing enough outflows to alter the energy deficit of
outflows versus turbulence in Taurus. \citet{bontemps1996} list an
upper limit of $\sim 0.16\times 10^{-5}$ \Msun\kms$yr^{-1}$ for the
outflows IRAS 04181+2655 and IRAS 04302+2247, both of which are missed
in our study. We can estimate from Table~\ref{tbl:mass} a momentum
flux of $\sim 0.1\times 10^{-5}$ \Msun\kms${\rm yr}^{-1}$ for our
weakest candidate outflow, 045312+265655. We estimate that even if we
missed 20 of these lower momentum flux outflows in Taurus in our
survey, the resultant energy is equivalent to one of our brighter
sources, so the missing sources clearly do not have enough energy to
tip the imbalance of turbulent energy and outflow energy discussed
above.

While most of the Taurus Molecular Cloud has been known to be a region
of poor star-formation efficiency, the exception is the L1551 dark
cloud region just south of the main complex, which is known to be an
active region of star-formation. The L1551 dark cloud contains at
least one class 0 protostar (L1551 NE), the prototypical class I
outflow source L1551 IRS5, \citep{snell1980}, several T Tauri (class
II) stars including HL/XZ Tau, and weak T Tauri (class III) stars
including UX Tau. The 100 square degree survey of the Taurus Molecular
Cloud in \citet{narayanan2008} did not cover the L1551 dark cloud, but
this region has been studied with comparable resolution but with
better sensitivity by \citet{stojimirovic2006}. From the latter study,
the mass of the L1551 dark cloud is 110 \Msun, and its total turbulent
energy is $\sim 8.5\times 10^{44}$ ergs. Adding up the energy from the
outflow lobes in L1551, gives a total outflow energy of $\sim 2\times
10^{45}$ ergs \citep{stojimirovic2006}. These estimates have not been
even been multiplied up by the factor of 10 used in the main Taurus
cloud above. So at least in the L1551 dark cloud the outflows have
more than enough energy to account for the turbulence present in the
parent cloud. Assuming a cloud size of 0.8 pc for the L1551 dark
cloud, equation 1 gives a turbulent dissipation timescale of
$4.4\times 10^5$ years. So the rate of turbulent dissipation in L1551,
L$_{\rm turb}$ is $6.1\times 10^{31}$
ergs~s$^{-1}$. \citet{stojimirovic2006} do not provide dynamical
time-scales for the L1551 outflows, but it can estimated from the data
to be $\sim 8\times 10^4$ years. Hence the overall luminosity from
outflows in L1551, L$_{\rm flow}$ is $8\times 10^{32}$ ergs~s$^{-1}$,
a factor of 13 bigger than L$_{\rm turb}$. Again, the more energetic
outflows in the L1551 dark cloud not only have the energy, but have
sufficient luminosity to keep pumping up the turbulence in the L1551
cloud.

%% If we add up the total turbulent energy and luminosities of the Taurus
%% main complex to that from L1551 dark cloud, we get an overall
%% turbulent energy of $4.66\times 10^{47}$ ergs, Similarly the sum of
%% all outflow energies in Taurus main complex and L1551 dark cloud gives
%% $2.3\times 10^{45}$ ergs. Again multiplying up the outflow energy by
%% the factor of 10 as above, we see that outflows in all of Taurus are
%% still a factor of $\sim 20$ away from accounting for the turbulent
%% energy present in the entire cloud.  However, adding the turbulent
%% dissipation rate for the main Taurus cloud and L1551, we get L$_{\rm
%%   turb_{tot}}\sim 5.5\times 10^{33}$ ergs~s$^{-1}$, and for the
%% outflows, L$_{\rm flow_{tot}}\sim 8.8\times 10^{32}$
%% ergs~s$^{-1}$. Multiplying the outflow luminosities by a factor of 10,
%% overall outflows in Taurus may be able to sustain turbulent
%% dissipation. 
While the current generation of outflows in the 100 square degrees we
surveyed have an overall outflow luminosity that is a factor of 12
smaller than the turbulent dissipation rate seen in that region, L1551
just south of the main complex gives a counter example of a region
where outflows can more than account for the turbulent energy
rate. There are clearly uncertainties in these estimates of energy and
luminosities, but we can conclude overall that at the scale of large
molecular clouds, outflows are an important source to keep the
turbulence sustained. From this study and others, it is clear that the
scale length for turbulence from outflows as a source is of the order
of a parsec or more. As shown in this study, outflows have a
relatively short time-scale, of order a few $10^5$ years, while
turbulent dissipation rates are a few $\sim 10^6$ years. We could then
suggest that multiple generations of star-formation episodes with its
short-lived vigorous outflow episodes could keep the turbulence
sustained in molecular clouds. It is also clear that just one very
energetic outflow like L1551 can provide a substantial source of
turbulent injection into the parent cloud. Is L1551 an unique outflow?
If it is not, in molecular clouds like Taurus, one such outflow like
L1551 going off once every $10^5$ years appears to be sufficient as a
source of the turbulence.

%% We conclude that while the
%% imbalance gets less when the very active outflow region in the L1551
%% dark cloud is added to the mix, overall, the kinetic energy in
%% protostellar outflows of Taurus cannot account for the present
%% repository of turbulent energy present in the system.

Figure~\ref{overview_figure} shows that outflows in Taurus, for the
most part, are localised in regions of high column density. Such
clustering of outflows suggests the possibility of studying the
balance of turbulent energy and luminosity against that provided by
outflows in smaller regions. We could study the {\em local} effect of
outflows on the turbulence present in their immediate environments. We
define four main regions with multiple outflows in each region: L1495,
B213, Heiles Cloud 2, and B18.  The polygonal regions for these 4
star-forming clouds are highlighted in \citet{goldsmith2008}, and the
masses and sizes of these individual regions are listed in
\citet{goldsmith2008, pineda2010}. To these four Taurus regions, we
add the L1551 dark cloud from the study of
\citet{stojimirovic2006}. The physical properties of these five
regions including outflow energy and luminosity are calculated and
summarised in Table~\ref{tbl:energies}. It should be emphasised that
it is indeed more appropriate to gauge the effect of outflows in
contributing to cloud turbulence by using the whole Taurus cloud
instead of selected regions. There are many more YSOs in Taurus than
there are outflows (see Figure~\ref{overview_figure}), so it is
possible that many of the outflows from more evolved YSOs are slowing
down to ambient cloud velocities at the current epoch (thereby
rendering them undetectable), but still have contributed to the energy
budget of the Taurus cloud. However, it is still instructive to
analyse individual star-forming regions as listed in
Table~\ref{tbl:energies} in order to gauge the local effect of
outflows in pumping cloud turbulence, in disrupting the parent cloud,
and in providing feedback that might reduce star-formation efficiency.
%In
%Table~\ref{tbl:energies}, we also list the outflow luminosity and
%luminosity due to turbulent energy. 

\begin{table*}
\centering
%\rotate
%\tablewidth{0pt} 
%\tabletypesize{\scriptsize} 
\caption{Physical Parameters of Star-forming Regions in Taurus
\label{tbl:energies}} 
\begin{minipage}{175mm}
\begin{tabular}{lcccccccccc}
\hline
{Name} & {M$_{\rm reg}$~$^{a}$}
& {R$_{\rm reg}$~$^{b}$} & {$\Delta v$~$^{c}$} & 
{v$_{\rm esc}$~$^{d}$} &
{E$_{\rm grav}$~$^{e}$} & {E$_{\rm turb}$~$^{f}$} 
& {t$_{\rm diss}$~$^{g}$} & 
{L$_{\rm turb}$~$^{h}$} 
& {E$_{\rm flow}$~$^{i}$} & {L$_{\rm flow}$~$^{j}$} \\ 
{} & {(\Msun)} &
{(pc)} & {(\kms)} & {(\kms)} &
{($10^{46}$ ergs)} & {($10^{46}$ ergs)} &
  {($10^5$ yr)} & {($10^{32}$ ergs~s$^{-1}$)} &
  {($10^{44}$ ergs)} & {($10^{32}$ ergs~s$^{-1}$)} \\
\hline
L1495$^{k}$ & 1836 & 3.2 & 1.6 & 2.2 & 8.9 & 2.5
& 8.7 & 9.3 & 0.36 & 0.09\\ 
B213$^{l}$ & 723 & 2.0 & 1.9 & 1.8 & 2.2 & 1.4 & 7.0 & 6.4 & 0.53 &
0.16 \\ 
Cloud 2$^{m}$ & 1303 & 2.3 & 1.9 & 2.2 & 6.3 & 2.5 & 6.4 & 12.6 & 1.3 & 0.37
\\ 
B18$^{n}$ & 828 & 2.1 & 1.9 & 1.8 & 2.5 & 1.6 & 8.0 &
6.4 & 0.75 & 0.18 \\ 
L1551$^{o}$ & 110 & 0.8 & 1.2 & 1.1 & 0.1 & 0.09 & 4.4 & 0.6 & 20.0 &
8.0 \\
\hline
\end{tabular}

(a) Mass of star-forming region as listed in \citet{pineda2010}; (b)
Radius estimate from using the area listed in \citet{pineda2010} and
approximating cloud as a circle; (c) Fitted FWHM of average
\coa\ spectrum in region; (d) Escape velocity given by $\sqrt{2GM_{\rm
    reg}/R_{\rm reg}}$; (e) Gravitational binding energy, $GM_{\rm
  reg}^2/R_{\rm reg}$; (f) Energy in turbulence, given by
${{3}\over{16ln2}}M_{\rm reg}\Delta v^2$; (g) Turbulent dissipation
time, given by $0.39\pi\sqrt{{{R_{\rm reg}^3}\over{8GM_{\rm reg}}}}$
(see \S\ref{turbulence}); (h) Turbulent energy dissipation rate, given
by $E_{\rm turb}/t_{\rm diss}$; (i) Sum of energies of lobes of
outflows in region; (j) Sum of all Outflow luminosities, where each
outflow's luminosity is derived from $E_{\rm flow}/t_{\rm dyn}$, where
$t_{\rm dyn}$ is the dynamice timescale for the flow listed in
Table~\ref{tbl:mass}; (k) Outflows in L1495 are IRAS 041159+294236 and
IRAS 04113+2758; (l) Outflows in B213 are IRAS 04166+2706, IRAS
04169+2702, FSTAU B, and IRAS 04248+2612; (m) Outflows in Cloud 2,
also called Heiles Cloud 2 are HH 706 flow, HH 705 flow, TMR1, L1527,
IC 2087, TMC1A, TMC-1 North, and TMC1; (n) Outflows in B18 are IRAS
04239+2436, Haro 6-10, ZZ Tau, Haro 6-13, and IRAS 04325+2402; (o)
From the outflows from L1551 IRS, L1551 NE, and the EW jet from data
in the L1551 dark cloud presented in \citet{stojimirovic2006}.
\end{minipage}
\end{table*}

In Table~\ref{tbl:energies}, we also list the escape velocity,
gravitational binding energy in the five regions. The outflows
contributing to the outflow energies in each region are also
listed. It is to be remembered that the outflow energy and outflow
luminosity listed in Table~\ref{tbl:energies} are lower estimates, and
will need to be scaled up by a factor of $\sim 10$. In
Table~\ref{tbl:outflowimpact}, the ratios of outflow energy to
turbulent energy and gravitational energy, as well as ratio of outflow
luminosity to turbulent dissipation rate are presented (again, the
numbers in this table are raw numbers, and need to be multiplied up by
a factor of 10 as in Table~\ref{tbl:energies}). It can be seen that in
both Table~\ref{tbl:energies} and \ref{tbl:outflowimpact}, the L1551
dark cloud is very different from any other region in Taurus. The
outflows in the L1551 dark cloud are highly energetic, and are
disrupting the cloud core, not just pumping up turbulence in its
parent cloud. For the other regions in Taurus, it can be seen that
outflows provide only a fraction of the necessary energy present in
turbulence even in these selected regions that have multiple
outflows. However, we do see that the luminosity of outflows is a
significant contributor (greater than 25\% in B213, Cloud 2 and B18,
after multiplying the ratios by the canonical factor of 10) to the
expected turbulent dissipation rate. We also see from the ratio of
outflow energy to the gravitational binding energy of each region,
that other than in L1551, outflows do not have enough energy to
disrupt the cloud.

\begin{table*}
\centering
\begin{minipage}{90mm}
\caption{Impact of Outflows on Star-Forming Regions in Taurus
\label{tbl:outflowimpact}}
\begin{tabular}{lccc}
\hline
{Name} & {E$_{\rm flow}$/E$_{\rm turb}$} &
{L$_{\rm flow}$/L$_{\rm turb}$} & {E$_{\rm flow}$/E$_{\rm grav}$} \\
\hline
L1495 & 0.001 & 0.01 & 0.0004\\
B213 & 0.004 & 0.025 & 0.002 \\
Cloud 2 & 0.005 & 0.029 & 0.002\\
B18 & 0.005 & 0.028 & 0.003\\
L1551 & 2.35 & 13.1 & 2.0\\
\hline
\end{tabular}
\end{minipage}
\end{table*}

%% \subsection{Sensitivity Limits of This Survey}
%% \subsection{Comparison with Spitzer Data}
%% \subsection{Comparison with HARP Data}
%% \label{harp}

\section{Conclusions}
\label{conclusions}

An unbiased study of the entire 100 square degrees of the \co\ and
\coa\ data in the FCRAO Taurus Molecular Cloud Survey was performed in
order to identify high-velocity features that could be associated with
molecular outflows from YSOs. The FCRAO survey of the Taurus Molecular
Cloud was not designed to exhaustively detect all the outflows in
Taurus, but the sensitivity reached in \co\ and \coa\ in the survey
were better than the original goals of the project, and allows an
unbiased search for outflows in this nearby star-forming region. Our
procedure for identifying outflows in an unbiased way utilises a
combination of integrated intensity maps, position velocity images,
average spectra inside polygonal areas representing presumed outflow
regions.

Using our search strategy we identify 20 outflows in Taurus, of which
8 are new detections. Our survey fails to detect three other outflows
in the region that have been previously reported and confirmed in
literature. The weak nature of these three outflows as reported in
previous studies are consistent with them not being detected to the
limits of the sensitivity reached in our survey. Eight of these 20
(40\%) outflows are parsec-scale in length, and of them four are new
detections. Even amongst the known outflows that are seen to be
parsec-scale in our survey, the outflow lengths are much larger than
previously suspected.

We detect outflows in 30\% of Class I sources, and 12\% of Flat
Spectrum sources from the subset of embedded YSOs in the Taurus
Spitzer catalogue. Five of the outflows reported in this study have
driving sources which have no known counterparts in the Spitzer
catalogue, indicating that they are likely Class 0 objects. Our study
detects outflows in 75\% of known Class 0 objects in Taurus. Based on
dynamical timescales derived for our outflows, and non-detection of
outflows towards a large number of Spitzer Class I and Flat Spectrum
sources, we conclude that outflows are a very short-lived phase in
protostellar evolution, and that most embedded YSOs in Taurus are past
this stage of their lives.

We compare the combined energetics of the detected outflows to the
observed cloud-wide turbulence in Taurus, and conclude that in the
main 100 square degree region of Taurus covered in the survey,
outflows lack the energy needed to feed the observed turbulence. But
if we include the very active L1551 star-forming region that is just
south of the main Taurus complex, which also features some very
powerful outflows, we determine that energy from outflows are only a
factor of 20 lower than the energy present in turbulence. However,
when comparing the net luminosity of outflows from the Taurus region
studied and include the L1551 dark cloud, the luminosity in the
outflows is able to sustain the turbulent dissipation rate seen in
Taurus. The energetics of smaller sub-regions of Taurus and the L1551
dark cloud region are compared to the repository of turbulent and
gravitational energies in these sites. Our comparison region, the
L1551 dark cloud is anomalous in that the outflows in that region are
powerful enough not only to easily account for the turbulence in that
cloud, but it's outflows are also unbinding the cloud. The regions
with active outflows in the 100 square degrees of Taurus that we
studied do not have enough energy to disrupt the cloud, but do have
enough luminosity to be a major player in sustaining the turbulence in
their parent clouds. We also conclude that for molecular clouds like
Taurus, an L1551-like outflow episode occurring once every $10^5$ years
is sufficient to sustain the observed turbulence in the cloud.

%\acknowledgments 
\section*{Acknowledgements}

We thank Amy Langford for her help in analysing the data and
identifying the first batch of outflows towards Taurus. This work was
supported by NSF grant AST-0838222 to the Five College Radio Astronomy
Observatory.

%Facilities: \facility{FCRAO}

%\nocite{*}
%\bibliographystyle{apj}
%\bibliography{outflowpaper}

\clearpage

\footnotesize{
  \bibliographystyle{mn2e}
  \bibliography{outflowpaper}
}
\label{lastpage}
\end{document}